\documentclass[
prd,
floatfix,
aps,
superscriptaddress,
showpacs,
nofootinbib,
amsfonts,
amsmath,
tightenlines,
twocolumn
]{revtex4-1}
\usepackage{bm,graphics,graphicx,epsfig,color,epstopdf
}

\begin{document}

\def\gsim{\mathrel{
\rlap{\raise 0.511ex \hbox{$>$}}{\lower 0.511ex
\hbox{$\sim$}}}}
\def\lsim{\mathrel{
\rlap{\raise 0.511ex \hbox{$<$}}{\lower 0.511ex
\hbox{$\sim$}}}}

\newcommand{\Cardiff}{School of Physics and Astronomy, Cardiff University, Queens Building, CF24 3AA, Cardiff, United Kingdom}
\newcommand{\UIB}{Departament de F\'isica, Universitat de les Illes Balears, 
Crta. Valldemossa km 7.5, E-07122 Palma, Spain}

\title{Black-hole hair loss: learning about binary progenitors from ringdown signals}

\author{Ioannis Kamaretsos}
\affiliation{\Cardiff}

\author{Mark Hannam}
\affiliation{\Cardiff}

\author{Sascha Husa}
\affiliation{\UIB}

\author{B.S.~Sathyaprakash}
\affiliation{\Cardiff}

\begin{abstract}
Perturbed Kerr black holes emit gravitational radiation, which (for 
the practical purposes of gravitational-wave astronomy) consists 
of a superposition of damped sinusoids termed quasi-normal modes. 
The frequencies and time-constants of the modes depend only on
the mass and spin of the black hole --- a consequence of the
{\em no-hair} theorem. It has been proposed that a measurement of two 
or more quasi-normal modes could be used to confirm that the source
is a black hole and to test if general relativity continues to hold
in ultra-strong gravitational fields.  In this paper we propose a 
practical approach to testing general relativity with quasi-normal modes. 
We will also argue 
that the relative amplitudes of the various quasi-normal modes encode 
important information about the origin of the perturbation that caused 
them. This helps in inferring the nature of the perturbation from an 
observation of the emitted quasi-normal modes. In particular, we will 
show that the relative amplitudes of the different quasi-normal modes 
emitted in the process of the merger of a pair of nonspinning black 
holes can be used to measure the component masses of the progenitor binary.
\end{abstract}
\pacs{04.30.Db, 04.25.Nx, 04.80.Nn, 95.55.Ym}
\maketitle

\section{Introduction}
The black hole no-hair theorem states that a charged, 
stationary and axially symmetric black hole with an event horizon of spherical 
topology can be described by the Kerr-Newman geometry \cite{Kerr:1963,Newman:1965}, 
where it is characterized by just three quantities -- its mass, spin and 
electric charge \cite{Israel:1967,Israel:1968,Carter:1971,Hawking:1972,Robinson:1975}. 
Astrophysical black holes are believed not to have any charge and so
are described by the Kerr geometry \cite{Kerr:1963}, characterized by just 
their mass and spin angular momentum. An important consequence of the 
no-hair theorem concerns the behaviour of Kerr
black holes when subjected to an external perturbation. There is strong
evidence that black holes are stable against such perturbations
\cite{Vishu:1970a,ReggeWheeler:1957,Vishu:1970b,Press:1971,Teukolsky:1972my,Teukolsky:1973ha,Press:1973zz} 
(see, however, Ref.~\cite{Whiting:1989}).
Perturbed black holes regain their axisymmetric configuration by emitting
gravitational radiation. The radiation observed by a detector takes the form
\begin{equation}
h(t) = \sum_{\ell,m,n} A_{\ell m n} e^{-t/\tau_{\ell m n}}\, 
\cos\left(\omega_{\ell m n}t+\phi_{\ell  m  n} \right ),
\label{eq:infinitesum}
\end{equation}
and it consists of a superposition
of quasi-normal modes (QNMs) with characteristic mode frequencies $\omega_{\ell mn}$
and time-constants $\tau_{\ell mn}.$ Here $\ell =2,\ldots,$ and $m=-\ell ,\ldots,\ell ,$ are 
the spheroidal harmonic indices and $n$ is an index corresponding to the overtones 
of each mode.
The amplitudes $A_{\ell mn}$ depend on the relative orientation of
the detector and the black hole as well as the nature of the perturbing agent and
$\phi_{\ell mn}$ are constants defining the initial phase of the various modes\footnote{For
a recent review on black hole quasi-normal modes see Ref.\,\cite{Berti:2009kk}}.
One of the goals of this paper is to determine the amplitudes of the most significant
modes excited during the merger of a pair of black holes.

The mode frequencies and time-constants of a black hole of mass $M$
are given by the general expressions
\begin{equation}
\omega_{\ell m n} = \frac{F_{\ell m n}(j)}{M},\quad
\tau_{\ell m n} = M G_{\ell m n}(j) 
\end{equation}
where $F_{\ell m n}(j)$ and $G_{\ell m n}(j)$ are functions of the dimensionless black 
hole spin magnitude, or Kerr parameter, $j$.  
All mode frequencies and time-constants then depend only on the mass $M$ and the 
spin magnitude $j$ of the black hole and no other parameter -- a consequence of the 
no-hair theorem. Several authors have noted that this aspect of the no-hair theorem could be used to test 
if massive compact objects at galactic cores are actually rotating black holes 
described by the Kerr metric of general relativity \cite{BHspect04,BCW05,Berti:2007a}; 
alternatively, it could be used as a strong field test of general relativity itself
\cite{BHspect04}. 

The key idea behind the proposed tests is the following: If one can reliably 
decompose the observed gravitational radiation from a ringing black hole into 
a superposition of different modes, then the frequencies and time-constants of 
each of the modes could be used to infer the mass and spin of the black hole. 
If the object is truly a black hole, then the masses and spins obtained
from the different modes should all be consistent within the measurement
errors. Inconsistencies in the values of the masses and spins inferred from different
modes would be an indication of the failure of general relativity or that the
radiation was emitted from an object that is not a black hole. If a merging
binary does not lead to a black hole then the inspiral phase may not
result in a superposition of QNMs that can be characterized by just two parameters.
Such signatures could be inferred by a model-independent analysis of the data,
e.g., a time-frequency transform, or by assessing the posterior probability
for alternative models.

In this paper we estimate the relative amplitudes $A_{\ell mn}$ of the various modes 
by fitting a superposition of QNMs to the radiation emitted by a merging
black-hole binary obtained from numerical-relativity simulations. These 
simulations involve the `simple' case of \emph{initially non-spinning} black holes in 
quasi-circular orbits and different mass ratios of the binary, in the range 1:1 to
1:11.  Analytical fits of the amplitudes so obtained are extrapolated 
to mass ratios of up to 25, so as to study a variety of different systems.
The validity of our extrapolation can only be confirmed by future numerical 
simulations of binary black holes with such large mass ratios.

From the analytical fits we construct a model waveform and calculate the 
signal-to-noise ratios (SNRs) in different modes during the merger of 
supermassive black-hole binaries of total mass in the range 
$\sim 10^6$-$10^8\,M_\odot$ observed with the Laser Interferometer Space 
Antenna (LISA)\footnote{Recently, the National Aeronautics an Space 
Administration in the United States opted out of the LISA mission. However, 
European Space Agency is pursuing an alternative that is similar in scope 
to LISA and we believe studying what science LISA could deliver is still 
very relevant.} \cite{Danzmann97} and of intermediate-mass black-hole 
binaries  of total mass in the range $\sim 100$-$10^3\,M_\odot$ observed with 
the Einstein Telescope (ET) \cite{Punturo:2010zz} and advanced configuration
of the Laser Interferometer Gravitational-Wave Observatory (aLIGO)
\cite{shoemaker2009,Smith2009,0264-9381-27-8-084006,advLIGO:2007}. The response of a 
gravitational-wave detector is, of course, not separately sensitive to 
the different modes but to only their superposition. However, it should
be possible to measure the relative strengths of the different modes by
fitting a generic model to the observed data. We will pursue this latter
approach in a forthcoming publication and restrict ourselves to a theoretical
study of the relative importance of different modes.

We find that over most of the parameter space explored, the modes with indices 
$(\ell,m,n)=(2,2,0), (3,3,0), (2,1,0)$ and $(4,4,0)$ have SNRs for the ringdown
phase larger than 500 in 
LISA provided the source is within a red-shift of $z=1$ and larger than 50 in 
ET provided the source is within a distance of $1\,\rm Gpc.$ For aLIGO, the SNRs
in $(2,2,0)$ and $(3,3,0)$ modes are larger than 10 in a significant
region of the parameter space. However, other sub-dominant modes will not be visible in 
aLIGO when the source is at a distance of $1\,\rm Gpc$ or greater. In all cases black 
hole ringdown signals that result from equal-mass binaries can have far larger SNRs.
The distance reach of LISA and ET to ringdown signals is large enough that
one can expect a few events per year with quite a large ($\gsim 100$) SNR 
\cite{ALISA06,AmaroSeoane:2009ui}.
One can, therefore, expect that future observations of black hole mergers 
will provide an excellent opportunity to test GR using several different QNMs. 

We will present a specific implementation of the test of the no-hair theorem and discuss
a {\em minimal} and a {\em maximal} set of parameters that could be used to carry out such a test.
The {\em chief} result of this paper is that the relative amplitudes of 
the modes depend on the mass ratio $q$ of the progenitor binary and that by 
measuring the relative amplitudes, in addition to the frequency and time-constant, 
it should be possible to measure the component masses of the binary that led to the 
QNMs. 

The rest of this paper is organized as follows.  Section \ref{sec:NR} is devoted 
to a discussion of the numerical relativity simulations used in this work, focusing
on their accuracy, so as to give an idea of how reliable are our estimates of 
the relative amplitudes of different modes.  The waveform 
model used in this study is given in Section \ref{sec:WF}, stating the conventions 
and assumptions made in constructing the model.  Section \ref{sec:AMP} constructs 
the amplitudes of the various modes in the ringdown signal using numerical 
simulations of Section \ref{sec:NR}. Section \ref{sec:AMP}
also deals with different options for identifying the 
ringdown phase, the method that was actually followed, the connection 
between the mode amplitudes and the mass ratio of the binary from which the black 
hole results and how this information was included in the signal model. 
In Section \ref{sec:SNR}, we discuss the detectability of the various modes 
with aLIGO, ET and LISA, and possible astrophysical information we can glean from such 
observations.  Sections \ref{sec:PE1} and \ref{sec:PE2} present the results from a
covariance matrix analysis of how well we are able to measure the parameters
of the ringing black hole and the progenitor binary.  In Section \ref{sec:GRTEST}, 
we propose a practical test of the no-hair theorem, making several remarks on which 
modes and parameters we could use for such a test. We make concluding remarks and outlook
for further work in Section \ref{sec:CON}. We use a system of units in which
the Newton's constant and the speed of light are both set to unity, $c=G=1.$

\section{Numerical simulations of merger and ringdown signals}
\label{sec:NR}

In this section, we shall briefly discuss how the numerical simulations were performed.
We used the BAM code~\cite{Brugmann:2008zz,Husa:2007hp}. 
The code starts with black-hole-binary puncture initial data 
\cite{Brandt:1997tf,Bowen:1980yu} generated using a pseudo-spectral 
elliptic solver~\cite{Ansorg:2004ds}, and evolves them with the 
$\chi$-variant of the
moving-puncture \cite{Campanelli:2005dd,Baker:2005vv,Hannam:2006vv}
version of the BSSN
\cite{Shibata:1995we,Baumgarte:1998te} formulation of the 3+1 Einstein 
evolution equations. We estimate initial momenta for low-eccentricity inspiral
using the post-Newtonian methods outlined in~\cite{Husa:2007rh,Hannam:2010ec}. 
Spatial finite-difference derivatives are
sixth-order accurate in the bulk \cite{Husa:2007hp}, Kreiss-Oliger
dissipation terms converge at fifth order and a fourth-order Runge-Kutta
algorithm is used for the time evolution. Time interpolation in the 
Berger-Oliger-like adaptive
mesh refinement algorithm converges at second order accuracy. 
In the limit of infinite resolution the code is thus expected to 
converge with second order accuracy.  However, in the regime of 
currently feasible simulations, the spatial finite differencing 
error dominates by far. Artificial dissipation has no measurable 
effect on the phase accuracy of the waves. For well-resolved 
simulations we thus find sixth-order accuracy, as expected.

The gravitational waves emitted by the binary are calculated from the
Newman-Penrose scalar $\Psi_4$, extracted at a distance $D_{\rm L}$ 
from the source. The details of our implementation of
this procedure are given in Ref.~\cite{Brugmann:2008zz}. Here we recall
those details that are important to this paper.
The quantity $D_{\rm L} \Psi_4$ is decomposed into spin-weighted spherical
harmonics, and related to the GW strain as \begin{eqnarray}
D_{\rm L} \Psi_4 & = & D_{\rm L} (\ddot{h}_+ - i \ddot{h}_\times ) \\
& = & \sum_{\ell,m} \Psi_{4,\ell m}\, {_{-2}Y^{\ell m}} (\iota, \phi).
\end{eqnarray} To calculate the radiated power, or luminosity, we require $\dot{h}$, 
which can be obtained by one time-integration of the spherical-harmonic coefficients 
$\Psi_{4,\ell m}.$ In principle we need to fix only one constant of integration to
obtain $\dot{h}$ from $\Psi_4$, but in practice the integration is contaminated by
numerical noise, see e.g.,~\cite{Berti:2007b,Damour:2008te,Reisswig:2010di}. As part
of our analysis of the data, we produced both $h$ and $\dot{h}$, by (a) fixing the
integration constants to ensure that $h$ oscillates around zero, and rings down to 
zero, (b) removing low-and high-frequency noise via FFTs, and (c) removing further
spurious noise effects by subtracting low-order polynomial fits through the strain.

We use results from simulations of non-spinning binaries with mass ratios $q = \{1,2,3,4\}$
that were previously presented in Refs.~\cite{Hannam:2007ik, Hannam:2010ec} and an additional 
simulation of a $q=11$ binary that was carried out as a part of this study. 

Detailed error analyses for the first four simulations were presented in 
Refs.~\cite{Hannam:2007ik, Hannam:2010ec}, although those works focused on the inspiral phase,
while here we focus on the ringdown phase. We find that for the $q = \{1,2,3,4\}$-simulations,
the error in the amplitude of the ringdown signal
is dominated by the error due to wave extraction at a finite distance from the source.
The wave extraction was performed at $R_{\rm ex} = 70M$ for the data used in this paper.
Waves were extracted at larger radii ($80M$ and $90M$), but the numerical resolution 
was lower at these radii and numerical errors began to dominate the uncertainty of the
ringdown waveform in the subdominant modes.
We also found that the \emph{ratios} of the waveform 
luminosities were remarkably robust with respect to the wave-extraction radius, and for
these the errors are even lower; a similar effect was found in Ref.~\cite{Babak:2008bu}, where
the amplitude ratios between harmonics also played a major role in the study.

The new $q=11$ simulation included only two orbits of inspiral before merger, and was
produced primarily to calculate the ringdown signal. The sizes of the mesh-refinement levels
were varied to optimize both memory usage and numerical accuracy of the wave extraction,
which was now performed at $R_{\rm ex} = 100\,M$. 
Three simulations were performed to validate the accuracy of the results; the resolutions at the
wave extraction sphere were $\{0.533,0.427,0.356\}/M$, and the finest resolution at each black
aim to capture the inspiral GW phase with high accuracy; our focus was on the accuracy of
the ringdown. Here the amplitude accuracy of all the modes we consider in this paper was within
0.5\%, and this uncertainty was dominated by the error due to extraction at a finite distance from the
source. 

The spin of the final black hole is $0.25\pm0.01$, and $(0.3\pm 0.01)$\% of the energy of 
the system is 
radiated during the last two orbits and merger and ringdown. The final black hole recoils
by $55\pm5$\,km/s. The large uncertainty in the recoil is due mostly to having the waveform from 
only a small number of orbits before merger, which makes it difficult to remove the oscillatory 
inspiral recoil from the results, as was done in Ref.~\cite{Hannam:2010ec}. Our results for this system
are consistent with previous simulations of $q=10$ 
binaries~\cite{Gonzalez:2008bi,Lousto:2010tb,Lousto:2010qx}.

\section{Antenna response to a ringdown signal}
\label{sec:WF}
Quasi-normal modes are transients that live for a very short duration in
the detector band: In the case of intermediate-mass black holes that could
be observed in ET, the time constant is at best about 60 ms (for a BH of mass 
$10^3\,M_\odot$ and spin $j=0.7$), while for supermassive 
black holes that could be observed in LISA the longest time constant is about
100 min (for a BH of mass $10^8\,M_\odot$ and spin $j=0.7$) (see Table
\ref{tab:mode frequencies}).
Consequently, it is not necessary to consider 
the motion of LISA or ET during the observation of a quasi-normal mode, at 
least not in the current evaluation of what science one might extract from 
their observation. 

Let us consider the response of an interferometric detector
to a ringdown signal. We assume that the radiation is
incident from a direction $(\theta,\,\varphi)$ with respect to, say, 
a geo-centric coordinate system. Let $({\mathbf e}^R_{x},\,
{\mathbf e}^R_{y},\,{\mathbf e}^R_{z})$ be a set 
of orthonormal vectors representing a coordinate frame in which the 
ringdown modes take the transverse-traceless form; that is, the
metric perturbation $h_{ij}$ due to the ringdown modes can be written in
this frame as
\begin{equation}
h_{ij} = h_+\, e_+^{ij}  + h_\times\, e_\times^{ij}
\end{equation}
where $h_+$ and $h_\times$ are the plus and cross polarizations, whose
explicit expressions for a ringdown signal will be given below,
and ${\mathbf e}_{+,\times}$ are the polarization tensors given by 
\begin{eqnarray}
{\mathbf e}_+ & = & {\mathbf e}^R_x \otimes {\mathbf e}^R_x 
- {\mathbf e}^R_y \otimes {\mathbf e}^R_y, \\
{\mathbf e}_\times & = & {\mathbf e}^R_x \otimes {\mathbf e}^R_y 
+ {\mathbf e}^R_y \otimes {\mathbf e}^R_x.
\end{eqnarray}
The response $h^A(t)$ of an interferometer, labelled by $A,$ can be written as:
\begin{equation}
h^{A}(t) = F^{A}_+(\theta,\varphi,\psi) \, h_+(t) + 
F^{A}_\times(\theta,\varphi,\psi) \, h_\times(t). 
\label{eq:response}
\end{equation}
Here $\psi$ is the polarization angle, 
\begin{equation}
\cos\psi = {\mathbf e}_\theta \cdot {\mathbf e}^R_x,
\end{equation}
and $F^{A}_{+,\times}(\theta,\varphi,\psi)$ are the antenna pattern functions
of the detector given by,
\begin{equation}
F^{A}_+ = {\mathbf D}^A_{ij} {\mathbf e}_+^{ij}, \quad  F^{A}_\times = {\mathbf D}^A_{ij} {\mathbf e}_\times^{ij}, 
\end{equation}
where 
${\mathbf D}^A$ is the detector tensor. If ${\mathbf e}^A_{1,2}$ are unit 
vectors (not necessarily orthogonal to each other) along the two arms of an 
interferometer, then the detector tensor is given by 
\begin{equation}
{\mathbf D}^A = {\mathbf e}^A_1 \otimes {\mathbf e}^A_1 
- {\mathbf e}^A_2 \otimes {\mathbf e}^A_2.
\end{equation}

For our purposes it is most useful to express the radiation from a source 
in the source frame, 
in terms of its expansion in spin-weighted spherical harmonics of weight
$-2$, namely $_{-2}Y^{\ell m}$:
\begin{equation}
h_+-ih_\times = \sum_{\ell=2}^\infty \sum_{m=-\ell}^\ell h^{\ell m}\, 
_{-2}Y^{\ell m}(\iota,\,\phi).
\label{eq:projection}
\end{equation}
Here, $(\iota,\,\phi)$ refer to the co-latitude and the azimuth angle 
at which the radiation is emitted from the source; $\iota$
is also the angle between the line-of-sight and
the orbital (spin) angular momentum of the binary (black hole). The complex 
coefficients $h^{\ell m}$ in the expansion are referred to as $\ell m$
modes. Explicit expressions for the first few modes in the post-Newtonian
approximation for the inspiral phase of a binary's evolution can be found in 
Kidder \cite{Kidder:2008}. It is useful to write the modes explicitly in
terms of their real and imaginary parts
\begin{equation}
h^{\ell m}=A_{\ell m} e^{-i\Phi_{\ell m}} = h_+^{\ell m} - i h_\times^{\ell m}.
\end{equation}
This helps in extracting the amplitude and phase of each mode in terms of its 
plus and cross modes obtained in numerical simulations:
\begin{equation}
A_{\ell m} = \sqrt{(h_+^{\ell m})^2 + (h_\times^{\ell m})^2},\quad 
\Phi_{\ell m}=\tan^{-1} \left [ -\frac{h_\times^{\ell m}}{h_+^{\ell m}}\right].
\label{eq:real and imaginary}
\end{equation}
Noting that $ _{-2}Y^{\ell m}(\iota,\,\phi) =\, _{-2}Y^{\ell m}(\iota,\,0)\,  e^{i m \phi} $, 
we can rearrange the sums in Eq.\;(\ref{eq:projection}) using 
Eq.\;(\ref{eq:real and imaginary}) to get \cite{Berti:2007b}
\begin{eqnarray}
h_+ & = & \sum_{\ell,m>0} A_{\ell m}\, Y^{\ell m}_+\,\cos(\Phi_{\ell m} - m\phi), \label{eq:hplus}\\
h_\times & = & -\sum_{\ell,m>0}
A_{\ell m} Y^{\ell m}_\times\,\sin(\Phi_{\ell m} - m\phi), \label{eq:hcross}
\end{eqnarray}
where we have dropped the ``memory-effect'' $m=0$ terms, for which the amplitude is low
(see, e.g., a recent numerical study \cite{Pollney:2010hs}).
Note that while these modes are non-oscillatory during inspiral, 
they do exhibit ringdown, which has been studied in some detail 
with numerical codes in axial symmetry, where they are the only 
non-zero modes (see, e.g., \cite{1997PhRvD..55..829B}).
In the above expressions, the angular functions $Y^{\ell m}_{+,\times}(\iota)$ are 
the following combinations of the spin-weighted spherical harmonics:
\begin{eqnarray}
Y^{\ell m}_+(\iota) & \equiv & {_{-2}Y^{\ell m}}(\iota,0) + (-1)^\ell\,  {_{-2}Y^{\ell -m}}(\iota,0),\nonumber\\
Y^{\ell m}_\times(\iota) & \equiv & {_{-2}Y^{\ell m}}(\iota,0) - (-1)^\ell\,  {_{-2}Y^{\ell -m}}(\iota,0).
\end{eqnarray}

For the inspiral phase of a binary when the two compact bodies are widely separated,
post-Newtonian (PN) approximation gives the amplitudes $A_{\ell m}(t)$ and phases $\Phi_{\ell m}(t)$
as expansions in $v/c,$ where $v$ is the velocity of the bodies (see, e.g., \cite{Kidder:2008}).
Numerical relativity simulations can be used to extract them when the PN approximation
breaks down. In the case of perturbed black holes, which a binary will result in,
black hole perturbation theory predicts that the modes are damped sinusoids with their
amplitudes and phases given by:
\begin{equation}
A_{\ell m}=\frac{\alpha_{\ell m}\,M}{D_{\rm L}}\, e^{-{t}/{\tau_{\ell m}}}, \quad \Phi_{\ell m}(t) =\omega_{\ell m}\, t,
\end{equation}
where $M$ is the mass of the black hole and $D_{\rm L}$ is its luminosity distance from Earth. Time-constants $\tau_{\ell m}$
and frequencies $\omega_{\ell m}$ can be computed from black hole perturbation theory
(see, e.g., Ref. \cite{BCW05} for a recent comprehensive listing of frequencies
and time-constants).  However, the amplitudes $\alpha_{\ell m}$ depend on the nature of 
the perturbation and are not analytically accessible in the case of black holes that form
from the coalescence of a binary. We shall ``measure" them later in this paper using 
results of our numerical simulations.

Using the above equations, the output of the numerical simulations for the plus and cross polarizations 
corresponds to the following expressions:
\begin{eqnarray}
h_{+}(t) & = &
\sum_{\ell ,m>0} \frac{\alpha_{\ell m}\,M}{D_{\rm L}}\, Y_{+}^{\ell m}(\iota)\, e^{-{t}/{\tau_{\ell m}}}\, 
\cos\left ( \omega_{\ell m} t - m \phi \right ), \nonumber \\
h_\times(t) & = &
\sum_{\ell ,m>0} \frac{\alpha_{\ell m}\,M}{D_{\rm L}}\, Y_{\times}^{\ell m}(\iota)\, e^{-{t}/{\tau_{\ell m}}}\, 
\sin\left ( \omega_{\ell m} t - m \phi \right ). \nonumber\\
\label{eq:gwmodes}
\end{eqnarray}
We have dropped the overtone index $n$ from all the relevant quantities 
(amplitudes, frequencies and time-constants) of quasi-normal modes, as we are assuming 
that higher (i.e., $n>0$) overtones, quickly become negligible in amplitude, compared to the 
fundamental $n=0$ overtone. Only the fundamental $n=0$ overtone of the various
modes is considered in this paper.

The amplitudes $A_{\ell m}$ of the various modes depend on the nature 
of the perturbation. For ringdowns resulting from the merger of two 
non-spinning black holes, $A_{\ell m}$ depend on the mass $M$ of the 
final black hole and mass ratio $q=m_1/m_2\, (q\ge 1)$ of the progenitor binary.

In the next section we will estimate the amplitude of dominant modes by fitting 
the late time signal from a numerical relativity simulation to
a superposition of ringdown modes. For binaries with non-spinning
black holes considered in this paper we find that
modes with $(\ell,m)=(2,2), (2,1), (3,3)$ and $(4,4)$ are the
ones that are excited with amplitudes large enough to be interesting.
For these modes, the angular functions $Y_{\ell m}^{+,\times}(\iota)$ are given by 
\begin{eqnarray}
Y_+^{22}(\iota) & = & \sqrt{\frac{5}{4\pi}}\, \frac{\left(1 + \cos^2\iota\right)}{2} ,\nonumber\\
Y_\times^{22}(\iota) & = & \sqrt{\frac{5}{4\pi}}\, \cos\iota,
\label{eq:y22}
\end{eqnarray}
\begin{eqnarray}
Y_+^{21}(\iota) & = & \sqrt{\frac{5}{4\pi}}\, \sin\iota,\nonumber\\
Y_\times^{21}(\iota) & = & \sqrt{\frac{5}{4\pi}}\, \cos\iota\, \sin\iota,
\label{eq:y21}
\end{eqnarray}
\begin{eqnarray}
Y_+^{33}(\iota) &=& -\sqrt{\frac{21}{8\pi}} \,
\frac{\left ( 1+\cos^2\iota \right )}{2}\, \sin\iota ,\nonumber\\
Y_\times^{33}(\iota) &=& -\sqrt{\frac{21}{8\pi}}\,\cos\iota \sin\iota,
\label{eq:y33}
\end{eqnarray}
\begin{eqnarray}
Y_+^{44}(\iota) &=& \sqrt{\frac{63}{16\pi}}\, 
\frac{\left ( 1+\cos^2\iota \right )}{2}\, \sin^2\iota ,\nonumber\\
Y_\times^{44}(\iota) &=& \sqrt{\frac{63}{16\pi}} \cos\iota \sin^2\iota.
\label{eq:y44}
\end{eqnarray}

In the case of binaries comprising of black holes with generic spins, the relative 
amplitudes of the various modes will also depend on the magnitude and direction of 
the spin vectors of the progenitor black holes. A detailed study of the dependence 
of the relative amplitudes of the various modes on the initial spin configurations 
and mass ratio of the progenitor binary is necessary to assess how accurately 
one might be able to use quasi-normal modes to measure a progenitor binary's 
parameters. For this, a more exhaustive set of simulations covering the full 
parameter space of binary black holes is required and will be taken up in the future.

\begin{figure*}[t]
\includegraphics[width=0.95\textwidth]{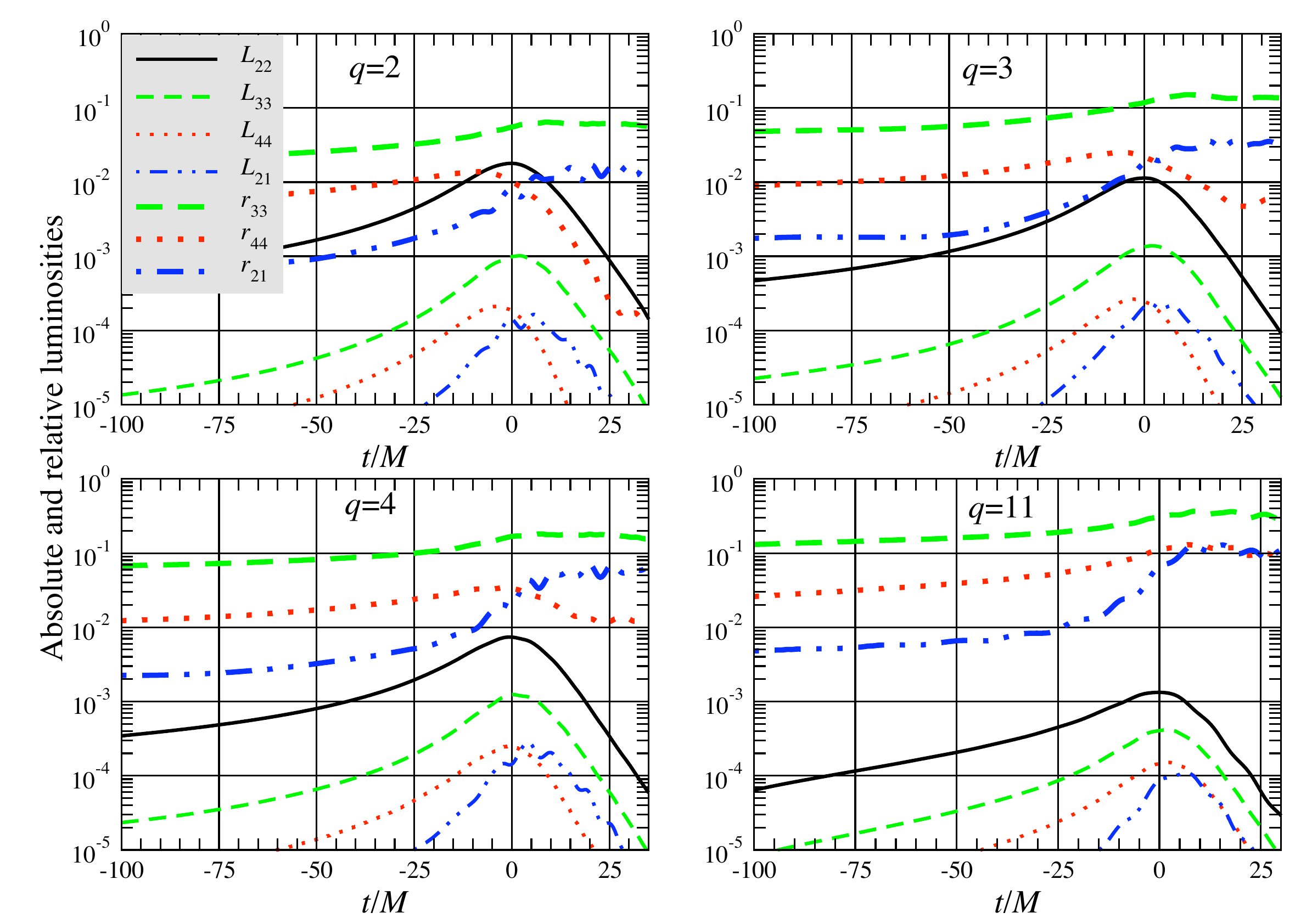}
\caption{This plot shows the relative luminosities, or radiated power, in 
modes $(2,2),$ $(3,3),$ $(4,4)$, and $(2,1).$ $L_{lm}$ represent the luminosities
(in units $c=G=1$ in which luminosity is dimensionless) and $r_{lm}$ 
denote the ratios $r_{lm}=L_{lm}/L_{22}.$
The different panels correspond to systems with different mass ratios
as indicated in the panel. Note that as the mass ratio increases, the luminosity
in each mode decreases but the amplitudes of all higher-order modes relative 
to the $(2,2)$-mode increase. We have omitted --- both in the figure and in 
this work --- the next most dominant modes, $(5,5)$, $(3,2)$, $(4,3)$, $(6,6)$ and 
$(5,4)$ as they are generally less than one percent as luminous as the $(2,2)$
mode. (see, however, Pan et al \cite{Pan:2011gk}). 
}
\label{fig:lumin}
\end{figure*}

We conclude this section by noting that using the expression for the two 
polarizations in Eq.\;(\ref{eq:gwmodes}) the detector response given in 
Eq.\;(\ref{eq:response}) can be written as:
\begin{equation}
h^A(t) = \sum_{\ell,m>0} B_{\ell m} e^{-t/\tau_{\ell m}}
\cos\left (\omega_{\ell m} t + \gamma_{\ell m}\right ),
\label{eq:reduced response}
\end{equation}
where the superscript $A$ is an index denoting the detector in question
(which is relevant when we have a network of detectors),
$B_{\ell m}$  ($\gamma_{\ell m}$) is the following combination of the 
amplitudes $A_{\ell m}$ (respectively, phases $m\phi$), antenna pattern 
functions $F^A_+$ and $F^A_\times$ and the inclination angle $\iota:$ 
\begin{eqnarray}
B_{\ell m} & = & \frac{\alpha_{\ell m}  M}{D_{\rm L}} \sqrt{ \left ( F^A_+\, Y^{\ell m}_+ \right )^2  
+ \left (F^A_\times\, Y^{\ell m}_\times \right )^2},
\label{eq:blm}\\
\gamma_{\ell m} & = & \phi_{\ell m} + m\, \phi + \tan^{-1} \left[ \frac{F^A_\times\, Y^{\ell m}_\times}
{F^A_+\, Y^{\ell m}_+} \right ].
\label{eq:gammalm}
\end{eqnarray}
Note that, for the sake of clarity, we have dropped the index $A$ on 
$B_{\ell m}$ and $\gamma_{\ell m}.$
Here, $\phi_{\ell m}$ are arbitrary constant phases of each 
quasi-normal mode \footnote{Specifically, the $\ell = m$ modes have a nearly consistent rotational 
phasing, while the $\ell \neq m$ modes seem to have somewhat distinct associated dynamics, with 
differentiated amplitude and phasing relationships during the merger process.\cite{Baker:2008d78}}
The effective amplitudes $B_{\ell m}$ are proportional to the intrinsic amplitudes
$\alpha_{\ell m}$ of the modes and vary inversely with the luminosity 
distance. Their magnitude also depends on the various angles 
$(\theta,\,\varphi,\,\psi,\,\iota)$ describing the position of the 
source on the sky and its orientation relative to 
the detector through the antenna pattern functions $F_+$ and $F_\times$ and
spherical harmonic functions $Y_+^{\ell m}$ and $Y_\times^{\ell m}$. 
The constant phases $\gamma_{\ell m}$ also depend on the angles and
the fiducial azimuth angle $\phi.$

The above form of the response is
more helpful in understanding which, or which combination, of the parameters can
be measured and how many detectors are required in solving the inverse problem, namely
to fully reconstruct the incident gravitational wave and the parameters of the source
that emitted the radiation. 
We shall use the above form of the waveform to compute the signal-to-noise ratios
and the covariance matrix.

\section{Amplitudes of modes excited during the ringdown phase of a black hole binary}
\label{sec:AMP}
In this section we will use numerical simulations to evaluate the amplitude
of the various modes excited as a function of the mass ratio $q.$ We will examine
how the amplitude of the dominant 22-mode varies as a function of $q.$ Of particular
interest would be the growth of 21-, 33- and 44-modes relative to the 22-mode as
the binary system becomes more asymmetric. 
We will provide simple analytical fits to the 
amplitudes of all the different modes and discuss how the inclination angle of the
black hole's final spin will affect the amplitude of the ringdown signal.
Our analysis is complementary to previous studies of the mode structure of 
unequal-mass nonspinning binaries, for example~\cite{Berti:2007b,Baker:2006kr}.
\begin{figure*}
\includegraphics[width=0.39\linewidth]{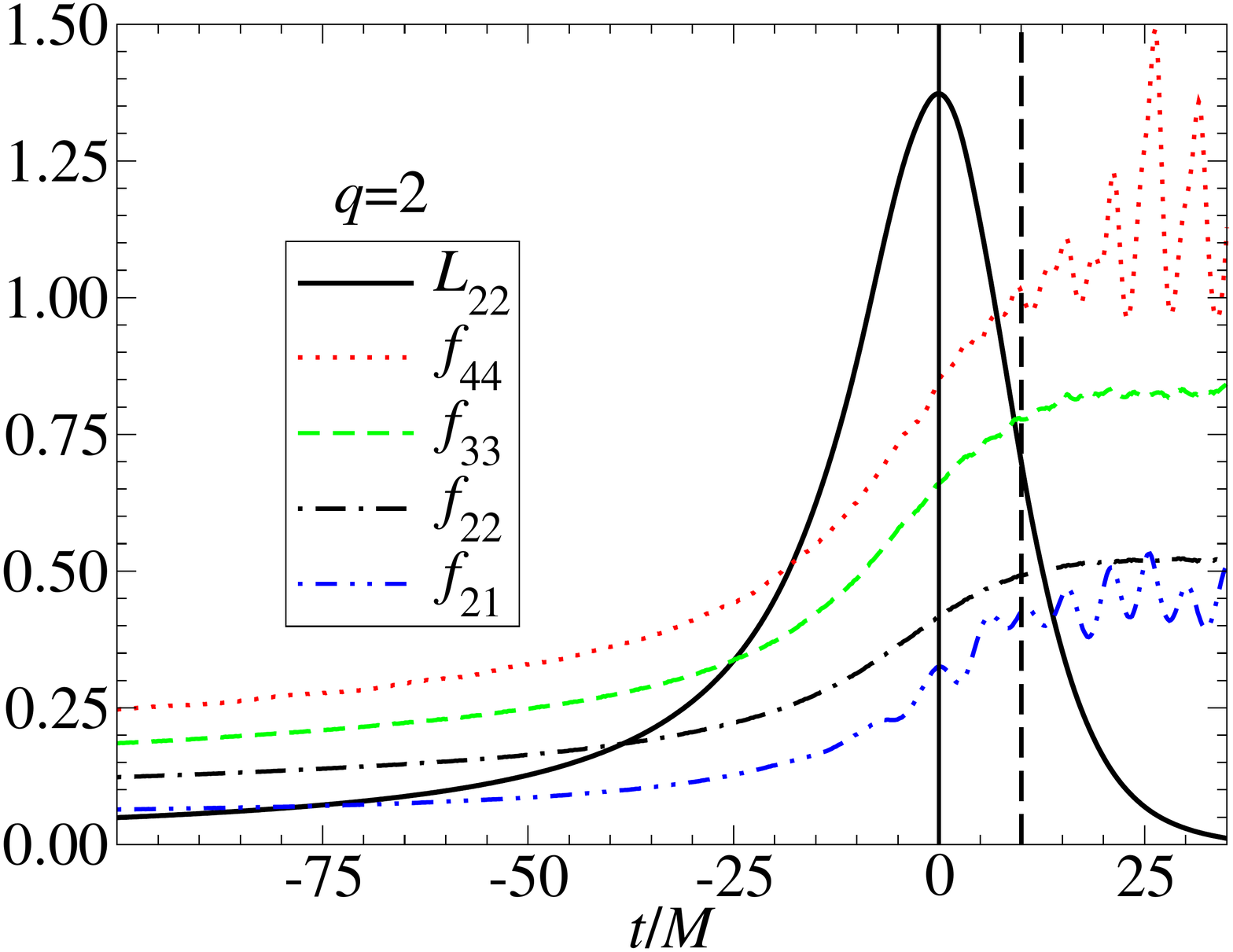}
\includegraphics[width=0.39\linewidth]{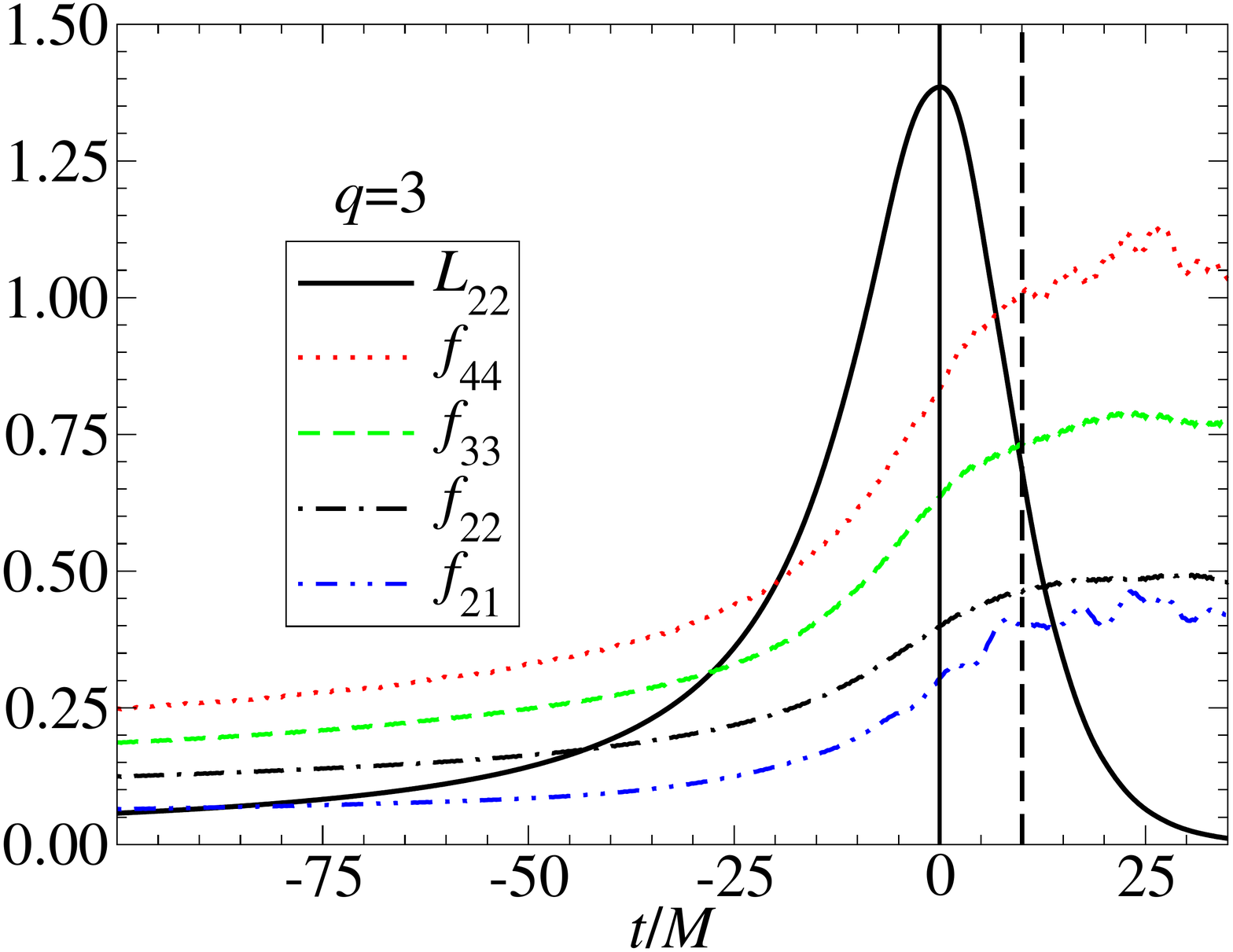}
\includegraphics[width=0.39\linewidth]{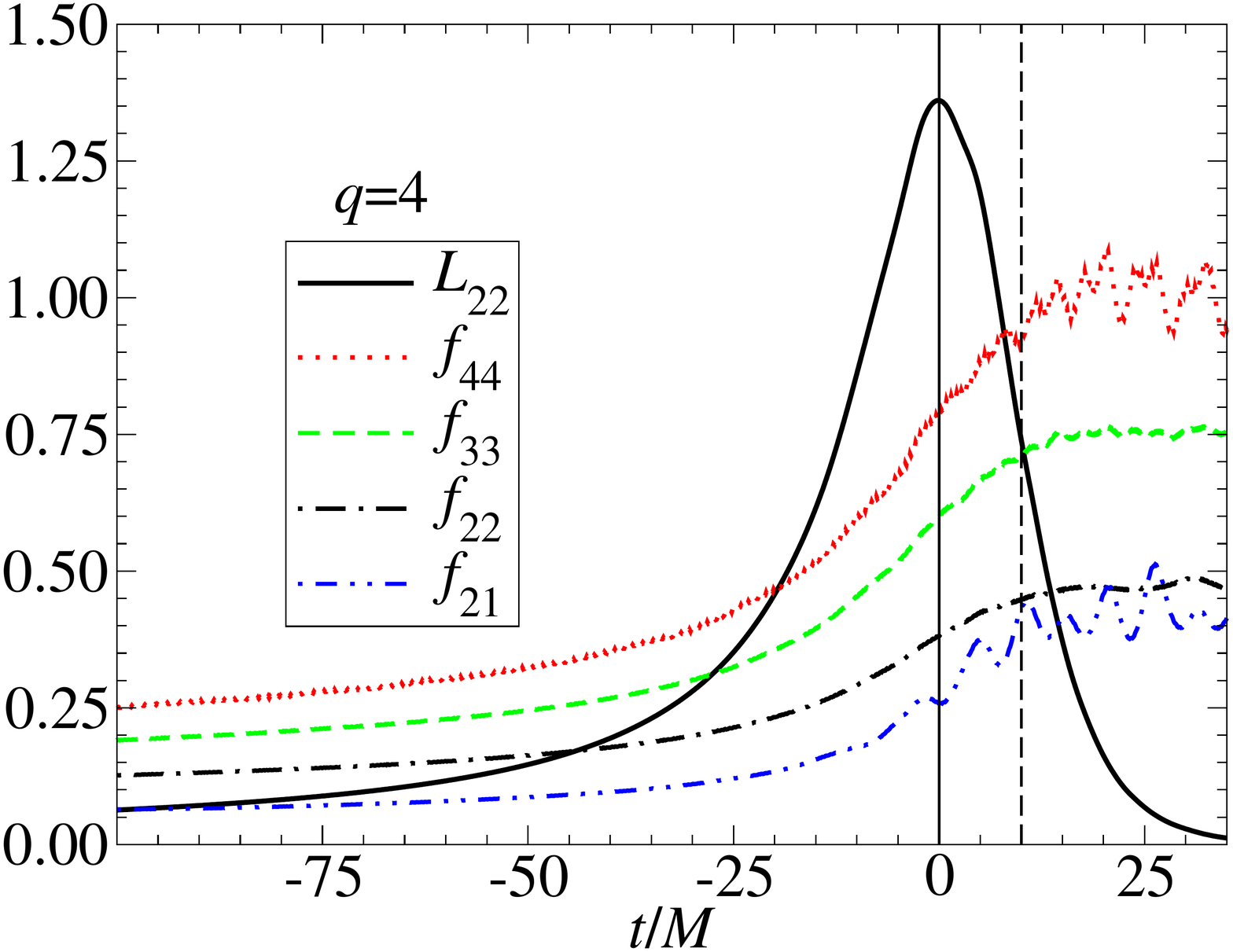}
\includegraphics[width=0.39\linewidth]{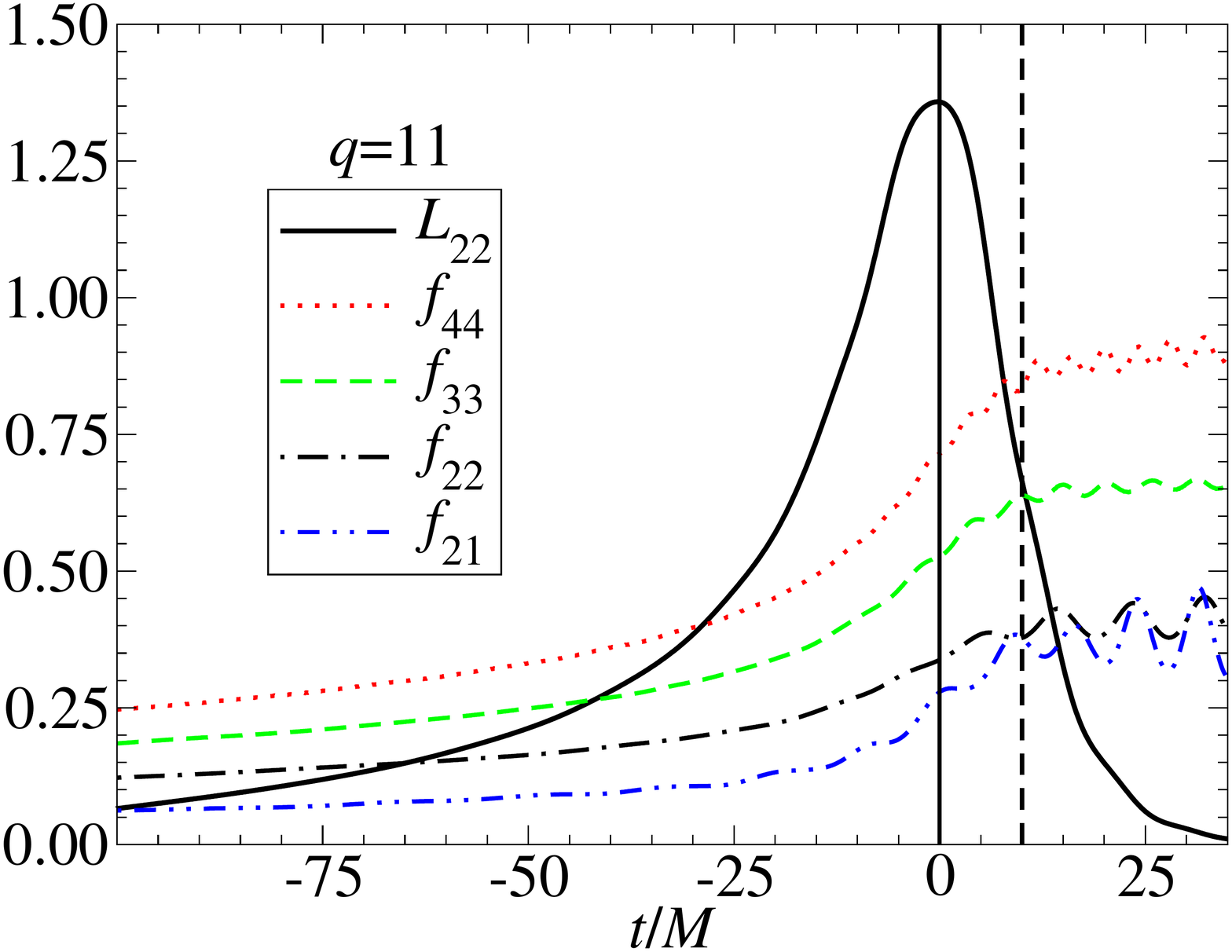}
\caption{Evolution of the first few dimensionless mode frequencies 
$f_{\ell m}=M\omega_{\ell m}$ as a function of the dimensionless 
time $t/M,$ for different values of the mass ratio $q$ of the 
progenitor binary. Also shown in arbitrary units is the luminosity 
in the 22 mode.  All mode frequencies, especially  $f_{22}$ and $f_{33},$ 
stop evolving and stabilise soon after the binary merges to form
a single black hole.  The waveform is assumed to contain a superposition
of only quasi-normal modes a duration $10M$ after the luminosity in
22 mode reaches its peak.
}
\label{fig:frequencies}
\end{figure*}

\subsection{Evolution of the luminosity}
An important question that arises in the study of QNMs excited during
the merger of a black hole binary is the determination of the 
most dominant modes in the infinite mode sum in  Eq.\;(\ref{eq:infinitesum})
for this particular kind of perturber. This analysis is necessary to construct 
a good model of the waveform to use in the analysis.  

Figure \ref{fig:lumin} plots the luminosity in gravitational waves, 
$L_{\ell m}=D_{\rm L}^2\,\left [ (\dot{h}_+^{\ell m})^2 + (\dot{h}_\times^{\ell m})^2\right],$ in 
the first four most dominant modes as a function of the dimensionless 
time $t/M$, where $M$ is the initial total mass of the binary. The 
luminosities are plotted for four values of the mass ratio $q=2,3,4$ and 11. 
We have left out the plot corresponding to the equal mass case $q=1,$ as in this case
the modes with odd values of $l$ or $m$ are not excited and hence not as interesting
as when the masses are unequal. The luminosity peaks just before the two black
holes collide but different modes peak at different times. The 21- and 33-modes peak
after the 22-mode reaches its maximum. However, the 44-mode shows the opposite 
behaviour. For a more thorough investigation on the different multipolar contributions
to the total radiated energy, see Ref.\,\cite{Baker:2008d78}.

Although the 33-mode is absent when $q=1,$ it is already more dominant
than the 44-mode when $q=2$ and remains the most dominant after the 22-mode, throughout
the inspiral and merger phase and for all mass ratios (except, of course, when $q=1$). 
The 44-mode remains more dominant than the 21-mode for the
most part, but the trend reverses after merger. This is because the 21-mode 
reaches its peak luminosity a little after the 44-mode. For $q=2,$ when the 22-mode 
reaches its peak, the luminosity in the 33-mode is an order-of-magnitude smaller 
than the dominant 22-mode; luminosities in 21- and 44-modes are 50 times smaller 
than the 22-mode. 

In addition to the luminosities, we have also plotted their ratios 
$r_{\ell m}=L_{\ell m}/L_{22}$ with respect to the 22-mode. It is clear that for
more symmetric systems (i.e. $q\gsim 1$) higher modes are hardly
excited. For instance, when $q=2$ the luminosity of the 21- and 44-modes remains below
$\sim$ few percent of the 22-mode, while the 33-mode is always less 
than $\sim$ 5\% of the 22-mode. As the mass ratio increases the higher order modes
are excited with greater amplitudes. The different mode amplitudes become comparable 
to the 22-mode and to one another as the mass ratio increases. In the next two sub-sections
we will give a more quantitative evaluation of the relative mode amplitudes in 
the ringdown part of the signal. 

\subsection{Identifying the ringdown phase}
An important task of our study is the identification of the point when the signal
is purely a superposition of various quasi-normal modes. This is necessary in order
that the proposed tests of general relativity are not corrupted due to the presence
of extraneous signals.  By assuming that the ringdown phase occurs sooner than it 
actually does, we are in danger of corrupting the waveform. Equally, by identifying the
ringdown phase to be much later than it actually is, we will significantly weaken the
tests since the signal amplitude falls off exponentially from the beginning of the
ringdown phase.  A proper identification of the beginning of the ringdown phase is
needed to correctly extract the amplitude of the quasi-normal modes and to compute 
the signal-to-noise ratio and other quantities. 

To this end, we shall use the
evolution of the frequency of the various modes $h_{\ell m}.$ As a binary evolves,
the frequency of each mode $h_{\ell m}$ increases, the rate of increase itself being
greater as the two black holes get closer. When the two black holes merge, a
common horizon forms and the frequency of each mode stabilizes, finally reaching 
the quasi-normal mode value as predicted by black hole perturbation theory. We shall 
identify the beginning of the ringdown phase to be (approximately) the epoch when the frequency
of the various modes begin to stabilize. 

We can compute the frequency of each mode from the evolution of its
phase given by the second of the equations in Eq.\;(\ref{eq:real and imaginary}).
Once the phase is known it is straightforward to write down the (dimensionless) frequency
$f_{\ell m} = M\,\omega_{\ell m}={\rm d}\Phi_{\ell m}(t)/{\rm d}(t/M).$ The ringdown 
phase can be assumed to begin when $f_{\ell m}$ computed from our numerical simulations
are close to those obtained from black hole perturbation theory. We will first take a 
look at the predictions from black hole perturbation theory and then compare those predictions
to the results obtained from our numerical simulations and plotted in Fig.\,\ref{fig:frequencies}.

There has been a lot of work on the computation of the frequencies and
time-constants of various modes of a perturbed Kerr black hole. Berti
et al \cite{BCW05} have found simple fits, as a function of the spin 
parameter $j,$ to the dimensionless mode frequencies\footnote{Note that
we only consider here, the least damped $n=0$ overtone for each mode and 
have therefore dropped the overtone index from mode frequencies, quality factors
and time-constants.} $f_{\ell m}=M\,\omega_{\ell m}$ and quality factors
$2\,Q_{\ell m}=\tau_{\ell m}\,\omega_{\ell m}.$ The fitting functions for 
the 22, 21, 33 and 44 modes are given by \cite{BCW05}
\begin{eqnarray}
f_{22} & = & 1.5251 - 1.1568 (1 - j)^{0.1292},\nonumber\\
Q_{22} & = & 0.7000 + 1.4187 (1 - j)^{-0.4990},
\label{eq:fit22}
\end{eqnarray}
\begin{eqnarray}
f_{21} & = & 0.6000 - 0.2339 (1 - j)^{0.4175},\nonumber\\
Q_{21} & = & -0.3000 + 2.3561 (1 - j)^{-0.2277},
\label{eq:fit21}
\end{eqnarray}
\begin{eqnarray}
f_{33} & = & 1.8956 - 1.3043 (1 - j)^{0.1818},\nonumber\\
Q_{33} & = & 0.9000 + 2.3430 (1 - j)^{-0.4810},
\label{eq:fit33}
\end{eqnarray}
\begin{eqnarray}
f_{44} & = & 2.3000 - 1.5056 (1 - j)^{0.2244},\nonumber\\
Q_{44} & = & 1.1929 + 3.1191 (1 - j)^{-0.4825}.
\label{eq:fit44}
\end{eqnarray}
These fits are quite robust and they differ from the actual values obtained 
for the frequencies and quality factors by no more than 3\% \cite{BCW05}. 
Table \ref{tab:frequencies} lists frequencies $f_{22},$ $f_{21},$ $f_{33}$ and
$f_{44},$ for several different values of the spin parameter $j.$ The chosen 
values of $j$ correspond to the final spins of black holes that result in 
our numerical simulations of binaries with different mass ratios $q.$ Values 
in columns labelled ``Fit" are those obtained using 
Eqs.\,(\ref{eq:fit22})-(\ref{eq:fit44}) and those in columns labelled 
``NR" are those obtained from our numerical relativity
simulations as follows.

\begin{table}[bt]
\caption{Dimensionless frequencies $f_{\ell m}$ for various
modes for different values of the black hole spin $j$ that 
results from the merger of a binary of mass ratio $q.$ We
assume that the ringdown phase begins when the frequencies
of various modes stop increasing and stabilize to a constant value.
For each mode frequency, the column labelled ``Fit'' gives the
values computed using Eqs.\,(\ref{eq:fit22})-(\ref{eq:fit44}) and that labelled
``NR" shows values at the ``beginning" of the ringdown mode,
a duration $10M$ after the gravitational wave luminosity in the
22 mode reaches its peak (dashed vertical line in Fig.\,\ref{fig:frequencies}).}

\label{tab:frequencies} 
\begin{tabular}{c|c|ccc||ccc||ccc||cc}
\hline
\hline
$q$    & $j$   & \multicolumn{2}{c}{$f_{22}$}  & & \multicolumn{2}{c}{$f_{21}$}  
& & \multicolumn{2}{c}{$f_{33}$} & & \multicolumn{2}{c}{$f_{44}$} \\
       &       & Fit  & NR      && Fit  & NR     && Fit  & NR && Fit & NR \\
\hline
 1     & 0.69  & 0.53 & 0.51    && 0.46 & $-$    && 0.84 & $-$  && 1.14 & 1.08\\ 
\hline
 2     & 0.62  & 0.50 & 0.49    && 0.44 & 0.42   && 0.80 & 0.78  && 1.09 & 1.05\\ 
\hline
 3     & 0.54  & 0.48 & 0.47    && 0.43 & 0.41   && 0.76 & 0.74 && 1.03 & 1.01\\
\hline
 4     & 0.47  & 0.46 & 0.45    && 0.42 & 0.43   && 0.73 & 0.72 && 0.99 & 0.97\\
\hline
 11   & 0.25 & 0.41& 0.39    && 0.39 & $0.41$     && 0.66 & 0.64  && 0.89 & 0.85\\
\hline
\hline
\end{tabular}
\end{table}

\begin{table*}[t]
\begin{center}
\caption{For different mass ratios (column 1), we list the 
final spin $j$ of the black hole (column 2), amplitudes of different 
modes at three different epochs: (i) when the 22 mode reaches its peak 
luminosity (columns 3-6), an epoch $10M$ (columns 7-10) and $15M$ after
the 22 mode reaches its peak luminosity. We list the absolute value of the
amplitude for the 22 mode and the ratio of amplitudes of the rest of the modes
to the 22 mode. The beginning of ringdown, taken as the point when the 
instantaneous frequency of each mode begins to stabilize, is typically 
found to be $\sim 1$-2 cycles after the peak luminosity. For concreteness
we take the beginning of the ringdown mode to be $10M$ after the peak
luminosity.}
\label{table:table1}
\begin{tabular}{c|c||ccccc||ccccc||cccc}
\hline
\hline
$\phantom{b}q$\phantom{b}&\phantom{b} $j$ \phantom{b}& \multicolumn{4}{c}{At peak luminosity}& & \multicolumn{4}{c}{At 10$M$ after peak} & & \multicolumn{4}{c}{At 15$M$ after peak}\\ 
\hline
&&\phantom{b} $\alpha_{22}$\phantom{b} & ${\alpha_{33}}/{\alpha_{22}}$ & ${\alpha_{44}}/{\alpha_{22}}$ & ${\alpha_{21}}/{\alpha_{22}}$ 
&&\phantom{b} $\alpha_{22}$\phantom{b} & ${\alpha_{33}}/{\alpha_{22}}$ & ${\alpha_{44}}/{\alpha_{22}}$ & ${\alpha_{21}}/{\alpha_{22}}$ 
&&\phantom{b} $\alpha_{22}$\phantom{b} & ${\alpha_{33}}/{\alpha_{22}}$ & ${\alpha_{44}}/{\alpha_{22}}$ & ${\alpha_{21}}/{\alpha_{22}}$\\
\hline
1&  0.69 & 0.365 & 0.000 &  0.052 & 0.000&& 0.217 & 0.000& 0.043&  0.000&& 0.152 & 0.000& 0.038& 0.000\\ 
\hline
2&  0.62 & 0.321 & 0.149 &  0.050 & 0.114&& 0.194 & 0.161& 0.030&  0.121&& 0.132 & 0.156& 0.020& 0.135\\ 
\hline
3&  0.54 & 0.266 & 0.216 &  0.070 & 0.178&& 0.158 & 0.247& 0.052&  0.195&& 0.108 & 0.244& 0.044& 0.212\\ 
\hline
4&  0.47 & 0.225 & 0.259 &  0.087 & 0.203&& 0.140 & 0.267& 0.069&  0.234&& 0.095 & 0.264& 0.057& 0.262\\ 
\hline
11& 0.25 & 0.100 & 0.349 &  0.156 & 0.312&& 0.063 & 0.377& 0.154&  0.407&& 0.048 & 0.347& 0.137& 0.478\\ 
\hline
\hline
\end{tabular}
\end{center}
\end{table*}

Figure \ref{fig:frequencies} plots the frequencies $f_{\ell m}$ for the 22, 21, 33, and
44 modes as (black) dot-dashed, (blue) dot-dot-dashed, (green) dashed and (red) dotted curves. 
As expected, the frequency of each mode increases, quite rapidly towards the
end, but stabilizes to a constant value -- the quasi-normal mode frequency of the
final black hole. The plots also show the (arbitrarily scaled) luminosity in the 22-mode 
as a (black) solid curve.  The epoch at which the luminosity reaches its peak (indicated 
by a solid vertical line) has been set to be $t/M=0.$  We see that the onset of the 
ringdown phase occurs significantly 
after the luminosity reaches its peak. For simplicity, we have chosen the beginning of the ringdown 
phase to be a duration $t=10\, M$ after the system reaches its peak luminosity, indicated by a dashed 
vertical line in Fig.\,\ref{fig:frequencies}.
In reality for the $\ell=m$ modes, it is $4M$-$5M$ earlier than for the $(2,\,1)$ mode.
The luminosity curves, Fig.\,\ref{fig:lumin}, exhibit a similar behaviour. See also \cite{Baker:2008d78}.

Most modes seem to stabilize at the onset of the ringdown phase. 
For $q\lsim 4$ the frequency of 22 and 33 modes stabilize, but it is less so with
21 and 44 modes. In our $q=11$ simulations all mode frequencies seem to oscillate around
a mean value, 22 and 21 more than 33 and 44. From Fig.\,\ref{fig:lumin}, we see that
after reaching the peak luminosity the amplitudes of 21 and 33 modes 
relative to the 22 mode (thick lines) are constant. This justifies why we might 
fit a $q$-dependent function to the relative amplitudes of various modes (see Sec.\,
\ref{sec:relative amplitudes} and Fig.\,\ref{fig:relat peakamp}) that is valid 
throughout the ringdown phase. The same cannot be said about the 44 mode.

Under columns labelled ``NR", Table \ref{tab:frequencies} gives 
frequencies $f_{22},$ $f_{21},$  $f_{33}$ and $f_{44}$ at the onset 
of the ringdown mode (i.e., an epoch $t=10M$ after the luminosity 
of the 22 mode reaches its peak) obtained from our numerical relativity 
simulations. Modes with odd $\ell$ or $m$ are not excited when a binary
comprising a pair of equal mass black holes merges, which is the reason 
why these entries are missing from the Table.  The mode frequencies at 
$10M$ after peak luminosity agree with the fits obtained from black 
hole perturbation theory to within 5\%.  Hence we believe that our 
method is quite robust in identifying the ringdown phase.

\subsection{Relative amplitudes in the ringdown phase}
\label{sec:relative amplitudes}
Our goal here is to estimate the amplitudes of the various modes
in the ringdown phase.  The transition from the inspiral to the ringdown phase 
is very smooth and it is not easy to pick a unique instant after which the transition 
occurs.  Amplitudes of the various modes in the ringdown phase are given 
in Table\;\ref{table:table1} at three different epochs: 
\begin{itemize}
\item at the epoch when the luminosity of the 22 mode reaches its peak,
\item a duration $10M$ after the 22 mode reaches its peak luminosity, and
\item a duration $15M$ after the 22 mode reaches its peak luminosity.
\end{itemize}
The Table lists the {\em absolute} amplitude $\alpha_{22}$ of the 22 mode and 
{\em relative} amplitudes $\alpha_{\ell m}/\alpha_{22}$ of the rest of the modes. 
These amplitudes are plotted in Fig.~\ref{fig:relat peakamp}.

First let us note that although, as expected, the absolute amplitude of the 22 mode
depends on the epoch that we identify as the start of the ringdown phase, the 
amplitudes of the sub-dominant modes relative to the 22 mode are not too sensitive
to that identification. This is especially true for the 33 and 44 modes whose peak
luminosity is at the same epoch as that of the 22 mode [thin (green) dashed and 
(red) dotted curves in Fig.~\ref{fig:lumin}], but less so for the 21 mode 
[thick (blue) dot-dot-dashed curve in Fig.~\ref{fig:lumin}, bottom right panel 
of Fig.~\ref{fig:relat peakamp}], whose peak luminosity occurs significantly 
after that of the 22 mode. As mentioned before, for concreteness 
we shall take the starting point of the ringdown mode as an epoch $10M$ after the 
22 mode reaches its peak luminosity (the dashed vertical line in Fig.~\ref{fig:frequencies}). 
All discussions in the remainder of this paper are based on this identification.
A duration of $10M$ corresponds to between 1 and 2 gravitational-wave cycles of
the merger signal. 

As the mass ratio of the progenitor binary increases, the amplitude of the 22 mode
rapidly decreases but the sub-dominant modes approach each other in
power (see Fig.~\ref{fig:lumin}) and their amplitudes increase (cf. Table 
\ref{table:table1} and Fig.~\ref{fig:relat peakamp}). For a mass ratio of
$q=4,$ the 33 and 21 modes have amplitudes 1/4 that of 22 while at $q=11$ they 
are 40\% of the 22 mode.  Of course, the overall luminosity 
decreases as the mass ratio increases and one expects no radiation in the 
limit $q\rightarrow \infty$. 
Indeed, the emitted energy during the `merger' phase goes roughly as the square 
of the symmetric mass ratio $\nu = m_{1}m_{2}/M^{2}$ of the progenitor 
binary \cite{Berti:2007b}. 

Let us note that the  values in Table \ref{table:table1} do not all
refer to the same final black hole spin. All our black holes are initially non-spinning 
and the final spin is simply the residual angular momentum of the progenitor binary. 
The final spin, therefore, 
depends on the mass ratio, and is greatest when the two black holes are 
of the same mass. In principle, it should be possible, but in practice very difficult,
to produce numerical data for different mass ratios all with the same final spin.
To do so we require an accurate mapping between the mass ratio and initial black-hole 
spins, and the spin of the final black hole. For example, configurations that lead to 
a nonspinning Schwarzschild black hole are suggested in Ref.~\cite{Buonanno:2007sv} and
more generic cases are considered in Refs.~\cite{Rezzolla:2007rz, Rezzolla:2007xa, 
Rezzolla:2008, Berti:2007nw}. However, fine-tuning the spin of the 
final black hole requires that the component black holes are also spinning, and 
in this paper we consider only binaries with non-spinning components.

\subsection{Fitting functions for relative amplitudes}
For the purpose of computing the signal-to-noise ratio and the covariance
matrix it is convenient to have analytical expressions for the
relative amplitudes of the various modes. Our fits are meant to capture
the dependence of the amplitudes on the mass ratio in the range we have
considered in this paper. They are not meant to explore the complex dynamics 
of a ringing black hole and are not necessarily valid outside the region 
we have explored. 

\begin{figure}[t]
\centering
\includegraphics[width=0.95\columnwidth]{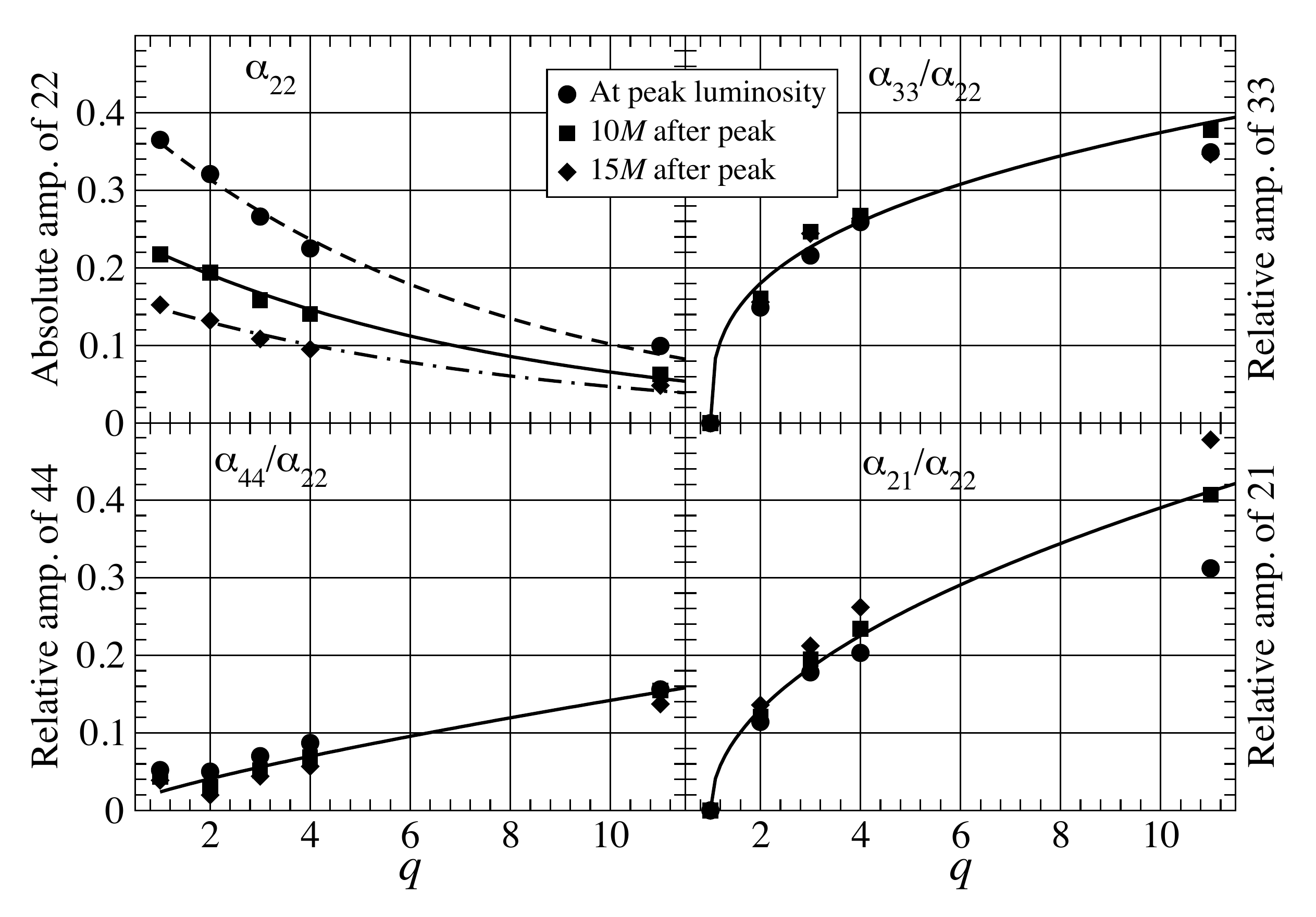}
\caption{This plot shows the amplitudes as a function of the mass ratio 
for different modes at the peak of the luminosity of the 22 mode (circles),
an epoch $10M$ and $15M$ after the peak (respectively, squares and triangles). 
We have plotted the {\em absolute} amplitude $\alpha_{22}$ of the 22 mode 
and ratio of the sub-dominant mode amplitudes $\alpha_{\ell m}/\alpha_{22}$ 
relative to 22 (cf.~Table \ref{table:table1}). The solid lines are the best
fits [cf.~Eqs.\;(\ref{eq:a22})-(\ref{eq:a44})] to the amplitudes at $10M$ after 
the peak luminosity.}
\label{fig:relat peakamp}
\end{figure}

We have seen that the
amplitudes inferred from the simulations depend on how we identify
the beginning of the ringdown phase. As mentioned before,
in our calculations we have assumed that the ringdown phase begins
an epoch $10M$ after the 22 mode reaches its peak luminosity. We find
that the amplitude of the 22 mode at this epoch as a function of
the mass ratio is well fitted by 
\begin{equation}
\alpha_{22}(q) = 0.25\, e^{-q/7.5}.\label{eq:a22}
\end{equation}
The solid curve in the top left panel in Fig.~\ref{fig:relat peakamp} shows
that this is a good fit to the data points (filled squares). Dot-dot-dashed
and dotted lines in the same panel show the fits that describe 
$\alpha_{22}(q)$ at the peak of the 
luminosity and $15M$ after the peak, respectively. In all cases, the data
is pretty well approximated by an exponentially falling function. 

The relative amplitudes of the sub-dominant 21, 33 and 44 modes are well
fitted by the following functions 
\begin{eqnarray}
\alpha_{21}(q) & = & 0.13\, \alpha_{22}(q)\, (q-1)^{1/2},\label{eq:a21}\\
\alpha_{33}(q) & = & 0.18\, \alpha_{22}(q)\, (q-1)^{1/3},\label{eq:a33}\\
\alpha_{44}(q) & = & 0.024\, \alpha_{22}(q)\, q^{3/4}.\label{eq:a44}
\end{eqnarray}
They are plotted as solid curves in the relevant panels of Fig.~\ref{fig:relat peakamp}.
The fits are motivated by the fact that when $q=1$ only modes with even values
of $\ell$ and $m$ are excited, while those with odd $\ell$ or $m$ are absent.
Note that the relative amplitudes grow as a binary becomes more asymmetric
and so higher order modes should be more easily detectable if the mass
ratio is large.  In absolute terms, of course, all modes are exponentially 
damped as a function of $q.$

In this work we estimated the relative amplitudes of different modes at 
a single epoch. It might be more reliable to estimate their average over 
a small duration, for instance between $10M$-$15M$. This could diminish 
any `numerical noise' present and extract more accurate fits but we did 
not explore this alternative approach in this work.
 
\section{Visibility of ringdown modes}

\label{sec:SNR}

In this section we will study the signal-to-noise ratio (SNR) obtained by various 
detectors for the ringdown phase of the coalescence of a black hole binary. We will
begin by defining the matched filter SNR, followed by the noise power spectral densities
of aLIGO, ET and LISA and the choice of signal parameters used in the study.
We will then discuss the visibility of the different modes, focusing on the distance 
reach of the various detectors.

\subsection{Matched filter SNR}

The matched filter SNR $\rho$ obtained while searching for a signal of known shape
buried in Gaussian background is given by (see, e.g., \cite{Th300})
\begin{equation}
\rho^2 = 4 \int_{0}^{\infty} \frac{|H(f)|^2}{S_h(f)} {\rm d}f
\end{equation}
where $S_h(f)$ is the detector noise power spectral density and $H(f)$ is the 
Fourier transform of the signal assumed (in this work) to be a superposition of the 22, 21 
and 33 modes. In the time-domain our waveform is given by
Eq.\,(\ref{eq:reduced response}) where the sum is over $(\ell,m)=(2,2),(2,1),(3,3)$
and we have ignored all higher order modes including the $(4,4)$ mode.  
The coefficients $\alpha_{\ell m}$ required to 
compute the waveform are assumed to be as in Eqs.\,(\ref{eq:a22})-(\ref{eq:a33}). 

The ringdown signal is a superposition of different modes and its visibility 
depends not only on the relative amplitudes but also on the relative {\it phases}
of the different modes. 
To this end, it is useful to define the SNR of a mode as
\begin{equation}
\rho_{\ell m}^2 = 4 \int_{0}^{\infty} \frac{|H_{\ell m}(f)|^2}{S_h(f)} {\rm d}f,
\end{equation}
with the caveat that the total SNR $\rho^2$ is {\em not} the quadrature sum of 
$\rho_{\ell m}^2$ since there are also interference terms that can be negative.
\begin{figure*}
\includegraphics[width=0.49\textwidth]{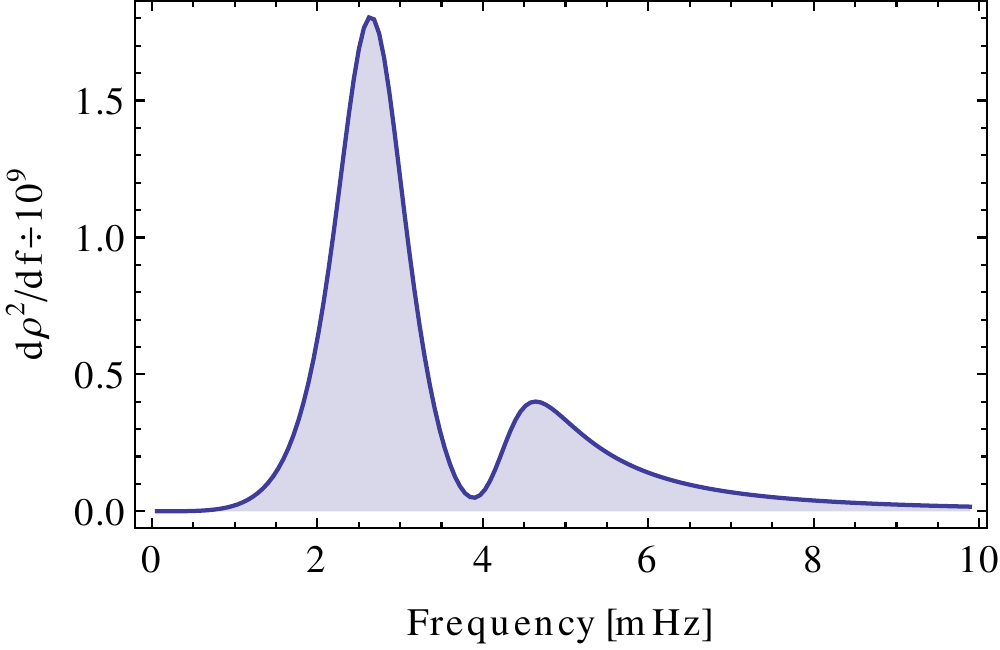}
\includegraphics[width=0.49\textwidth]{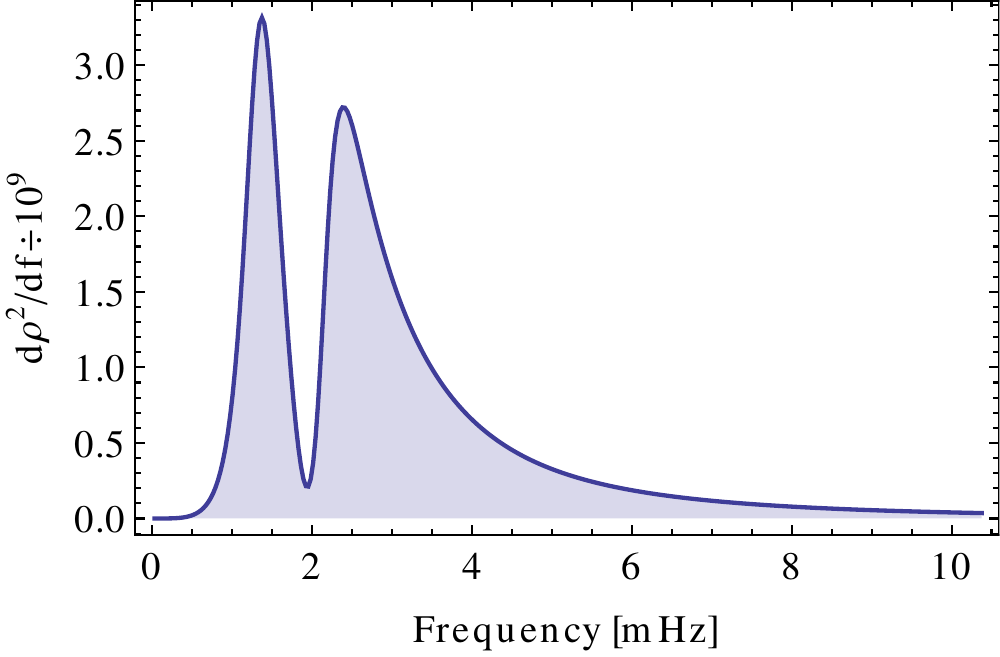}
\caption{The signal-to-noise ratio integrand for LISA for
a quasi-normal mode signal that is composed of 22, 21 and 33 modes 
--- the three most dominant ones.  The source is assumed to be
at a red-shift of $z=1$ and the various angles are as in Table \ref{tab:params}.
The left panel corresponds to a black hole of mass $M=5\times 10^6\,M_\odot$ and
the right panel to a black hole of mass $M=10^7\,M_\odot.$ In both
cases the mass ratio of the progenitor binary is taken to be 
$q=10.$}
\label{fig:snr integrand}
\end{figure*}

\subsection{Sensitivity curves}
In our study, we will consider the performance of three detectors: the aLIGO ET 
and LISA.  A fit to the aLIGO noise spectral density tuned to detect binary neutron 
stars is\footnote{The fit was provided by C. Capano, Syracuse University.}
\begin{eqnarray}
S_h(f) & = & 10^{-49} \Biggl [ 10^{16-4\,(f-7.9)^2} 
+ 0.08\, x^{-4.69} \nonumber \\ & + & 123.35\, 
\frac{1-0.23\, x^2+0.0764\,x^4}{1+0.17\, x^2}\Biggr ]\,{\rm Hz}^{-1}
\end{eqnarray}
where $x=f/215\,{\rm Hz}.$ 
In the case of ET we consider the sensitivity curve 
designated ET-B \cite{Hild:2009ns} whose noise power spectral density is given by 
$S_h(f) = 10^{-50} h_n(f)^2\,{\rm Hz}^{-1}$
\begin{eqnarray}
h_n(f) & = & 2.39\times 10^{-27}\, x^{-15.64} + 0.349\, x^{-2.145} \nonumber \\ 
       & + & 1.76\, x^{-0.12} + 0.409\, x^{1.10},
\end{eqnarray}
where $x = f/100\,{\rm Hz}.$ We take LISA noise spectral density to be the one 
that was used by the LISA Parameter Estimation Taskforce in Ref.\, \cite{Arun:2008zn}, 
which also corresponds to the noise curve from the second round of the Mock 
LISA Data Challenge \cite{Arnaud:2007jy,Babak:2007zd}. 

\subsection{Choice of various parameters}

The SNR depends on a number of source parameters as well 
as the location of the source on the sky. We have limited our investigations 
to studying the SNR and covariance matrix as a function
of the black hole's (observed) mass $M$ and the mass ratio $q$ of the progenitor 
binary, for fixed values of the distance to the black hole and various angles. 
In the case of LISA the black hole is assumed to be at a red-shift of $z=1$ 
which corresponds (in our cosmological model) to a luminosity distance of 
$D_{\rm L}\simeq 6.73\,\rm Gpc.$ In the case of aLIGO and ET we set $D_{\rm L}=1\,\rm Gpc.$ 
In all cases the angles are fixed to be $\theta=\psi=\iota=\varphi=\pi/3.$ For a statistical 
analysis of the effect of these angular parameters on the detectability of a ringdown signal see \cite{Kamaretsos:2011aa}. 

The black hole mass is varied over the range 
$[100,\,10^3]M_\odot$ in the case of aLIGO, 
$[10,\,10^3]\,M_\odot$ in the case of ET and $[3\times 10^6,\,10^8]\,M_\odot$ in the case of
LISA. These choices are dictated by the frequency sensitivity of the instruments
which further dictates the range of black hole masses whose ringdown radiation
they are most sensitive to.

Our choice of parameters is summarized in Table \ref{tab:params}. We reiterate
that our masses are {\em observed masses} which means that the intrinsic mass of 
the black hole is smaller by a factor $1+z\simeq 1.2,$ for aLIGO and ET and by a 
factor of $1+z=2$ in the case of LISA.  Although the signal visibility simply 
scales as the inverse of the 
distance, the fact that the mass is blue-shifted means that we cannot easily
scale our results to another (say, a greater) distance for the same intrinsic 
masses. Such a scaling will be valid if at the same time the intrinsic masses 
are also scaled up/down by the appropriate red-shift factor.

\begin{table}[h]
\caption{This Table lists the values of the various parameters used in our study.
In all cases the angles are all set to $\theta=\varphi=\psi=\iota=\phi=\pi/3.$} 
\label{tab:params}
\begin{tabular}{cccc}
\hline
\hline
Detector & $D_{\rm L}/\rm Gpc$ & $M/M_\odot$  & $q$ \\
\hline
aLIGO    & 1.00 & $[100,\,10^3]$  & 2-10 \\
\hline
ET       & 1.00 & $[10,\,10^3]$   & 2-10 \\
\hline
LISA     & 6.73 & $[3\times 10^6,\,10^8]$ & 2-25 \\
\hline
\hline
\end{tabular}
\end{table}

\subsection{Visibility of different modes}
\begin{figure*}
\begin{tabular}{|c|c|}
\hline
\includegraphics[width=0.42\textwidth]{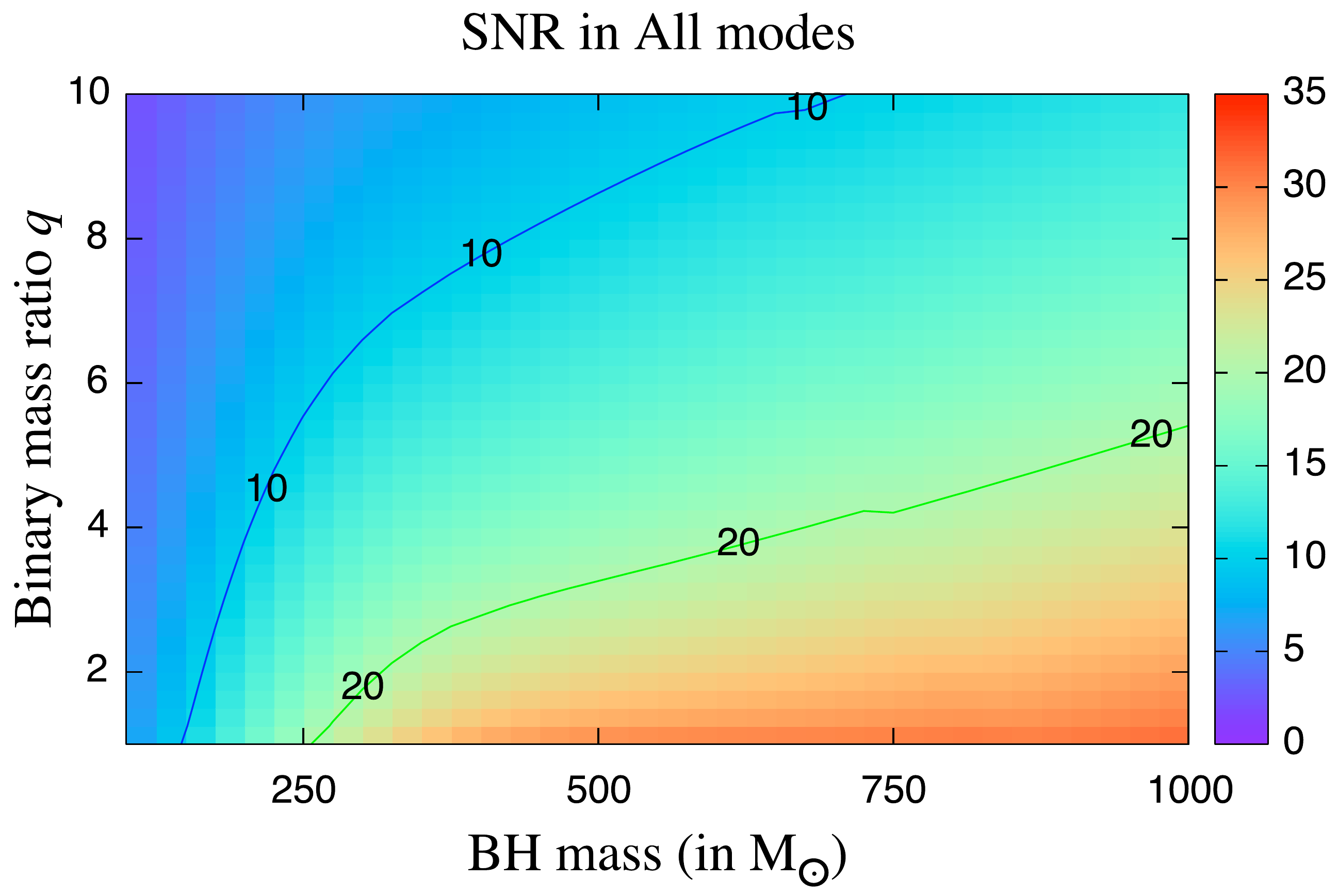}  &
\includegraphics[width=0.42\textwidth]{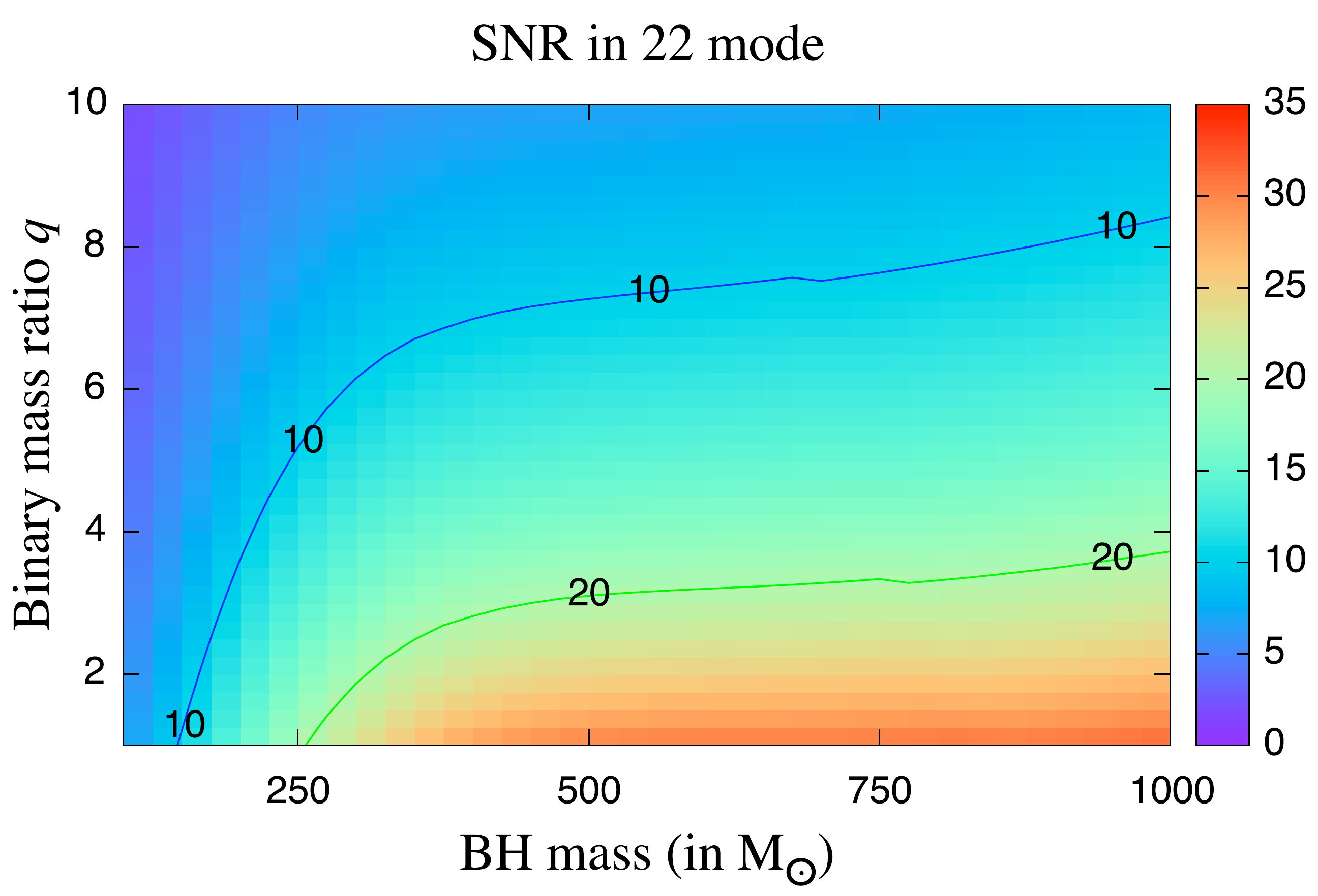} \\
\includegraphics[width=0.42\textwidth]{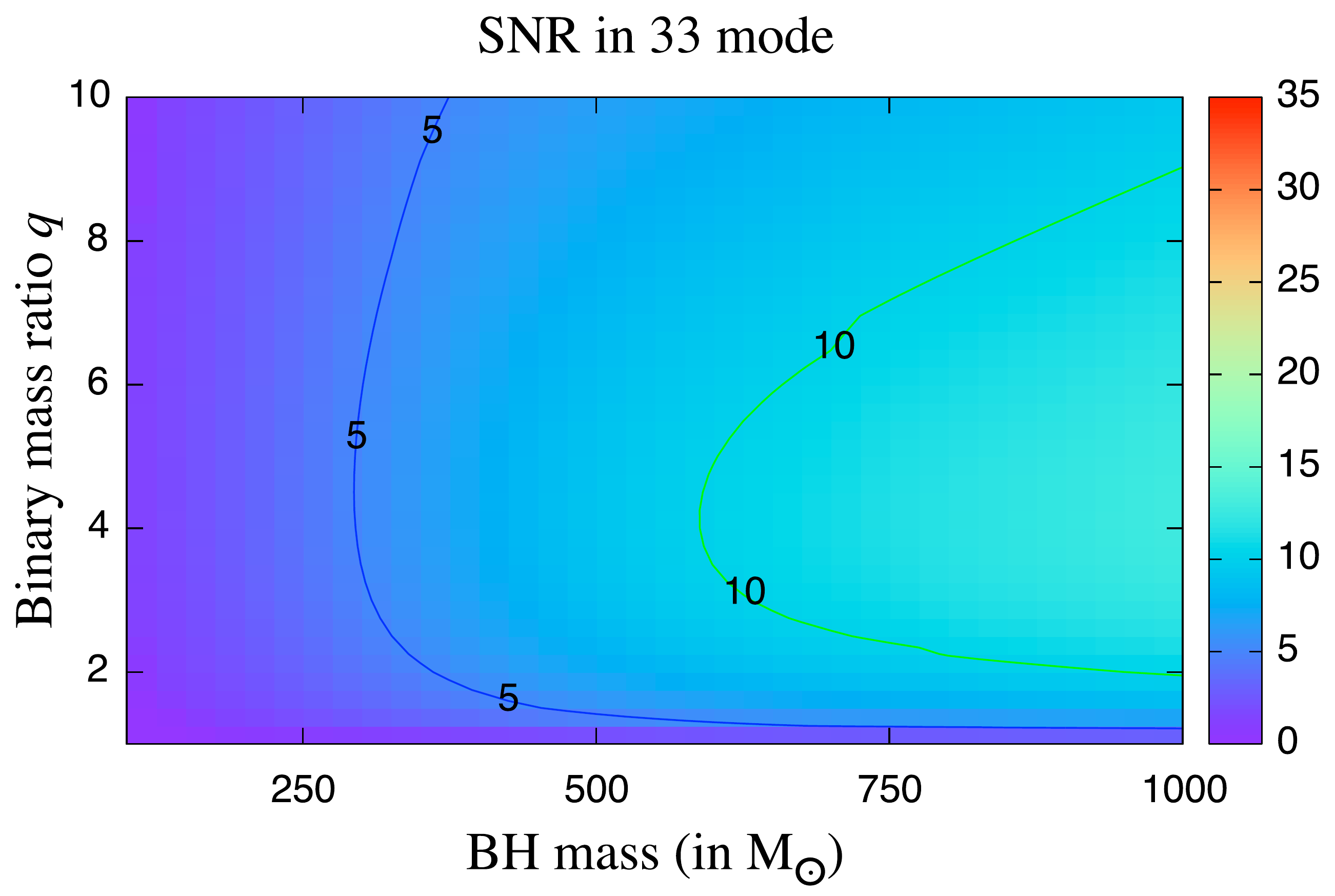} &
\includegraphics[width=0.42\textwidth]{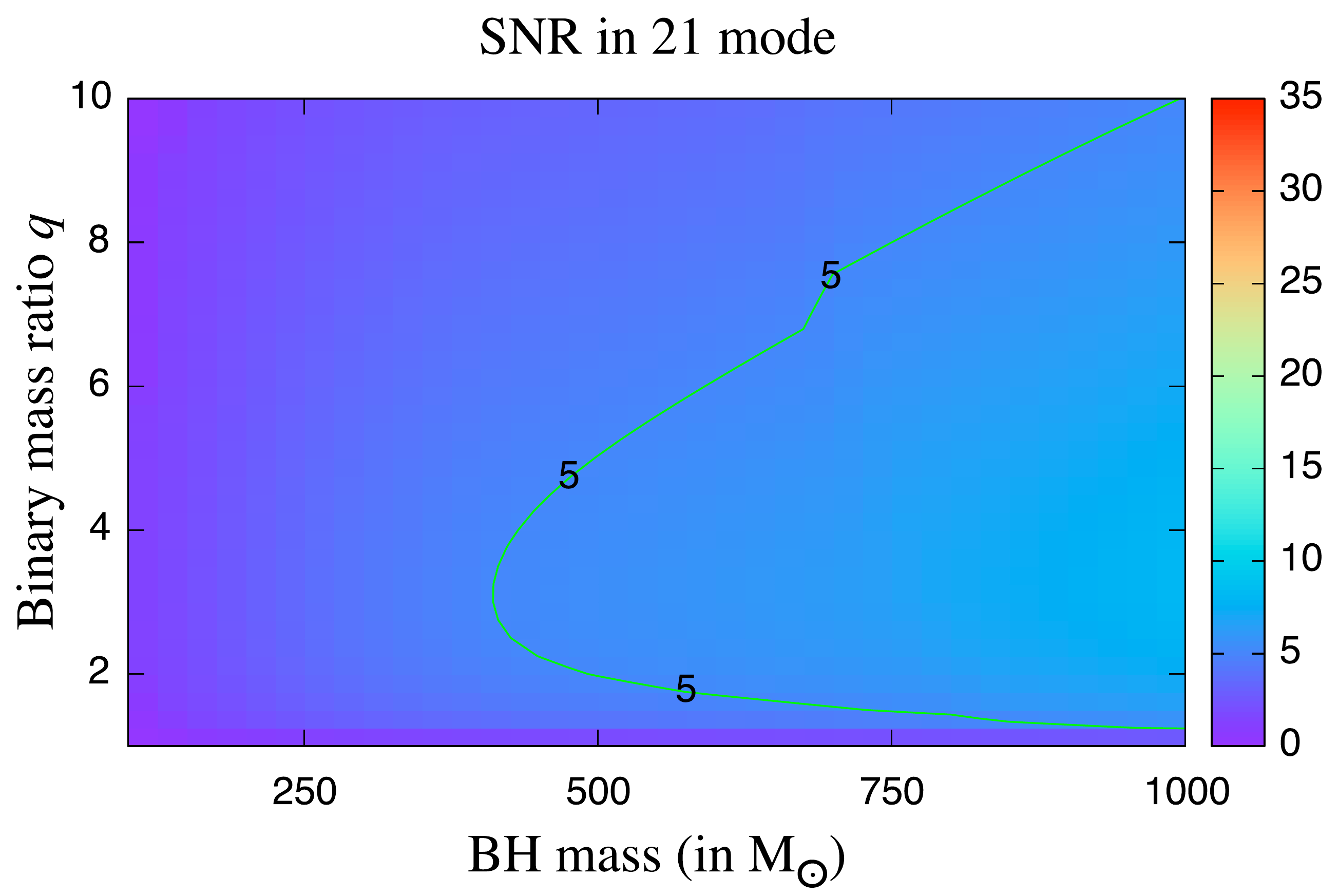} \\
\hline
\includegraphics[width=0.45\textwidth]{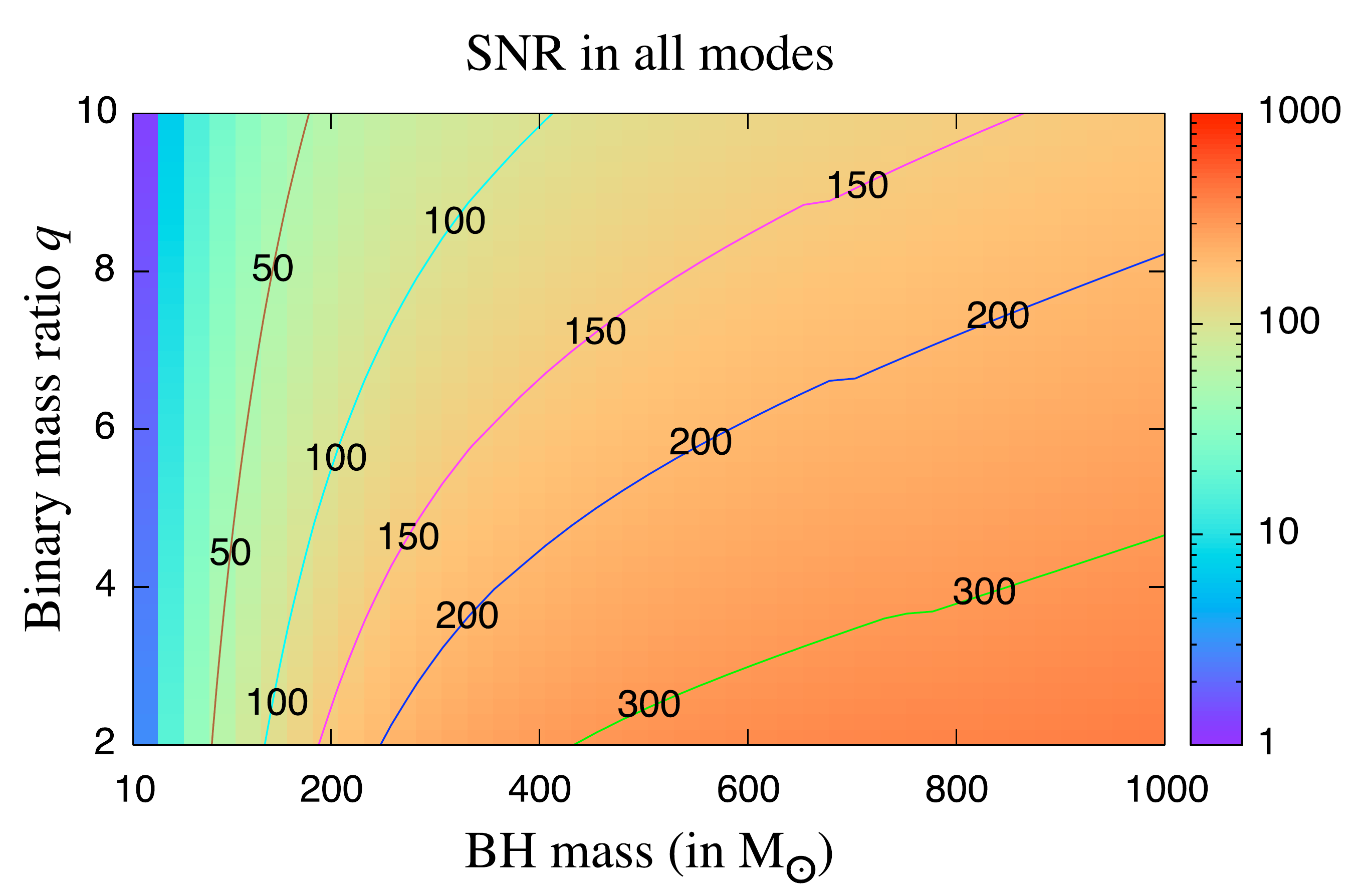}  &
\includegraphics[width=0.45\textwidth]{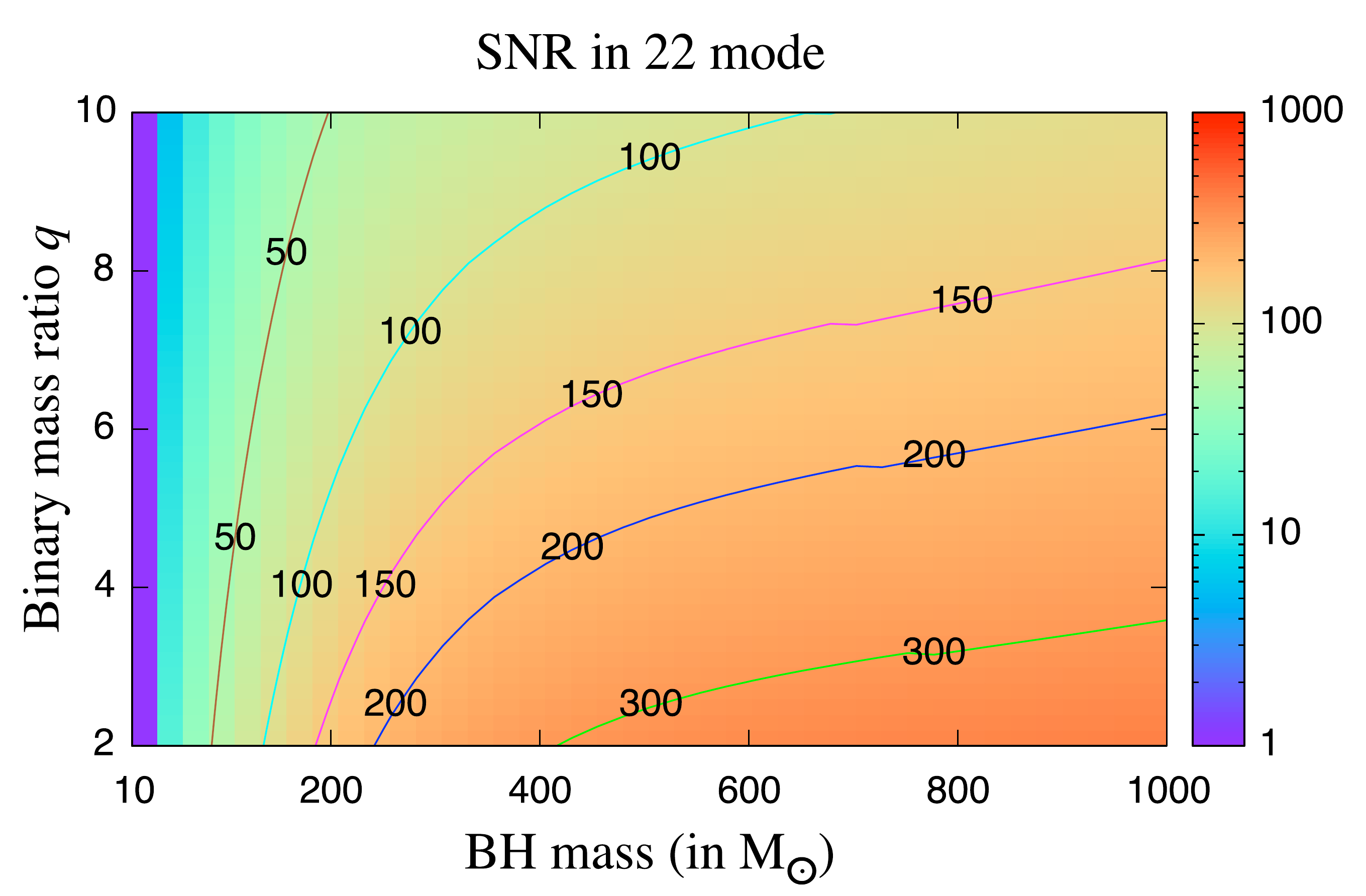} \\
\includegraphics[width=0.45\textwidth]{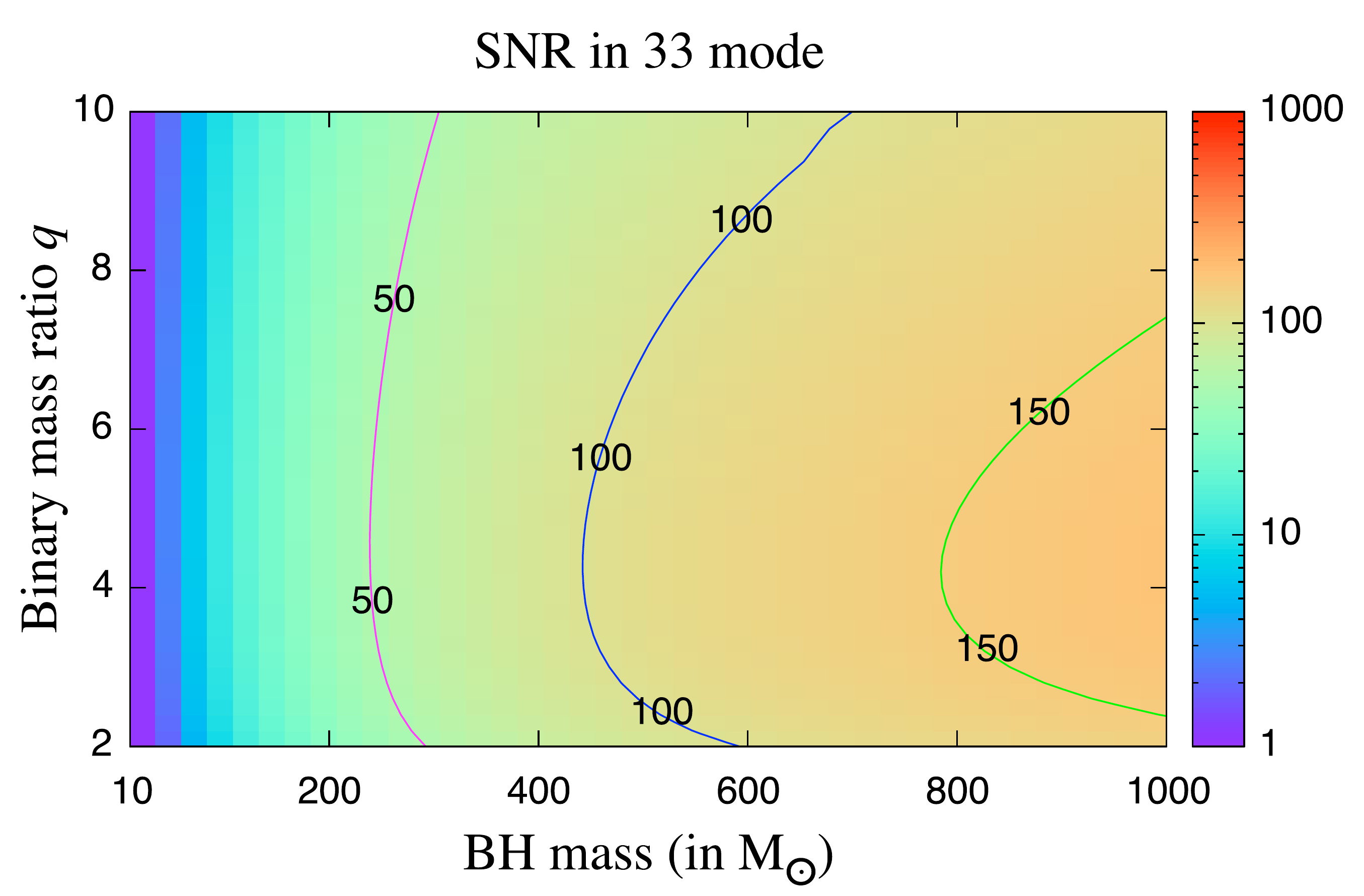} &
\includegraphics[width=0.45\textwidth]{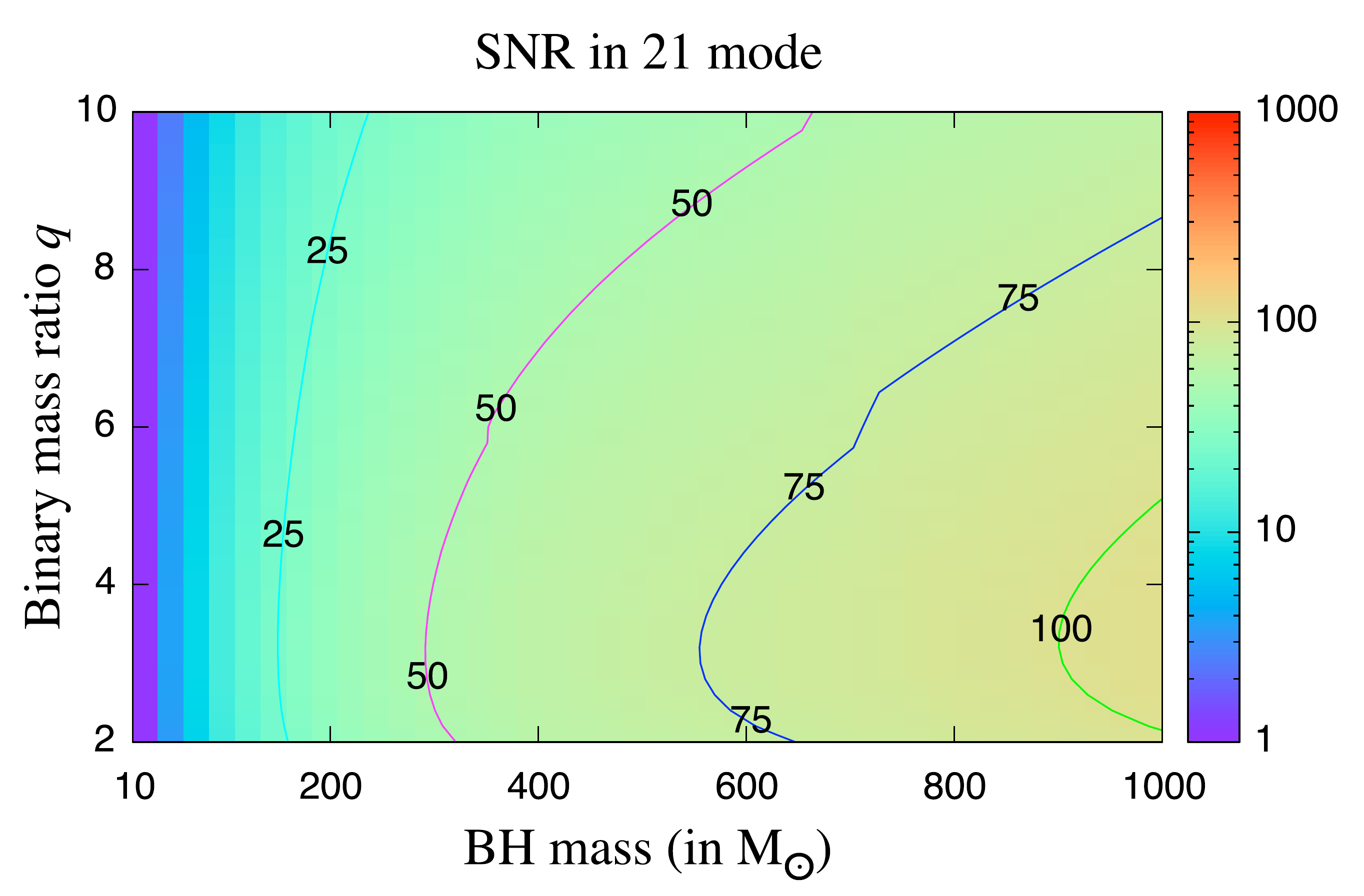} \\
\hline
\end{tabular}
\caption{Signal-to-noise ratio (SNR) in Advanced LIGO (top
set of four panels) and Einstein Telescope (bottom set of four panels) as a 
function of the black hole's mass $M$ and progenitor binary's 
mass ratio $q$ for different modes. Most of the contribution to the 
SNR comes from the 22 mode but other modes too have significant 
contributions, 33 being more important than 21. The source is
assumed to be at a distance of 1 Gpc and various angles are as in
Table \ref{tab:params}.}
\label{fig:snr ground}
\end{figure*}

It is instructive to plot the SNR integrand 
${\rm d}\rho^2/{\rm d}f=|H(f)|^2/S_h(f)$ as it depicts how the different 
modes become important for systems with different masses. 
Fig.\,\ref{fig:snr integrand} plots this quantity for two systems as
seen in LISA. The left panel corresponds to a black hole of total
mass $M=5\times 10^6\,M_\odot$ and the right panel to $M=10^7\,M_\odot.$
The mass ratio $q$ is $q=10$ in both cases and the angles are as in 
Table \ref{tab:params}.

The various mode frequencies $F_{\ell m}=\omega_{\ell m}/(2\pi)$ of the two 
systems are $F_{22} \simeq 2.74\,\rm mHz,$ $F_{21} \simeq 2.54\,\rm mHz$ and
$F_{33} \simeq 4.26\,\rm mHz$ for the lighter black hole and 
$F_{22}\simeq1.37\,\rm mHz,$ $F_{21}\simeq1.27\,\rm mHz$ and 
$F_{33}\simeq2.13\,\rm mHz$ for the heavier black hole. 
Let us first note that the 22 mode of the lighter black hole 
and 33 mode of the heavier black hole are close to the region 
where LISA has best sensitivity. This will be relevant in the 
discussion that follows.

The intrinsic amplitudes of the 21 and 33 modes are a little more than a 
third of the 22 mode for $q=10.$  However, since the SNR integrand depends on the 
signal power weighted down by the noise power, for a given black hole
mass the SNR integrand could be as large as, or even dominated by,
modes different from the 22 mode. This does not happen for the 21 mode since the
frequencies of the 22 and 21 modes are very close to each other and
so the 21 mode is always far smaller than the 22 mode. For a black 
hole mass of $10^7\, M_\odot$ the 33 mode is as strong as the 22 mode
and for masses even larger, the 33 mode overwhelms the 22 mode.
The total SNR for the $5 \times 10^6\,M_\odot$ black hole is 
$\rho=1670$, with the different modes contributing 
$\rho_{22}=1500$, $\rho_{21}=625$, and $\rho_{33}=950$. 
The SNR is clearly dominated by the 22 mode.

In the case of the heavier black hole, the total SNR is $\rho=2520,$ with
the different modes contributing $\rho_{22}=1940,$ $\rho_{21}=920,$ $\rho_{33}=1860$.
In this case, the 33 mode is as strong as the 22 mode but the 21 mode, as
expected, is sub-dominant.
\subsection{Exploring black hole demographics with ET and LISA}

\begin{figure*}
\begin{tabular}{|c|c|}
\hline
\includegraphics[width=0.45\textwidth]{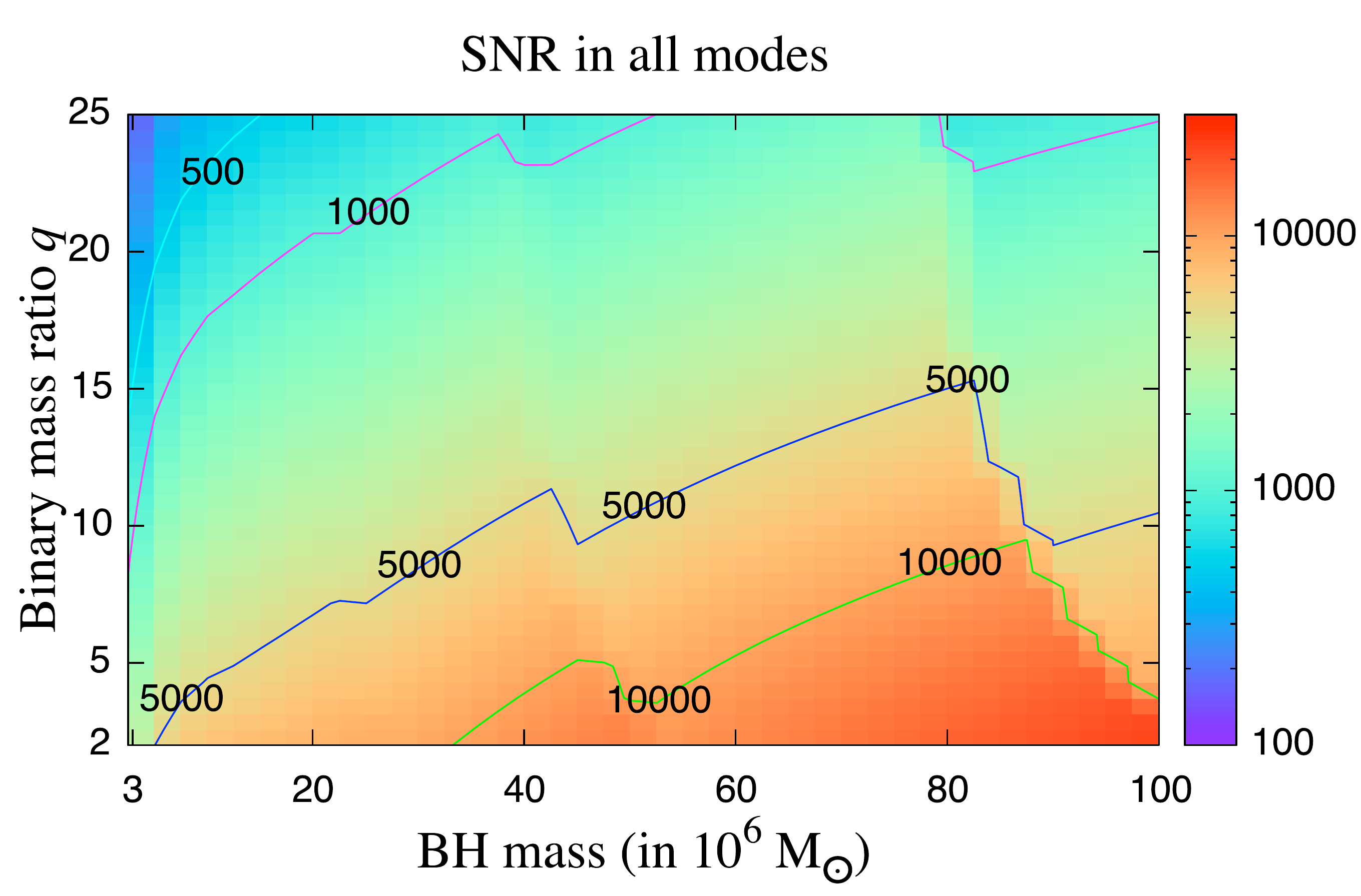} &
\includegraphics[width=0.45\textwidth]{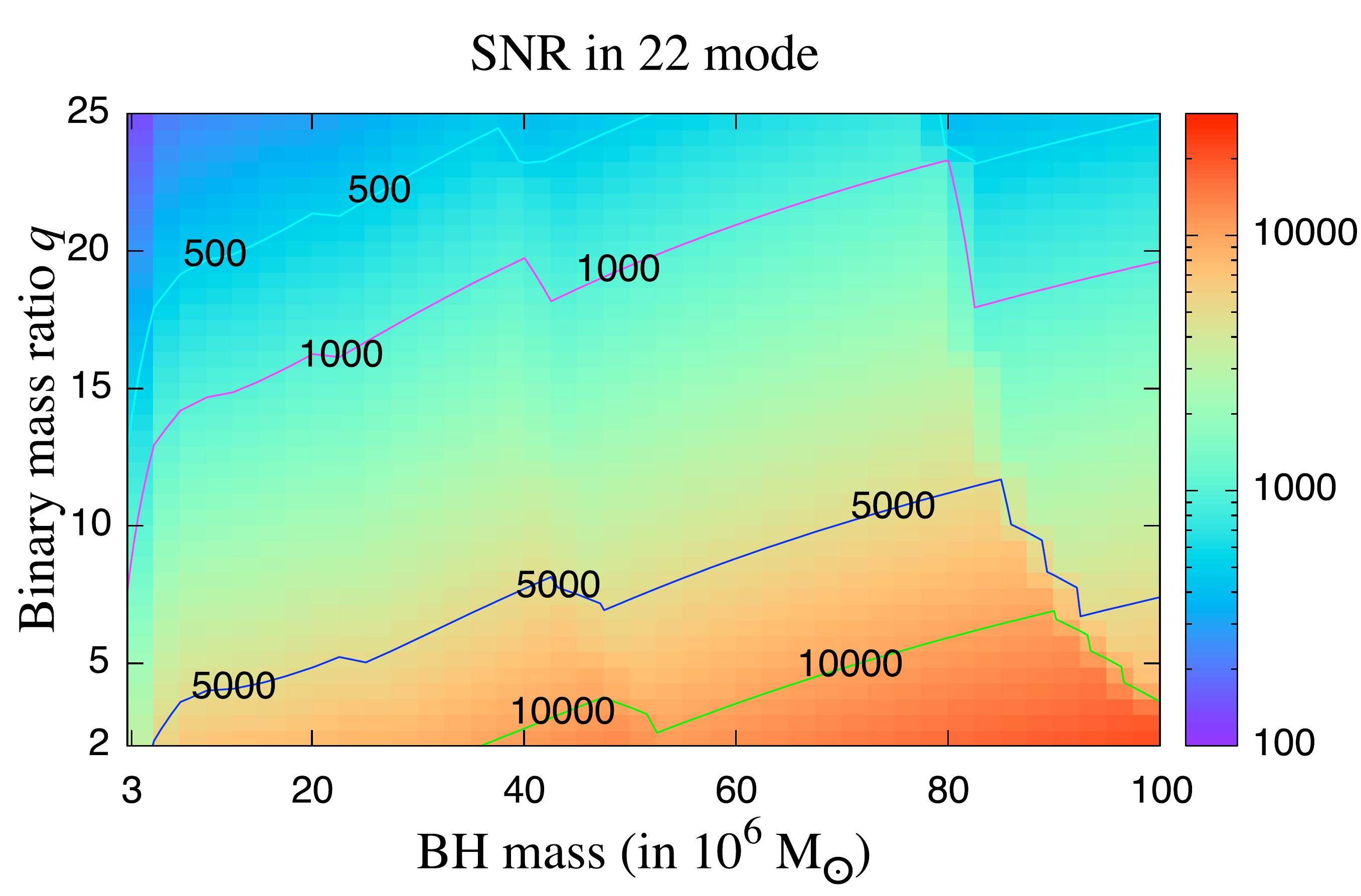} \\
\includegraphics[width=0.45\textwidth]{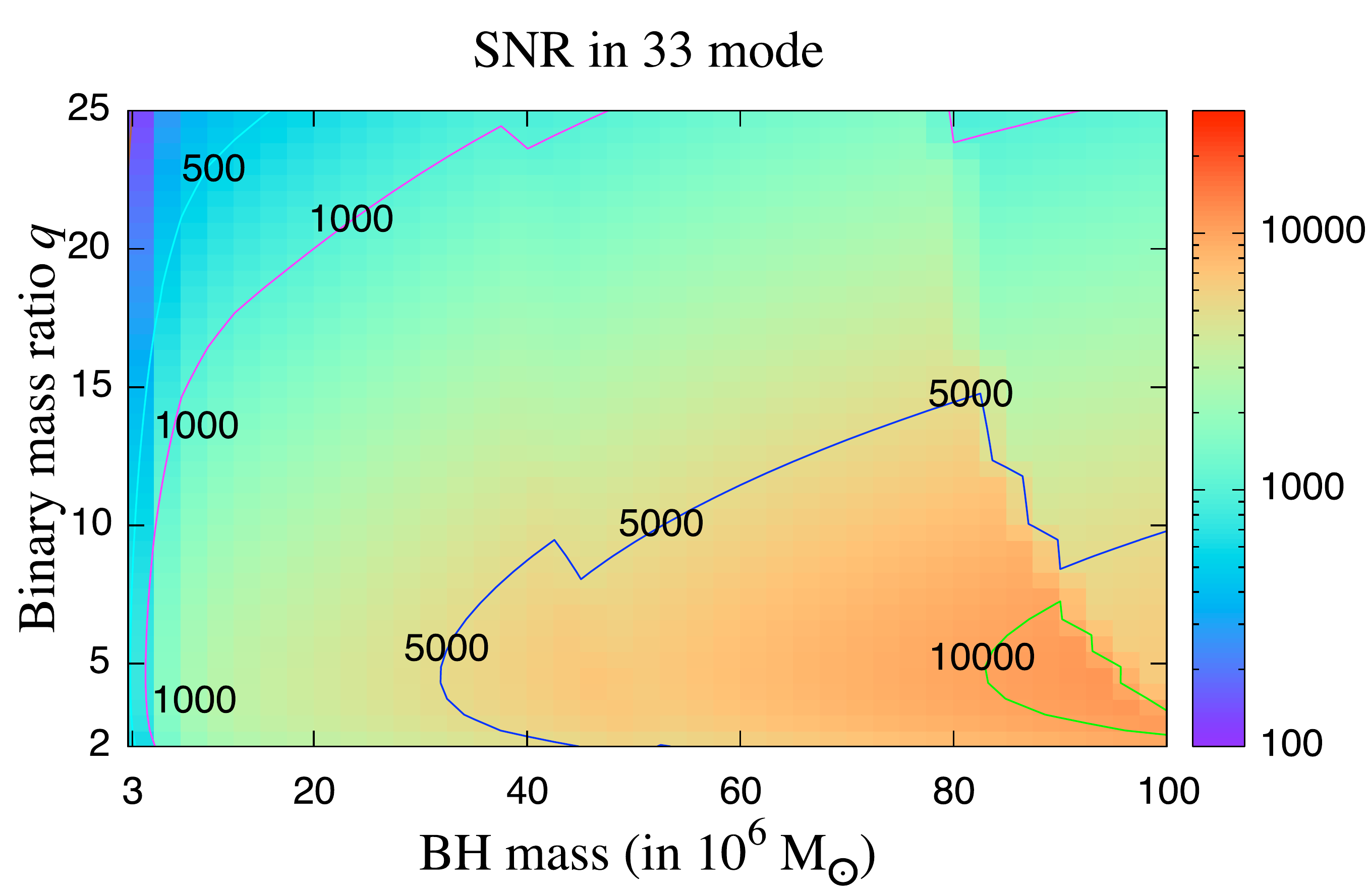} &
\includegraphics[width=0.45\textwidth]{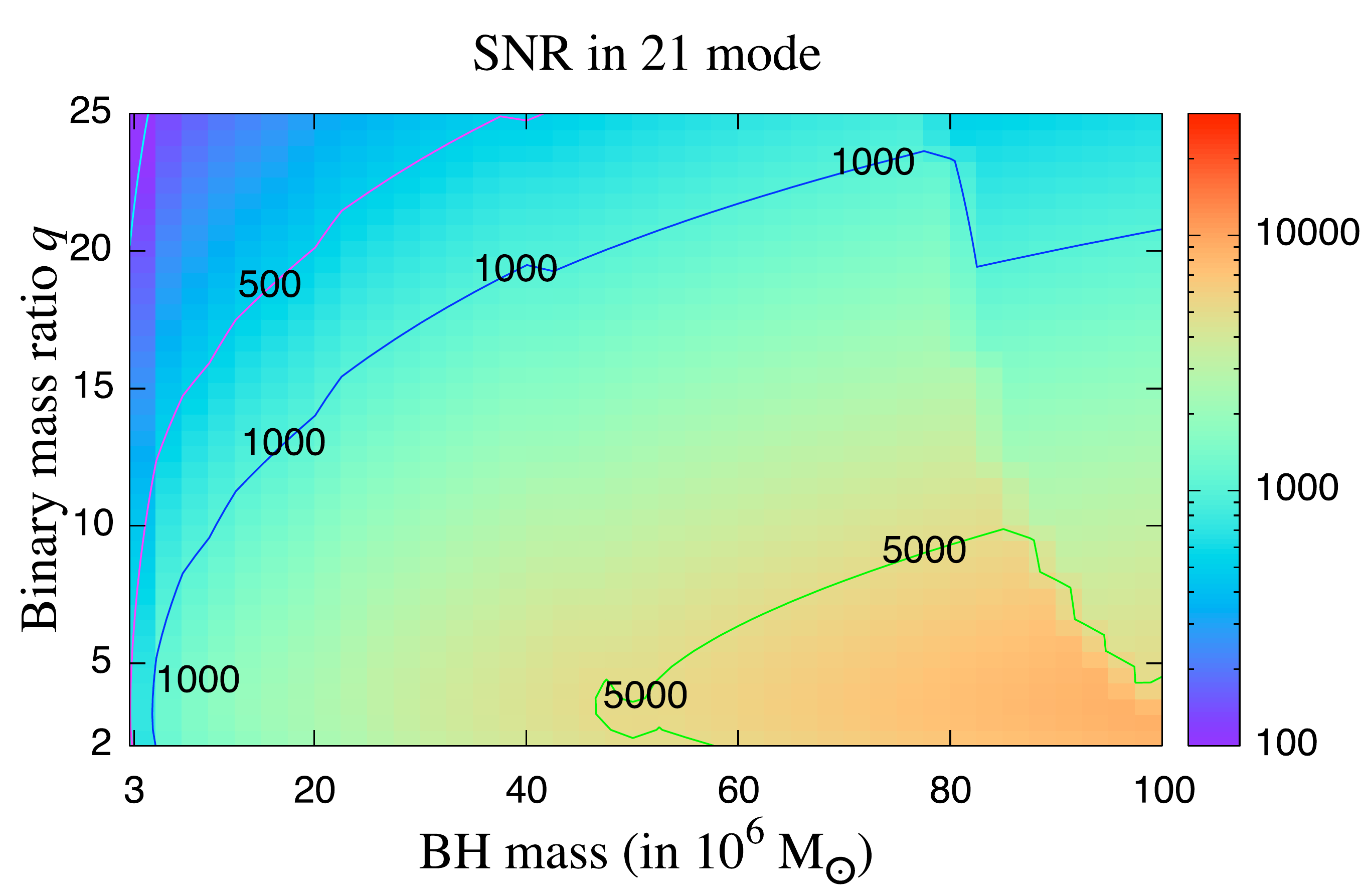}\\
\hline
\end{tabular}
\caption{Same as Fig.\,\ref{fig:snr ground} but for LISA and the source
is assumed to be at a red-shift of $z=1.$ Also, the mass ratio is allowed 
to vary over a larger range from 2 to 25 instead of 2 to 10.  The `steps' that can
be seen around $40\times 10^{6}M_{\odot}$ and $80\times 10^{6}M_{\odot}$ are mostly
due to the LISA noise curve.}
\label{fig:snr space}
\end{figure*}

Figures \ref{fig:snr ground} and \ref{fig:snr space} plot the SNR in the ringdown
signal (plots titled ``SNR in all modes") and contribution from 
the 22, 21 and 33 modes (plots titled accordingly) as a function of the
black hole mass $M$ and mass ratio $q$ of the progenitor binary for
aLIGO, ET and LISA; $M$ and $q$ are varied over the range as in 
Table \ref{tab:params}.  Most of the contribution
to the SNR comes from the 22 mode followed by 33 and 21. Let us recall
that SNRs from different modes do not add in quadrature.

In the case of aLIGO, the 22 and 33 modes will be visible in a significant
fraction of the parameter space explored provided the source is within
a distance of 1 Gpc. The 21 mode will not be visible in aLIGO at this 
distance except perhaps for the heaviest systems explored.

In the case of ET, assuming a SNR threshold of 10 for detection, 
the signal is visible to a red-shift of $z\sim 0.8$ in most of the 
parameter space explored. Black holes of total mass $M>400 M_\odot$ 
that form from the coalescence of binaries whose mass ratio is less 
than 4 will be visible at red-shifts $z\sim 2$-$3.5$ 

In the case of LISA, ringdowns produce a very large SNR. Even assuming an
SNR threshold of 40, LISA should see the formation of supermassive black 
holes in the range $[10^6,\, 10^8]\,M_\odot$ up to a red-shift of at
least $z\sim 6$ but if the progenitor black holes have mass ratio
$q < 10$ they should be visible from the earliest moments of their
formation in the Universe. 

Our results unambiguously demonstrate that ET and LISA can together
probe black hole demographics, ET exploring the lower end of the mass 
spectrum of seed black holes and LISA the higher end of that spectrum.
The two detectors together cover a large mass range
from $\sim 10^2\,M_\odot,$ all the way to $\sim 10^8\,M_\odot.$ (Although
out of the range of masses explored, note that ET could observe heavier
black holes of mass $10^4\,M_\odot$ and LISA could explore lighter black 
holes of mass $10^5\,M_\odot.$)
The distance reach will be different depending on the mass ratio of 
the progenitor binary and the total mass of the black hole. Even so, LISA
and ET will make it possible to explore the formation of black holes 
and trace their merger histories and possibly help understand the role of
black holes as seeds of galaxies and large scale structure in the Universe.

\section{What can a ringdown signal measure?} 
\label{sec:PE1}

By measuring the ringdown signal and resolving it into different modes, we should
be able to learn a great deal about the merger dynamics and test general
relativity. For instance, by determining the total mass of the binary from 
the inspiral phase and comparing it to the mass of the black hole obtained from
the ringdown phase we can measure, quite precisely, how much mass is converted
into radiation in the process of merger. LISA can typically measure
the total mass of a binary from its inspiral phase to a fraction of a percent.
We shall see in this section that the ringdown modes can determine a black
hole's mass to a percent or tenth of a percent depending on the mass ratio.
Therefore, the inspiral and ringdown phases together can shed light on how
much mass is lost in the process of merger and how that depends on the mass
ratio of the binary and, not probed in this study but expected to 
depend on, the spin magnitudes and orientations of progenitor black holes.

When black holes merge, some of the orbital angular momentum goes into the
final black hole. Therefore, the final black hole will spin in a direction
different from either of the progenitor black holes. Independent measurements
of the orbital angular momentum from the inspiral phase and black hole spin 
from the ringdown phase, could unravel the spin-orbit dynamics of black hole
merger. While this is an exciting possibility, in this paper we have focussed
only on binaries with non-spinning components.
For such systems, the relative amplitudes of the different modes depend 
on only the mass ratio of the progenitor binary.
For binaries with spinning black holes it is hard to guess how many more
parameters might be required to characterize the relative amplitudes
and hence their measurability. We will address this question in a forthcoming
publication.
We shall show in this section that one can exploit this fact to determine the 
mass ratio of the progenitor binary from the measurement of the ringdown mode 
alone. Consistency of the mass ratios from the inspiral and ringdown phases 
could offer further tests of general relativity. 

In this section we will explore what measurements might be possible by using
ringdown signals alone and what we might learn by combining the information
obtained from the inspiral phase of the signal with that obtained from
the ringdown phase. To this end, we shall assume that the signals are
loud and compute the covariance matrix to get an estimate of the measurement
uncertainties in the various parameters of a ringdown signal $h^A(t)$. 
The covariance matrix $C^A_{km}$ is the inverse of the Fisher matrix $F^A_{km}$ 
given by \cite{Wainstein,Finn92,BalSatDhu96},
\begin{equation}
F^A_{km} = \left < \frac{\partial h^A}{\partial \lambda^k},\, \frac{\partial h^A}{\partial \lambda^m}
\right >, \quad \lambda^k=\{M,\,j,\,q,\,\ldots\},
\label{eq:fisher}
\end{equation}
where for any two functions $g(t)$ and $h(t)$ the angular bracket 
$\left <g,\,h\right >$ denotes their {\em scalar product} defined by
\begin{equation}
\left <g,\,h\right >= 4\, \Re \int_{0}^\infty G(f)\,H^*(f) 
\frac{{\rm d}f}{S_h(f)}. 
\label{eq:scalar product}
\end{equation}
Here, as before, $A$ is an index denoting the detector in question,
$S_h(f)$ is the one-sided noise power spectral density of the
detector, $G(f)$ and $H(f)$ are the Fourier transforms of the
time-domain functions $g(t)$ and $h(t),$ respectively, and a $*$ denotes 
the complex conjugate of the quantity in question. The above integrals 
are often performed numerically
and it is essential then to appropriately choose the lower and upper limits in
the integral so that outside this limit the integral has negligible contribution.
Note that the detector noise power spectral density rises steeply outside a certain
frequency range often assuring the convergence of these integrals. If a network
of detectors is used to estimate the parameters then the Fisher matrix for the
network is simply the sum of the Fisher matrices for the individual detectors:
\begin{equation}
F_{km} = \sum_A F^A_{km},
\end{equation}
where the sum is over all the detectors in the network.

\subsection{The full parameter set}

In the case of a binary consisting of non-spinning black holes 
on a quasi-circular orbit, the effective amplitudes $B_{\ell m}$ in Eq.\;(\ref{eq:reduced response}) 
of the quasi-normal modes of the final black hole, depend on a set of eight 
parameters\footnote{Recall that the final spin of the black hole is determined
by the mass ratio of the progenitor binary and so it is not necessary to treat
both $q$ and $j$ as independent. However, such a treatment allows us to check if
the final black hole spin is consistent with the mass ratio as predicted by
numerical relativity simulations, which would indeed be another test of
general relativity.} $(M,\, j,\, q,\, D_{\rm L},\, \theta,\, \varphi,\, \psi,\, \iota):$
the mass $M$ and spin magnitude $j$ of the black hole, the mass ratio $q$ 
of the progenitor binary, the position vector $(D_{\rm L},\,\theta,\,\varphi)$ of the 
black hole with respect to Earth, the polarization angle $\psi$ and the inclination 
$\iota$ of the black hole's spin angular momentum with respect to the line-of-sight.
The phases $\gamma_{\ell m}$  are given by Eq.\, (\ref{eq:gammalm}) and they depend on 
the angles $(\theta,\, \varphi,\, \psi,\, \iota,\,\phi,\,\phi_{\ell m}),$ where 
$(\iota,\,\phi)$ are the spherical polar coordinates giving the direction in 
which the black hole quasi-normal mode is emitted in a frame fixed to the black 
hole and $\phi_{\ell m}$ are the initial phases of different quasi-normal modes.
Thus, if we consider a superposition of three quasi-normal modes
then there will be 12 parameters, including $\phi_{22},$ $\phi_{33}$ and $\phi_{21}.$
The amplitudes $B_{\ell m}$ depend on eight of these parameters (exceptions are $\phi$,
$\phi_{22},$ $\phi_{33}$ and $\phi_{21}$) and phases $\gamma_{\ell m}$
also depend on a (different) set of eight parameters (exceptions are $D_{\rm L},$ $M,$ $q,$ and $j$).

\subsection{Measurements with a network of detectors}
Measuring all the parameters of a ringdown signal will require simultaneous observation
of the signal in two or more detectors. Let us first take a look at the configurations of
LISA, ET and advanced ground-based detectors.

ET and LISA both have a triangular topology and consist of three V-shaped interferometers,
with an opening angle of 60 degrees, rotated relative to each other by 120 degrees. The 
three interferometers are completely equivalent, in terms of sensitivity, to two L-shaped
interferometers \cite{Cutler98}, with arms that are only three-quarters in length
of the arms in the triangle. Thus, for the purpose of detection and measurement,
we can consider ET and LISA to be a network of two collocated detectors.  At least 
three advanced ground-based detectors (two LIGO detectors and Virgo) would be operating
by 2015, with the possibility of the Japanese Large Cryogenic Gravitational Telescope
joining the network soon after. Thus, there will be a global network of ground-based
detectors that will be operating for a number of years from around 2015.

Of all the parameters, the direction to the source $(\theta,\varphi)$ is the most
critical and difficult to measure from the ringdown modes alone. However, since the ringdown
modes we study are preceded by the inspiral phase of a binary coalescence, we can expect
the direction to the source to have been measured to some degree of accuracy. In the case
of LISA, the inspiral phase of supermassive black holes could last for several months to
years in the detector band. The modulation of the signal caused by LISA's motion relative
to the source over the observation period will be good enough to measure the sky position
(see, for instance, Ref.\,\cite{AISSV07}). ET, together with a network of other detectors,
advanced or 3rd generation detectors present at the time, should be able to
triangulate the source. This is also true with the network of advanced detectors.
Thus, we shall assume that the parameters $(\theta,\varphi)$
are known, leaving 10 parameters to be measured from the ringdown phase.
However, for very massive systems (depending on the detector in question), only
the ringdown phase might be visible and for such systems it will not be possible 
to infer the location of the source without a network of detectors. 
Such events will not be very useful for testing GR.

As expected, the relative contributions of the inspiral and ringdown phases
depend on the total mass of the system: For lighter masses the ringdown phase 
makes little impact on parameter estimation; for heavier systems just the 
opposite is true. For systems with intermediate masses the contributions 
could be roughly equal.  Such systems will be golden binaries with the best 
ability to test GR.

A single detector can measure the mass $M$ and spin $j$ of the black hole by 
simply inverting the QNM frequencies and damping times.  Additionally, each 
detector in a network would also measure three independent amplitudes $B^A_{22},$
$B^A_{21}$ and $B^A_{33}$ and three independent phases $\gamma^A_{22}$, $\gamma^A_{21}$
and $\gamma^A_{33}$ --- a set of 12 additional measurements from two detectors. 
Of course, the amplitudes and phases (as well as the time-constants and mode 
frequencies) are all expressed in terms of the 10 physical parameters and will 
not treated as independent.  The counting argument given here shows that a 
set of two or more detectors allows enough measurements to fully reconstruct 
the ringdown signal.

Therefore, one can, in principle, measure all the ten parameters of a QNM 
composed of three modes, using a network of two or more detectors.  We have not, 
however, explored the problem in its full generality as the 
Fisher matrix that includes both intrinsic and extrinsic parameters 
happens to be highly ill-conditioned. In such cases, fisher matrix is not 
the right approach for computing the errors incurred.  We will, in the near
future, investigate this problem by other means, for instance using Bayesian inference.
For now, our goal is to see how well a subset of interesting
parameters can be measured if, as mentioned earlier, we know some of the parameters
from the inspiral phase.

For the sake of simplicity, we shall assume that the phase of 
the different quasi-normal modes at the beginning of the ringdown are all
the same and equal to zero: $\phi_{22}=\phi_{21}=\phi_{33}=0.$ 
In this work, we have dropped them from further consideration so that 
we can focus our effort on the main goal of the paper, which is to show that 
one can infer the mass ratio of the progenitor binary by measuring two or more
quasi-normal mode amplitudes.
We shall, therefore, assume that the ringdown signal depends on the parameters 
$(M,\,j,\,q,\,D_{\rm L},\, \iota,\,\psi,\,\phi),$ seven parameters in all. 
A single detector can measure the mass and spin of the black hole from 
the different mode frequencies and damping times, as well as three
amplitudes and three phases. Consequently, in the case of a simplified signal
model, where we have dropped the constant phases and the location of the black
hole from the list of parameters, we do not need a network of detectors to resolve 
the signal parameters.

\section{Understanding mass loss, spin re-orientation and mass ratio from
ringdown signals}
\label{sec:PE2}

\begin{figure*}
\begin{tabular}{|c|c|}
\hline
\includegraphics[width=0.45\textwidth]{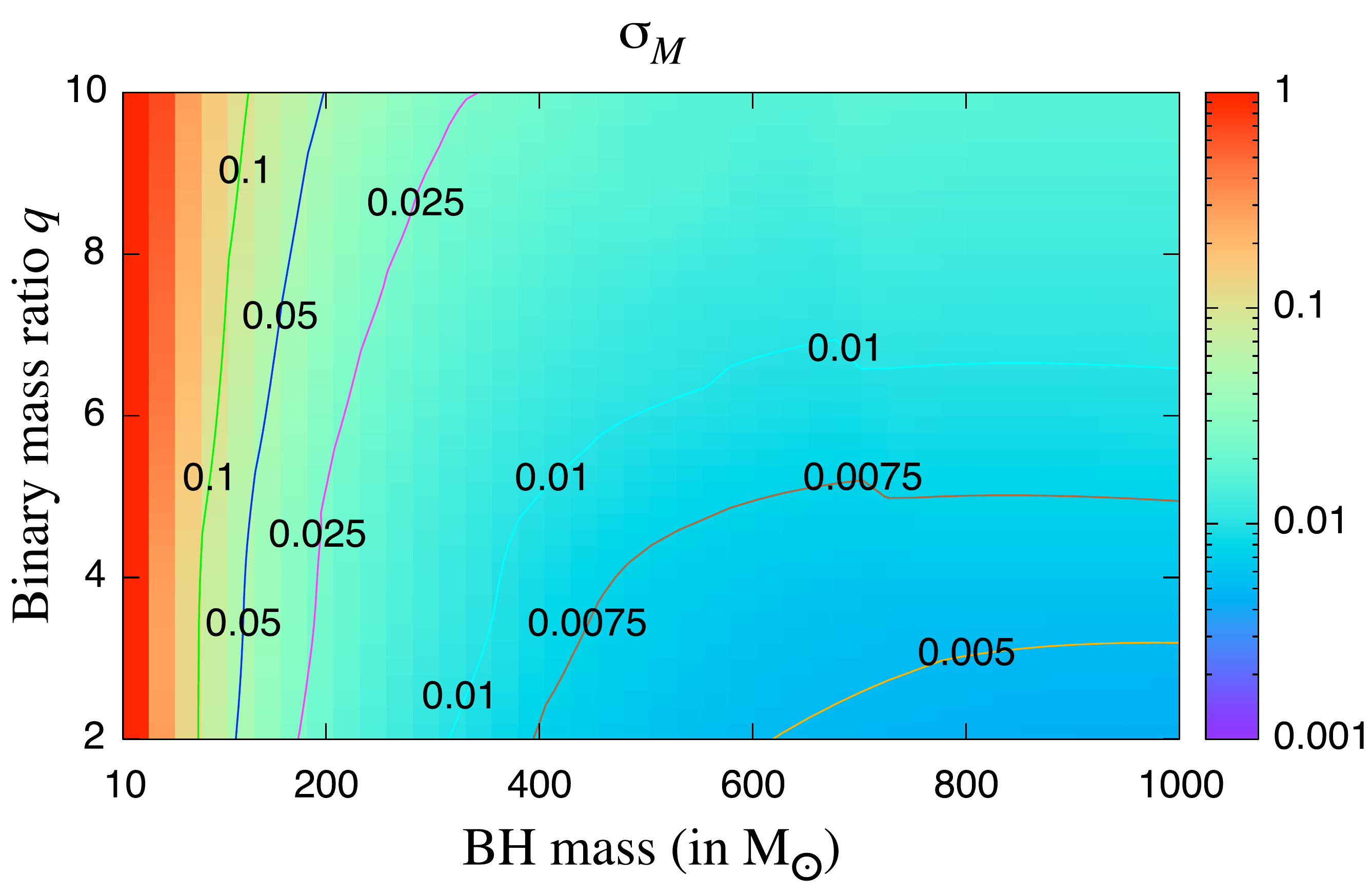}   &
\includegraphics[width=0.45\textwidth]{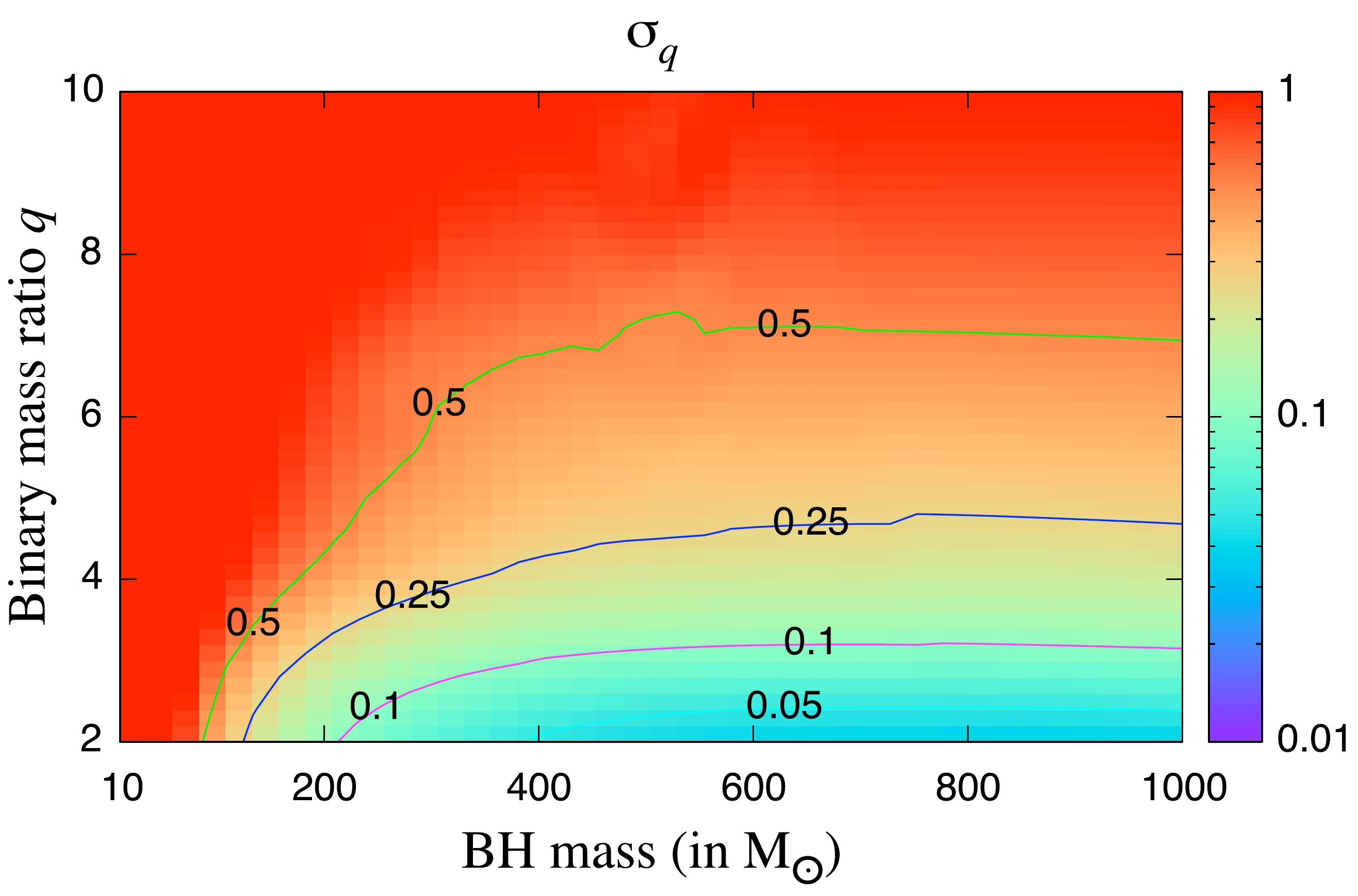}  \\
\hline
\includegraphics[width=0.45\textwidth]{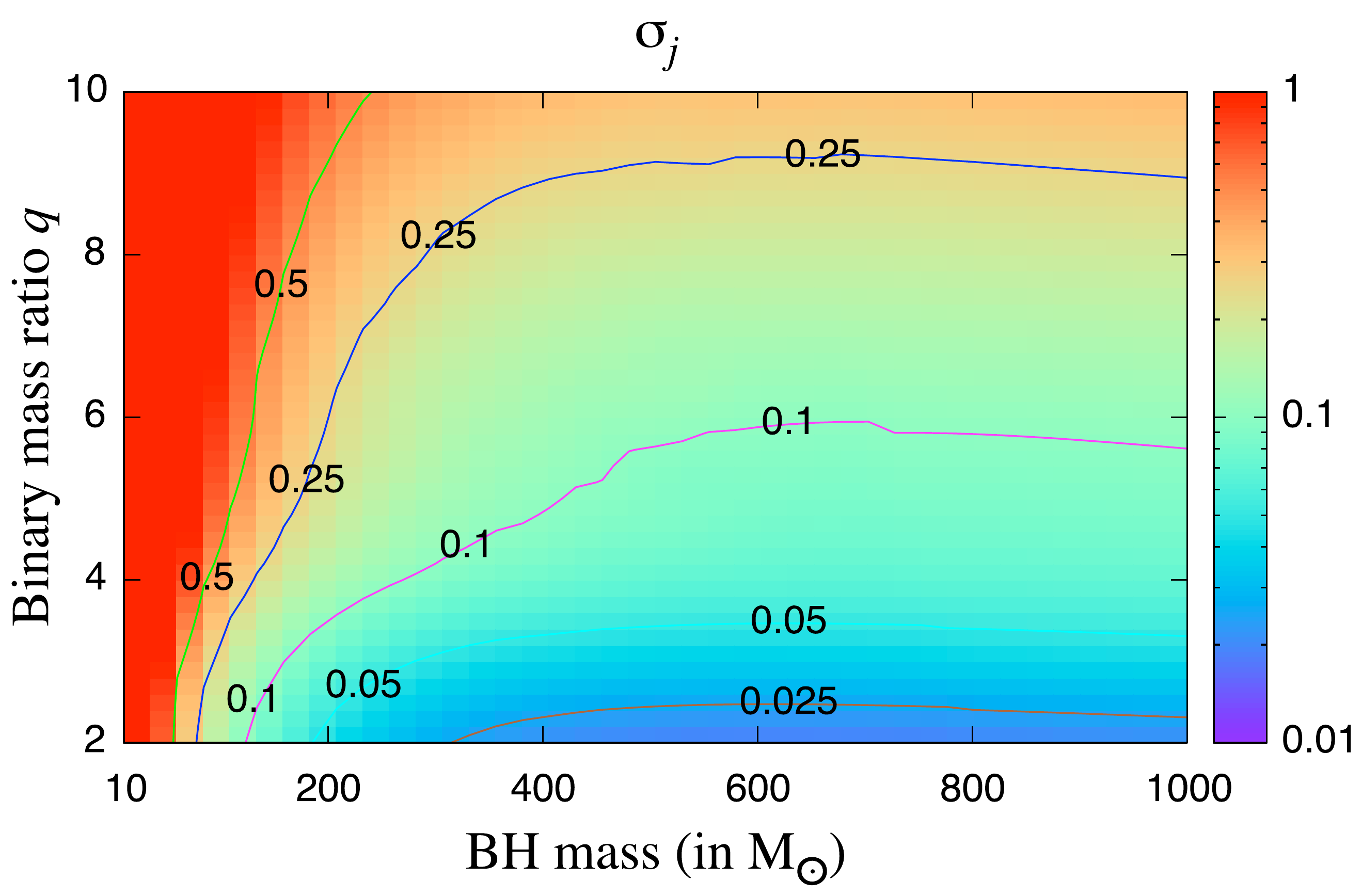}   &
\includegraphics[width=0.45\textwidth]{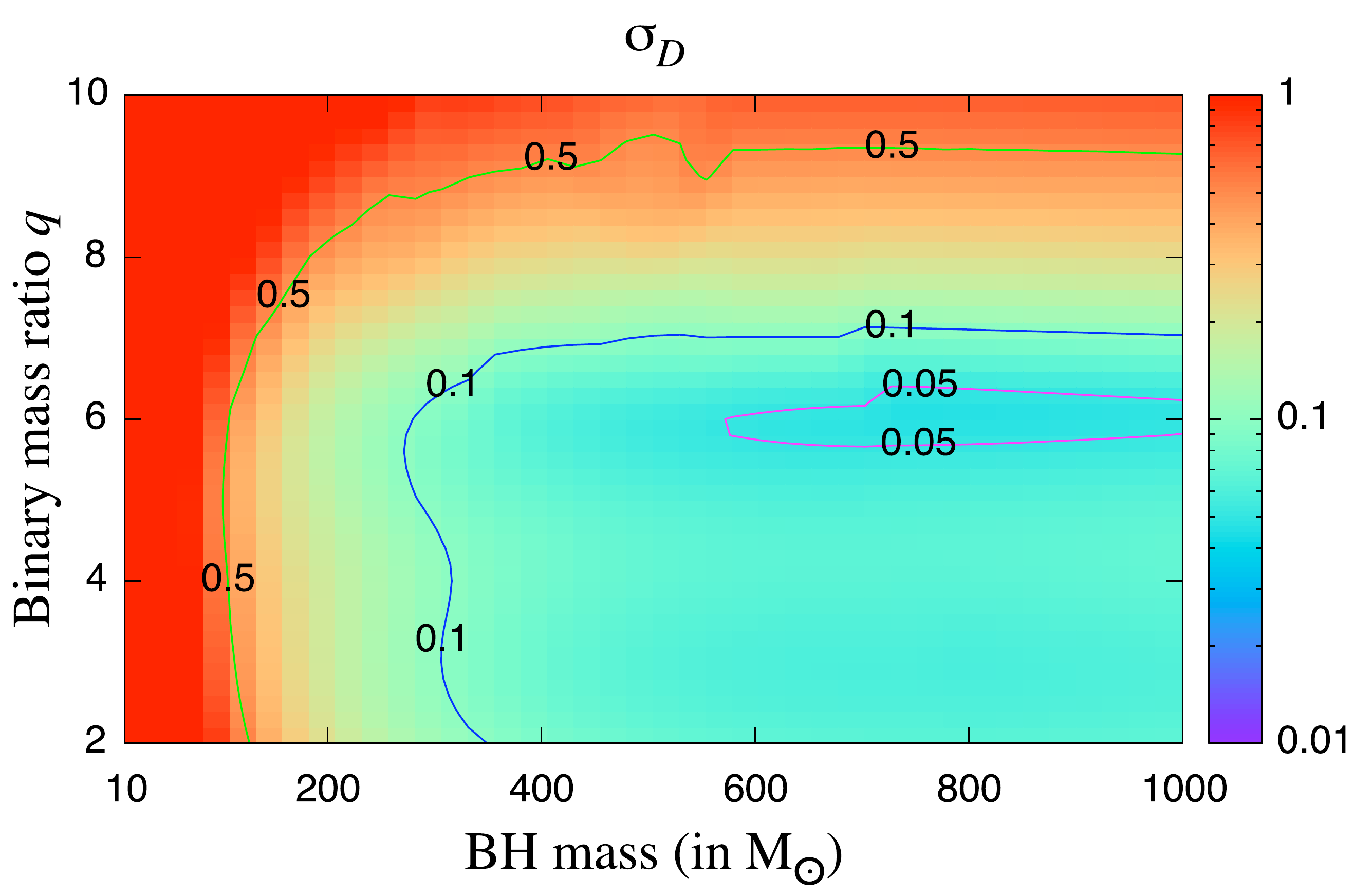}  \\
\hline
\includegraphics[width=0.45\textwidth]{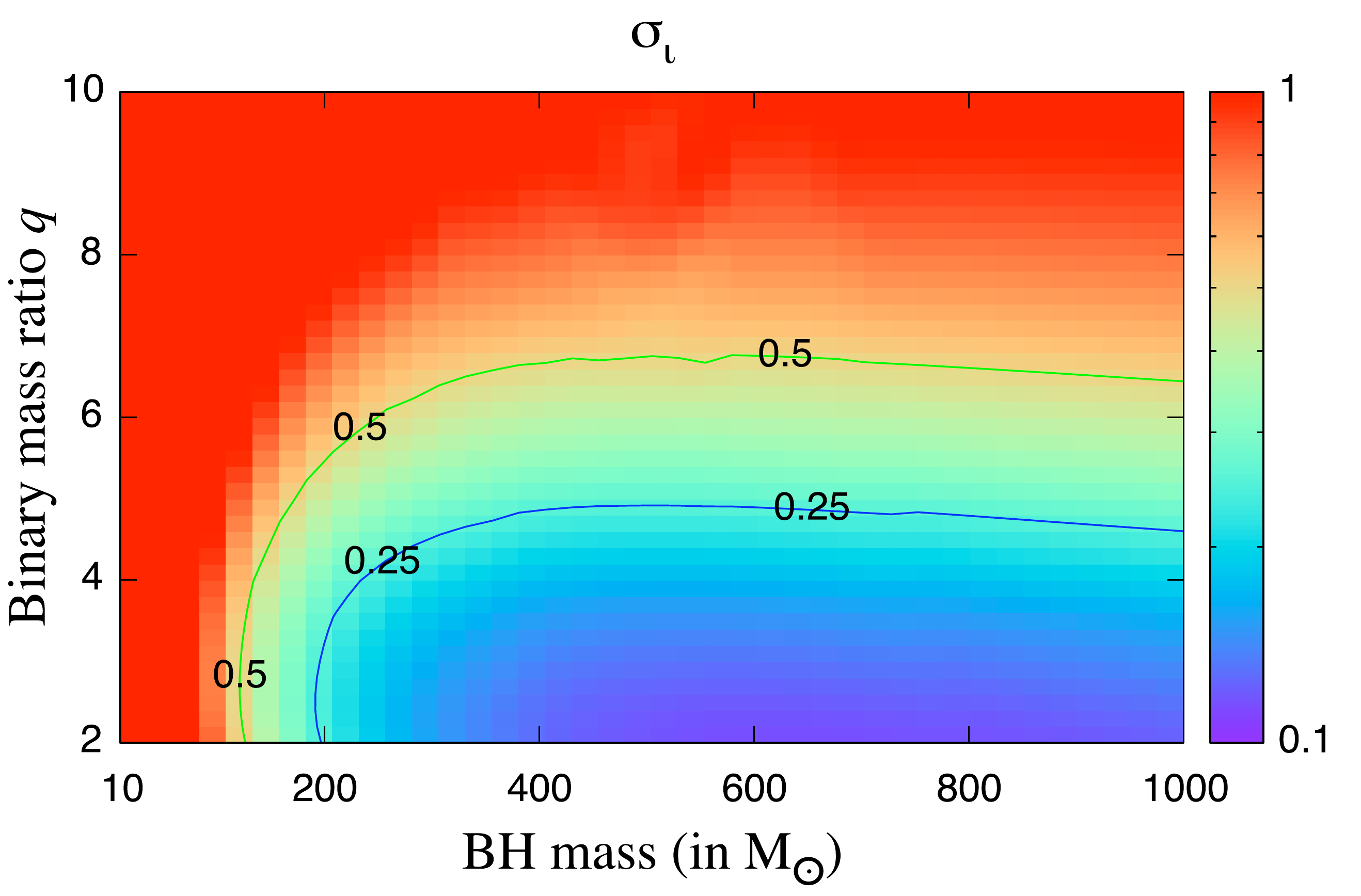}   &
\includegraphics[width=0.45\textwidth]{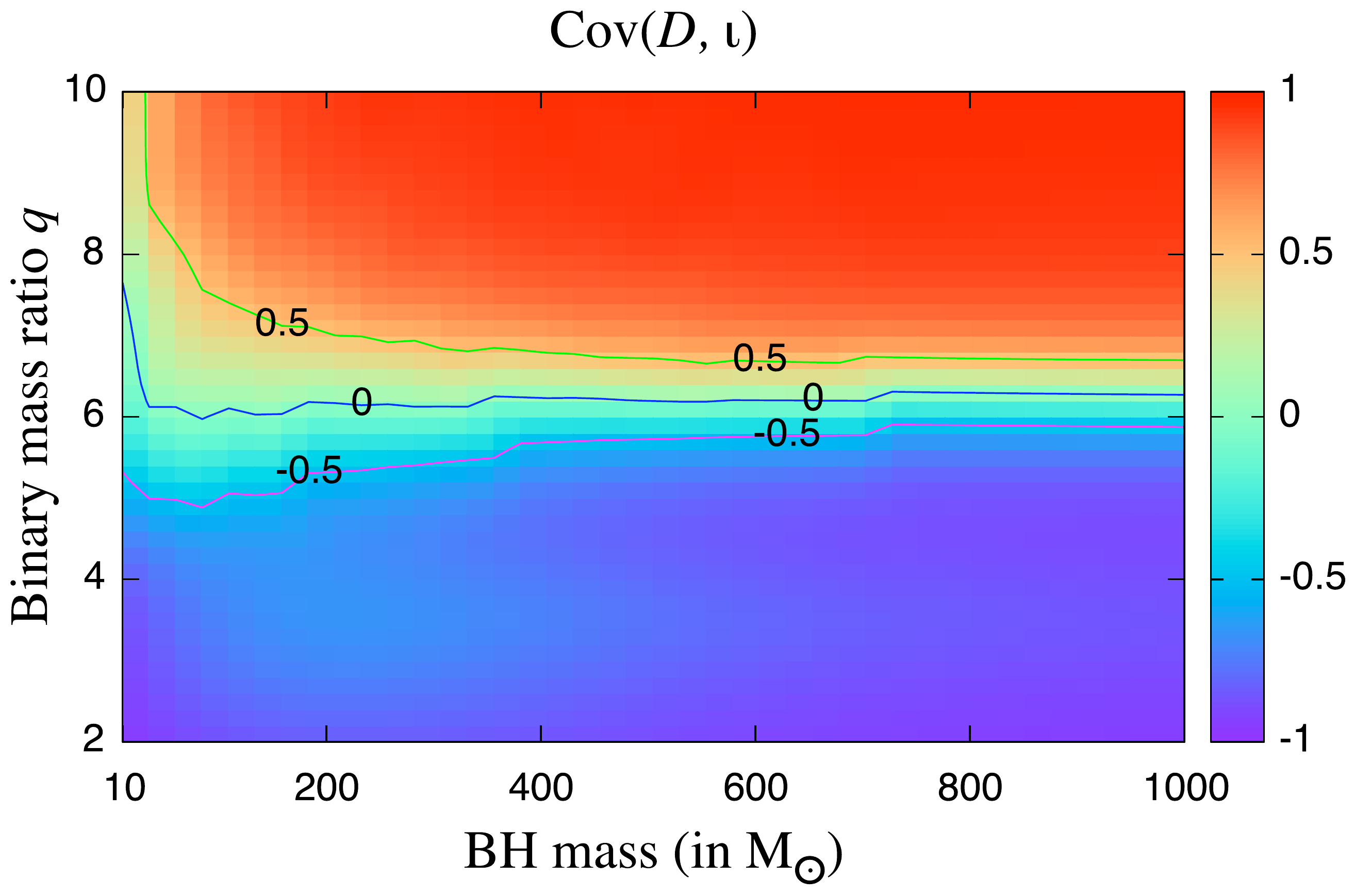}  \\
\hline
\end{tabular}
\caption{Dimensionless fractional errors $\sigma_k/\lambda_k$ in various parameters 
as a function of the black hole mass and progenitor binary's mass ratio.
The black hole is assumed to form at a luminosity distance of $D_{\rm L}
=1\,\rm Gpc$ and the various angles are assumed to be as in Table \ref{tab:params}.
The bottom right panel plots the correlation coefficient between the 
luminosity distance and the orientation $\iota$ of the black hole's spin 
with respect to the line-of-sight. The general trend for the errors is to 
increase with increasing mass ratio and decreasing mass, except for $D_{\rm L}$
for which there appears to be an `island' around $q=6$. We have used the ET-B 
sensitivity curve in computing the covariance matrix.}
\label{fig:error et}
\end{figure*}

\begin{figure*}
\begin{tabular}{|c|c|}
\hline
\includegraphics[width=0.45\textwidth]{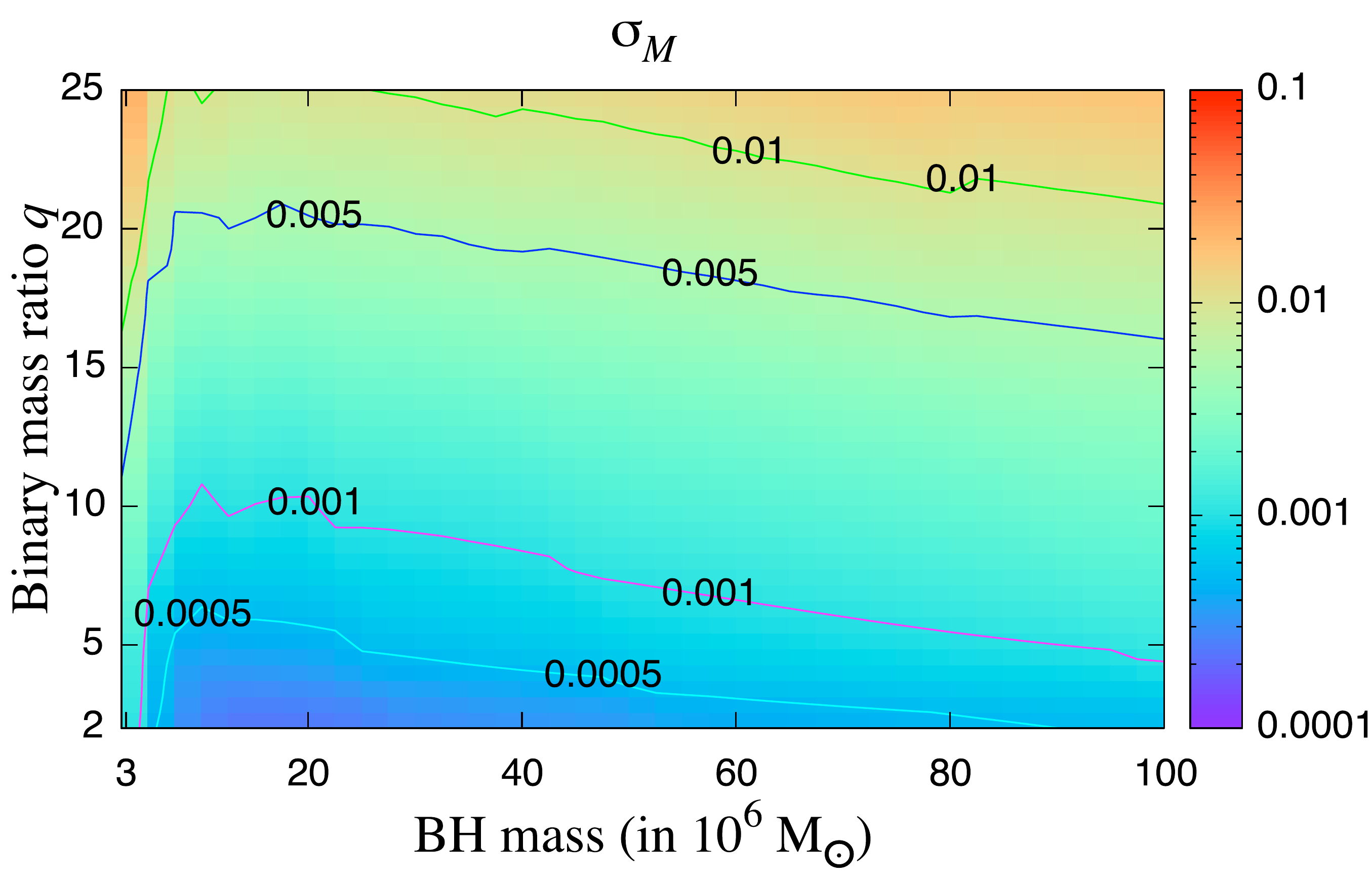} &
\includegraphics[width=0.45\textwidth]{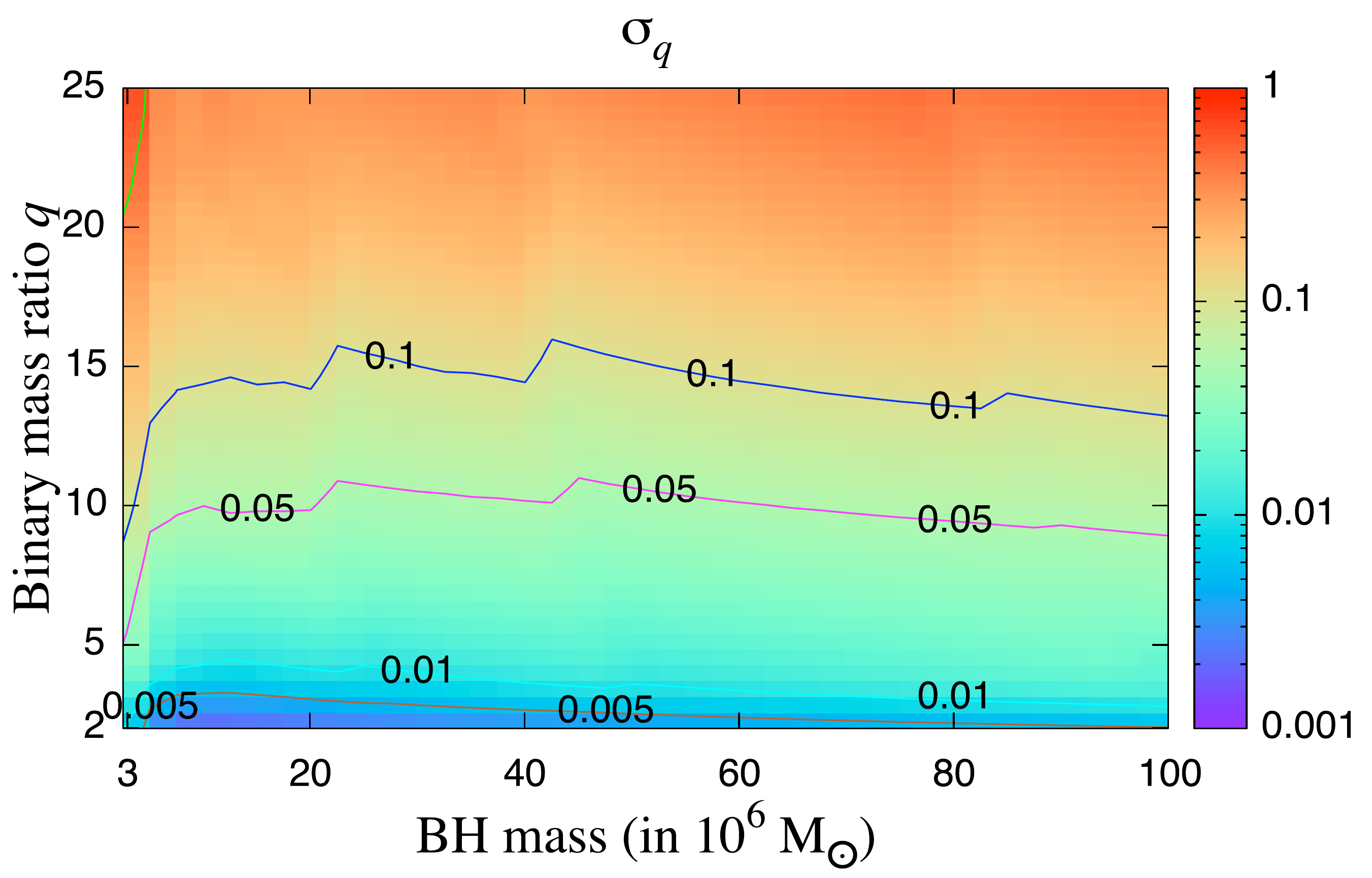} \\
\hline
\includegraphics[width=0.45\textwidth]{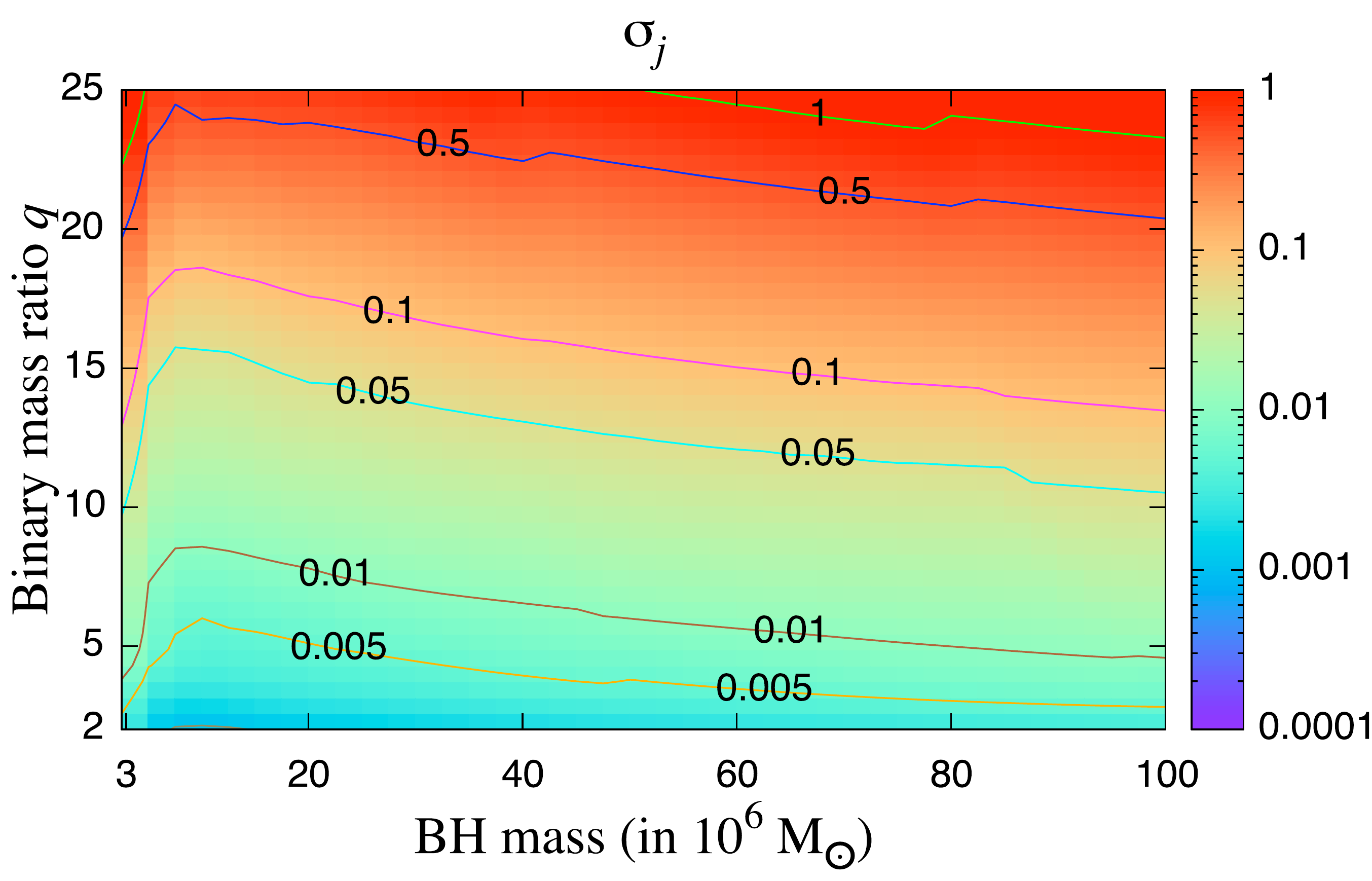} &
\includegraphics[width=0.45\textwidth]{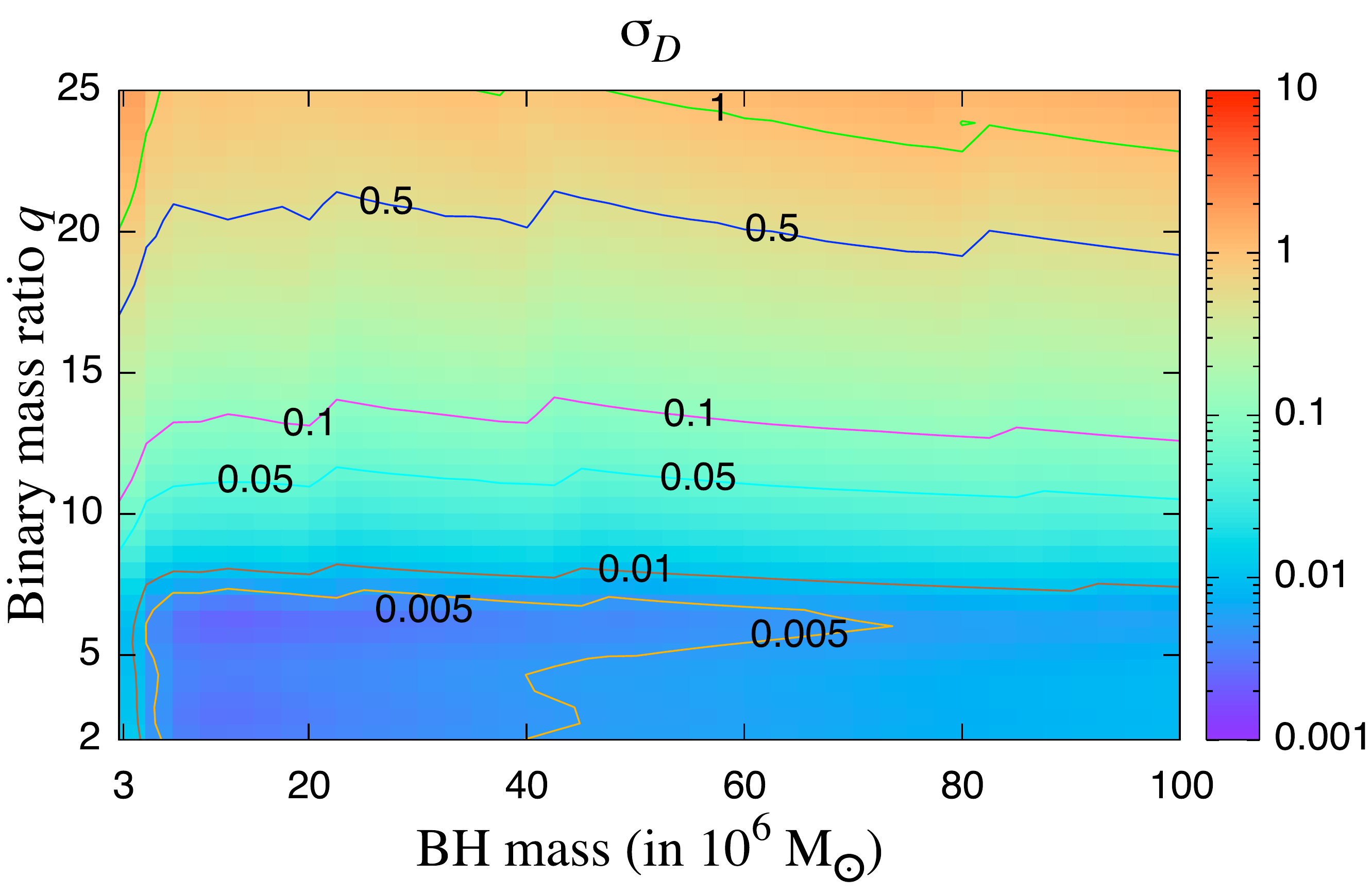} \\
\hline
\includegraphics[width=0.45\textwidth]{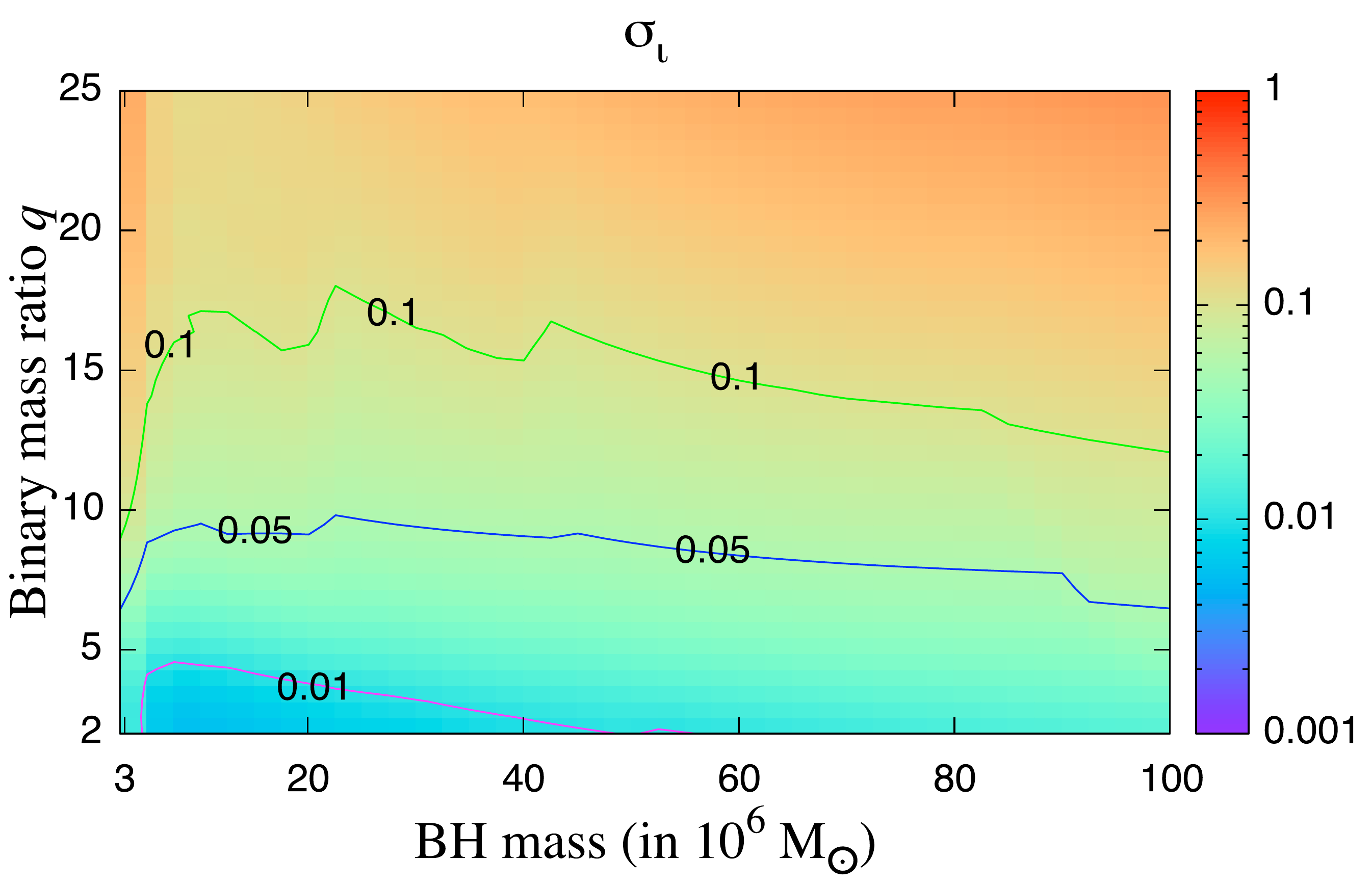} &
\includegraphics[width=0.45\textwidth]{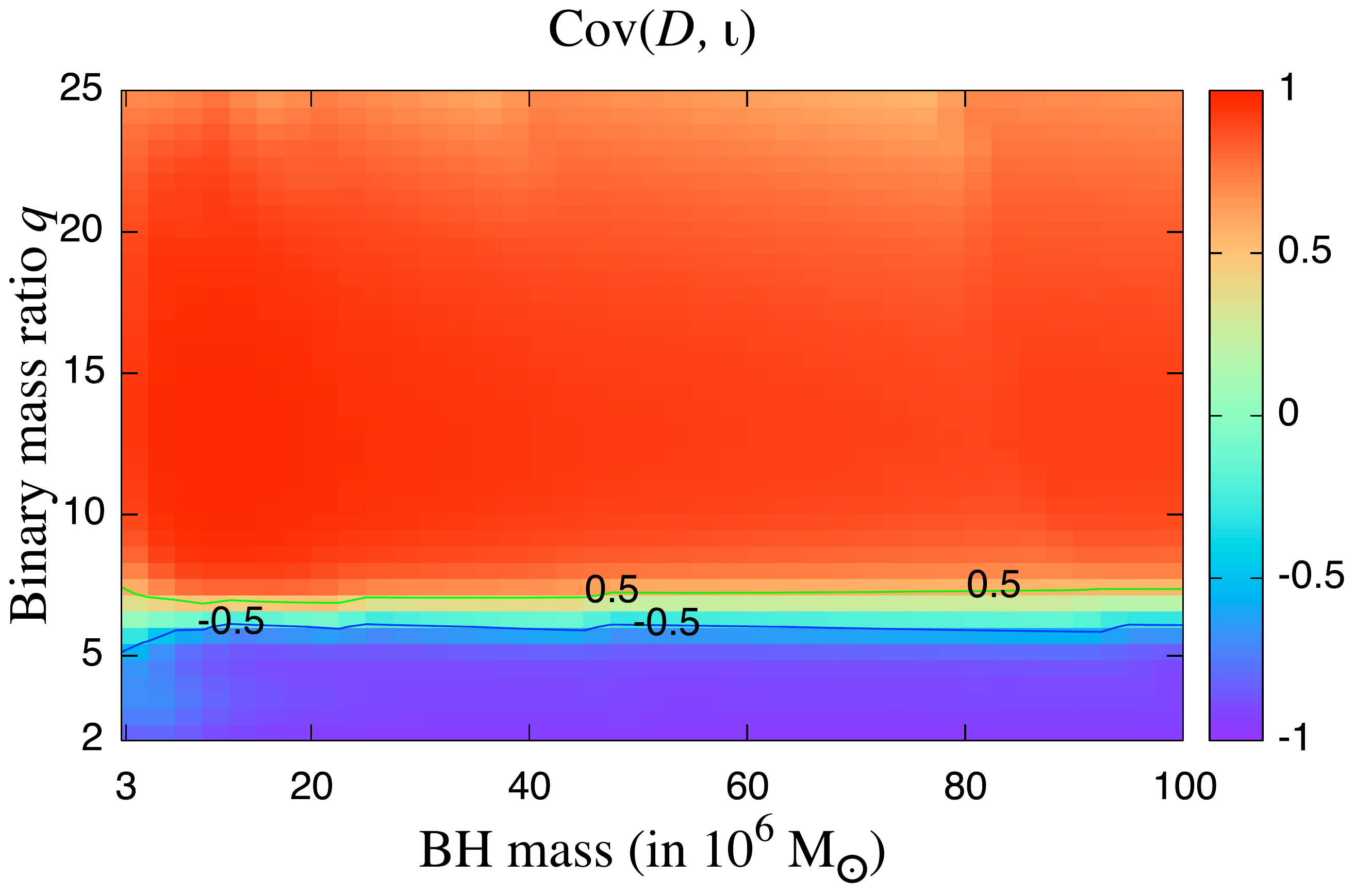} \\
\hline
\end{tabular}
\caption{As in Fig.~\ref{fig:error et} but for supermassive black holes
observed in LISA at a red-shift of $z=1.$ Note also that the mass ratio
$q$ in this case is allowed to vary from 2 to 25.}
\label{fig:error lisa}
\end{figure*}

We have computed the covariance matrix $C_{km}$ [cf.\ Eq.\,(\ref{eq:fisher})]
of ringdown signals as a function of the black hole mass $M$ and the mass 
ratio $q$ of the progenitor binary. In computing the covariance matrix, we
used, as described in the previous section, a signal model with seven 
parameters: $\lambda^k=\{M,\,j,\,q,\,D_{\rm L},\, \iota,\, \psi,\, \phi\}.$ 
The full covariance matrix contains 7 variances $C_{kk}$ and 21 covariances
$C_{mk}=C_{km},\, k\ne m$.  The full set of results is too large as our covariance 
matrix contains 28 independent elements at each point in the $(M,\,q)$ plane.  
To save space, we have chosen a subset of these for further discussion. 

We will discuss the error in the estimation $\sigma_k=\sqrt{C_{kk}}$
of 5 of the 7 parameters, $\lambda^k=\{M,\,j,\,q,\,D_{\rm L},\, \iota\},$ and
also include in our discussion the covariance between the luminosity distance $D_{\rm L}$
and the inclination $\iota$ of the black hole's spin axis with the line-of-sight.
Also, it is instructive to deal with the {\em correlation coefficient} defined as
$c_{km}\equiv C_{km}/(\sigma_k\,\sigma_m),$ instead of the covariances 
themselves. Correlation coefficients are bound to the range $[-1,\,1]$
and capture how variation in one parameter might be offset by varying another.
A correlation coefficient of 0 for a pair of parameters indicates that they 
are completely independent of each other and have ``orthogonally" different
effects on the waveform. For instance, the amplitude $A$ and phase $\varphi$
of a simple sinusoid function $s(t)=A\,\sin[\omega\,(t-t_0) +\varphi]$ will have a 
correlation coefficient of 0, while the phase and time offset $t_0$ are perfectly
anti-correlated and so have a correlation coefficient of $-1:$ A change in 
the phase cannot be mimicked by a change in the amplitude but it is completely 
replicated by a change in the time offset. Consequently, the amplitude is
completely independent of the phase and time offset, while only one of
phase or time offset can be considered to be an independent 
parameter\footnote{In fact, the Fisher matrix for the parameter set 
$(A,\,\varphi,\,t_0)$ will, as can easily be verified, be singular.}. When
covariances are close to $\pm 1,$ the parameters concerned will have large
uncertainties and this, as we shall see below, is a major source of error
for the parameters\footnote{This important point was noted recently by
Nissanke et al \cite{Nissanke:2009kt}.} $(D_{\rm L},\, \iota).$

Figures \ref{fig:error et} and \ref{fig:error lisa} plot fractional errors 
(i.e.~$\sigma_{\lambda_k}/\lambda_k$) incurred 
in the measurement of the parameters $(M,\,q,\,j,\,D_{\rm L})$ and the absolute
error in the parameter $\cos\iota.$ We have also plotted the 
correlation coefficient $c_{D_{\rm L}\iota},$ labelled in 
the figure as Cov($D,\iota$). Figure \ref{fig:error et} corresponds to
ET's observation of stellar and intermediate mass black holes in the range 
$[10,\,1000]\, M_\odot$ and the mass ratio $q$ of the progenitor binary
in the range $[2,\,10].$ Figure \ref{fig:error lisa} corresponds
to LISA's observation of supermassive black holes of mass $M$ in the range 
$[3\times 10^6,\,10^8]\, M_\odot$ and mass ratio $q$ in the range $[2,25].$ 
In the case of aLIGO, in most of the parameter space and for all parameters
(except the total mass), the fractional errors are larger than 50\% and so
they are not shown. Advanced LIGO will determine the total mass of an 
intermediate mass black hole that forms from the merger of two nearly 
equal mass black holes within 1 Gpc to
within a few percents and this could be very interesting for some of the tests
of general relativity to be discussed below. 

Let us recall that our results for mass ratios greater than 11 are 
based on the extrapolation of analytical fits to numerical simulations 
of binary black hole mergers that are only available up to a mass ratio 
of $q=11.$ In the next few sections we will discuss our results in the 
context of the science questions they can address.

\subsection{Mass ratio and component masses of the progenitor binary}

A key result of our study is that one can measure the mass ratio of a 
progenitor binary by observing the ringdown signals emitted by the 
black hole that forms from the merger. While the different mode frequencies 
and time-constants all depend only on the mass of the black hole and its 
spin magnitude, their relative amplitudes depend on the mass ratio. As we have
argued before, under certain circumstances it should be possible to measure
the relative amplitudes of the different modes, thereby measure the
mass ratio of the progenitor binary and hence deduce its component masses.

The top right panels of Figs.~\ref{fig:error et} and \ref{fig:error lisa} show
the fractional accuracy with which the mass ratio can be determined assuming
that the signal is composed of three  ringdown modes, namely (2,2), (3,3) 
and (2,1).  ET will not measure the mass ratio very well in most of
the parameter space. However, for equal mass mergers, ET should constrain
the mass ratio to within 5\%. Remarkably, LISA will be able to measure the
mass ratio to better than 10\% over 60\% of the parameter space studied 
and black holes that result from the merger of equal mass black holes enable 
the determination of $q$ to better than 1\%.

Since the mass ratio of a binary is easily determined from the inspiral phase,
its measurement also from the ringdown phase offers newer tests of general 
relativity.

\subsection{Mass loss to gravitational radiation}
In the process of inspiral and merger, a binary black hole emits a 
significant fraction (a few percent) of its mass as gravitational radiation. 
The total mass of a binary can be measured very accurately from the inspiral
radiation it emits.  Estimates\footnote{Note that most of the literature
quotes error in the measurement of the chirp mass ${\cal M}\equiv M
\,\nu^{3/5},$ where $M$ is the total mass and $\nu$ is the symmetric mass
ratio of the binary. To estimate the error in the total mass we have used
the error propagation formula 
$$
\left ( \frac{\sigma_M}{M} \right )^2 =
\left ( \frac {\sigma_{\cal M}}{\cal M} \right )^2
+ \frac{9}{25} \left ( \frac {\sigma_{\nu}}{\nu} \right )^2
-\frac{6}{5} c_{\cal M,\nu}\left ( \frac{\sigma_{\cal M}}{\cal M}\right ) \left ( \frac{\sigma_{\nu}}{\nu}\right ).
$$
where $c_{\cal M,\nu}$ is the correlation coefficient of the parameters
$\ln \cal M$ and $\ln \nu.$} range from a fraction of a percent (for 
an equal mass binary black hole of total mass 200 $M_\odot$ at a distance 
of 1 Gpc) in the case of Einstein Telescope \cite{ChrisAnand06b} to 50 
parts per million (for an equal mass binary black hole of total mass 
$2\times 10^6\,M_\odot$ at a distance of 3 Gpc) in the case of LISA 
\cite{ALISA06}.  For most binaries observed with ET and LISA, the total 
mass before merger can be measured with an error that is much smaller 
than the fraction of mass that is expected to be lost in gravitational radiation 
during merger. 

What we see from the top
left panels of Figs.~ \ref{fig:error et} and \ref{fig:error lisa} is that the
mass of the final black hole that results from a merger can also be measured
very accurately. ET cannot measure masses of stellar mass black holes very well
but if intermediate mass black holes (and binaries composed of such black holes)
exist, then ET will measure their masses to better than 1\%, assuming the source
is at 1 Gpc, over a significant
range of the parameter space ($q\lsim6$ and $M>400\,M_\odot$)
we explored. A black hole that results from the merger of two black holes
each of mass roughly about $500\,M_\odot$ could be measured to an accuracy
of better than half-a-percent.

LISA is able to measure the mass of a supermassive black hole that it observes
at a red-shift of $z=1$ with an accuracy of better than 1\% all over the parameter
space that we explored. Masses of supermassive black holes that form from the 
merger of two roughly equal-mass ($q\lsim 10$) black holes could be measured
to an accuracy of 0.1\%. This means that from the ringdown signal alone, we should
be able to measure masses of supermassive black holes even at a red-shift
of $z=5$ to better than 1\%. 

It would, therefore, be very interesting to compare the observed mass loss with 
the predictions of analytical and numerical relativity and verify if the mass loss is 
in accord with their predictions. Such comparisons will put general relativity
to new kinds of tests in the dissipative regime of the theory. Let us recall that
the luminosity of a binary black hole, close to merger, could be as large
as $10^{50}\,\rm J\,s^{-1},$ which is arguably the largest luminosity any
physical system could have. It would be very interesting to test the theory
when the luminosity is as large as this.

These phenomenal accuracies with which masses can be measured raise the 
question if it is prudent to treat the mass of a binary to be constant in 
the course of its inspiral and merger. It might be possible to deduce the
rate of mass loss by treating in our computation of the waveforms the mass 
of the system to be a function of time.

\subsection{Exploring naked singularities}
The magnitude and orientation of the spin of the final black hole 
that results from the merger of a black hole binary depend on a
number of parameters of the progenitor binary: magnitudes and 
orientations of the spins of the two component black holes relative
to the orbital angular momentum and the mass ratio of the progenitor 
binary.  A spinning black hole binary has a rather large parameter space, six 
parameters more than a non-spinning system. Limited studies have
been carried out in assessing how well one might be able to measure
black hole spins from the inspiral phase of the merger of a black hole 
binary \cite{Vecchio04,LangHughes06}. Numerical relativity simulations of the merger of spinning 
black holes are still in their early stages. In the coming years we
are likely to learn a great deal about spin dynamics of a binary
before and after merger. While these are important problems to be addressed
in the future, we recall that in this paper we have only studied binaries 
comprising of initially non-spinning black holes.

Figures \ref{fig:error et} and \ref{fig:error lisa},
middle left panels, show how accurately one might be able to measure 
the spin magnitude $j$ in ET and LISA, respectively. The accuracy here
is not as good as in the case of the black hole mass. ET can deduce the
final spin to within 10\% over 40\% of the parameter space and to better than
5\% in 20\% of the parameter space, for black holes that form within 1 Gpc.
LISA, on the other hand, can measure spin magnitudes to better than 5\%
over 50\% of the parameter space and to better than 1\% over 20\% of the parameter
space for black holes that form within a red-shift of $z=1.$ Spins of black holes
that form from the merger of nearly equal mass black holes can be measured to
0.5\%.  Thus, LISA should be able to reliably measure spin magnitudes that are 
only a few percent larger than 1.  Measuring spin magnitudes to such a 
high accuracy will be useful in testing whether a merging binary 
results in a black hole or a naked singularity \cite{BHspect04}.

\subsection{Spin orientations and the luminosity distance}
The measurement accuracy of spin orientation, given by $\cos\iota,$ is shown
in the bottom left panels of Figs.\,\ref{fig:error et} and \ref{fig:error lisa}.
$\cos\iota$ can be measured to within 10\% in about one-third of the parameter space
in the case of ET and to better than 5\% in about one-half of the parameter
space in the case of LISA.  The dynamics of spins before and after merger 
could be relevant in understanding the x-shaped radio galaxies \cite{Merritt:2002hc}.

The spin orientation of a black hole is very strongly correlated with the
luminosity distance. We see from the bottom right panels that the correlation
coefficient is close to either $+1$ or $-1$ in most of the parameter space.
The transition from negative to positive correlation between $D_{\rm L}$ and
$\cos\iota$ occurs when mass ratio $q\simeq 6.$ The significance of this number 
is not clear to us at the moment. This correlation completely destroys the
accuracy with which the luminosity distance can be measured. For instance,
we see that in the case of ET $\sigma_{D_{\rm L}}/D_{\rm L}$ is in the range
$5$-10\%, although the SNR in this region of the parameter space is $\sim 300.$
The parameter $\cos\iota$ in this region is also determined rather poorly, at
about 25\%. 

These numbers are better in the case of LISA: For 50\% of the parameter space,
LISA can measure the distance to within 10\% but black holes that result from
the merger of roughly equal mass systems might allow the luminosity distance to
be determined to 1\% or less. Using the inspiral phase alone, but with the help
of sub-dominant signal harmonics, luminosity distance can be measured to a fraction
of a percent in the case of LISA. Thus, it would be very interesting to see how
similar are the distances obtained from these different phases of the merger dynamics.

\section{Parametrization for testing the no-hair theorem}
\label{sec:GRTEST}
\begin{figure*}
\centering
\includegraphics[width=0.45\textwidth]{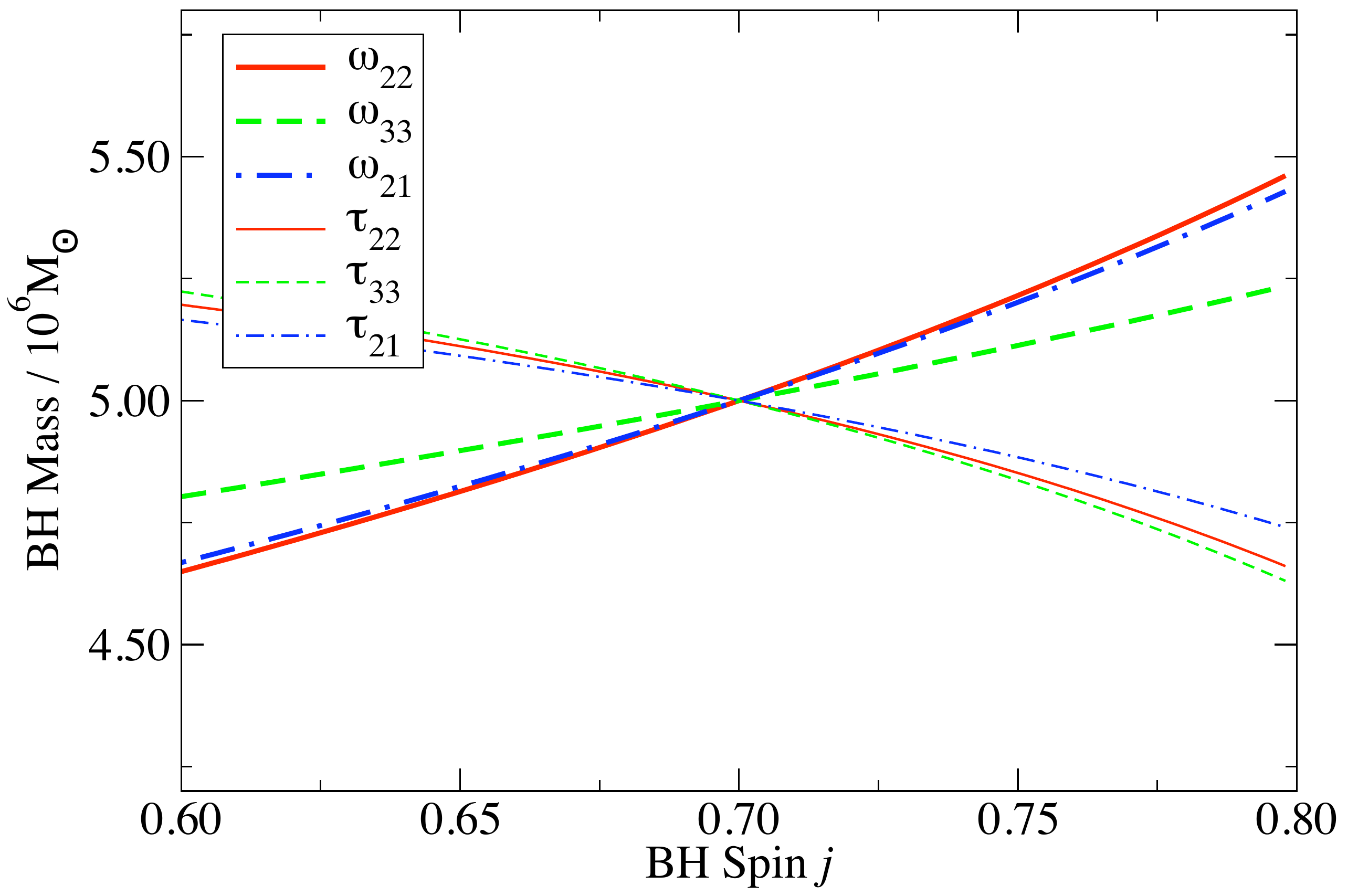}
\includegraphics[width=0.45\textwidth]{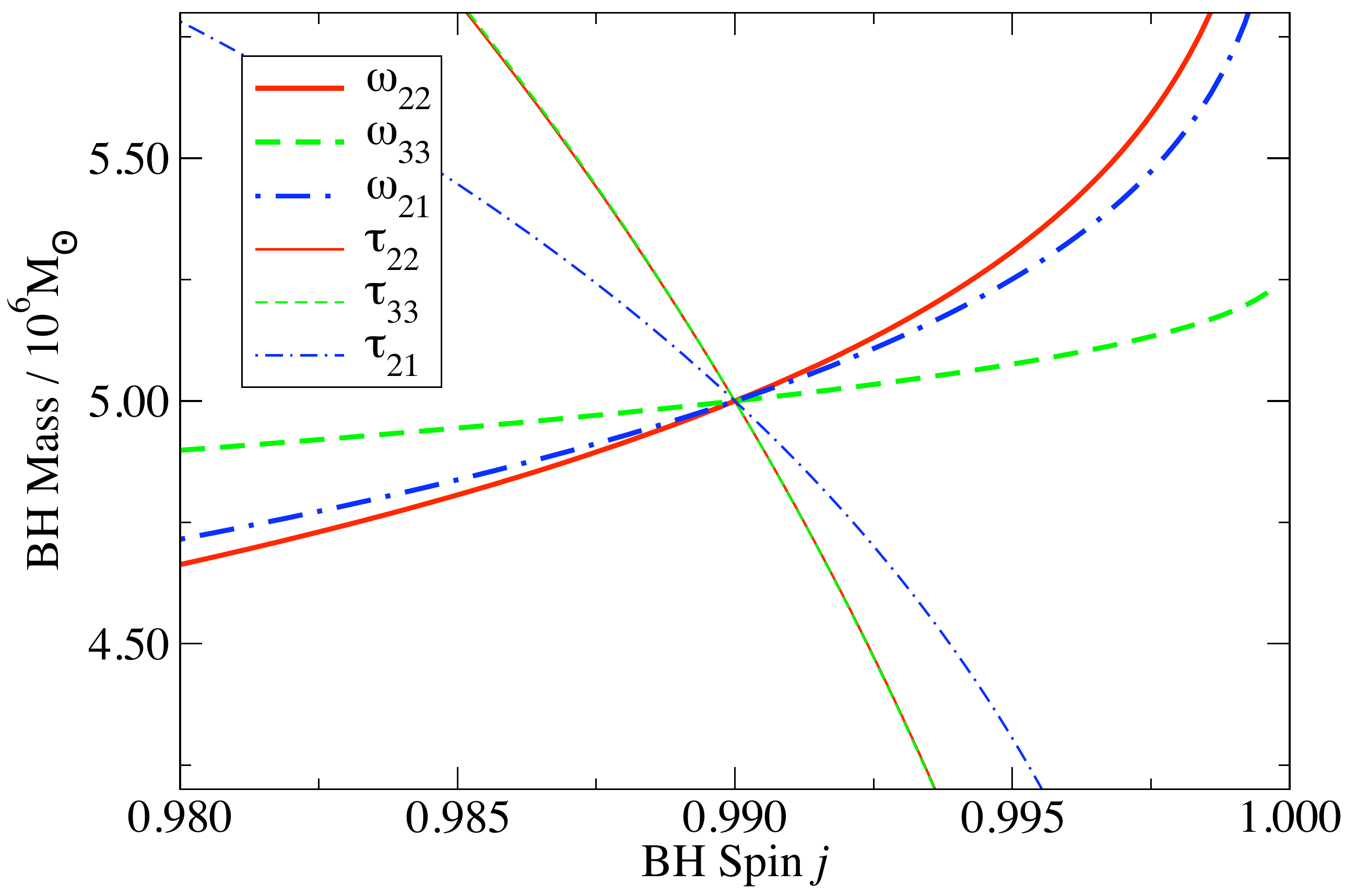}
\includegraphics[width=0.45\textwidth]{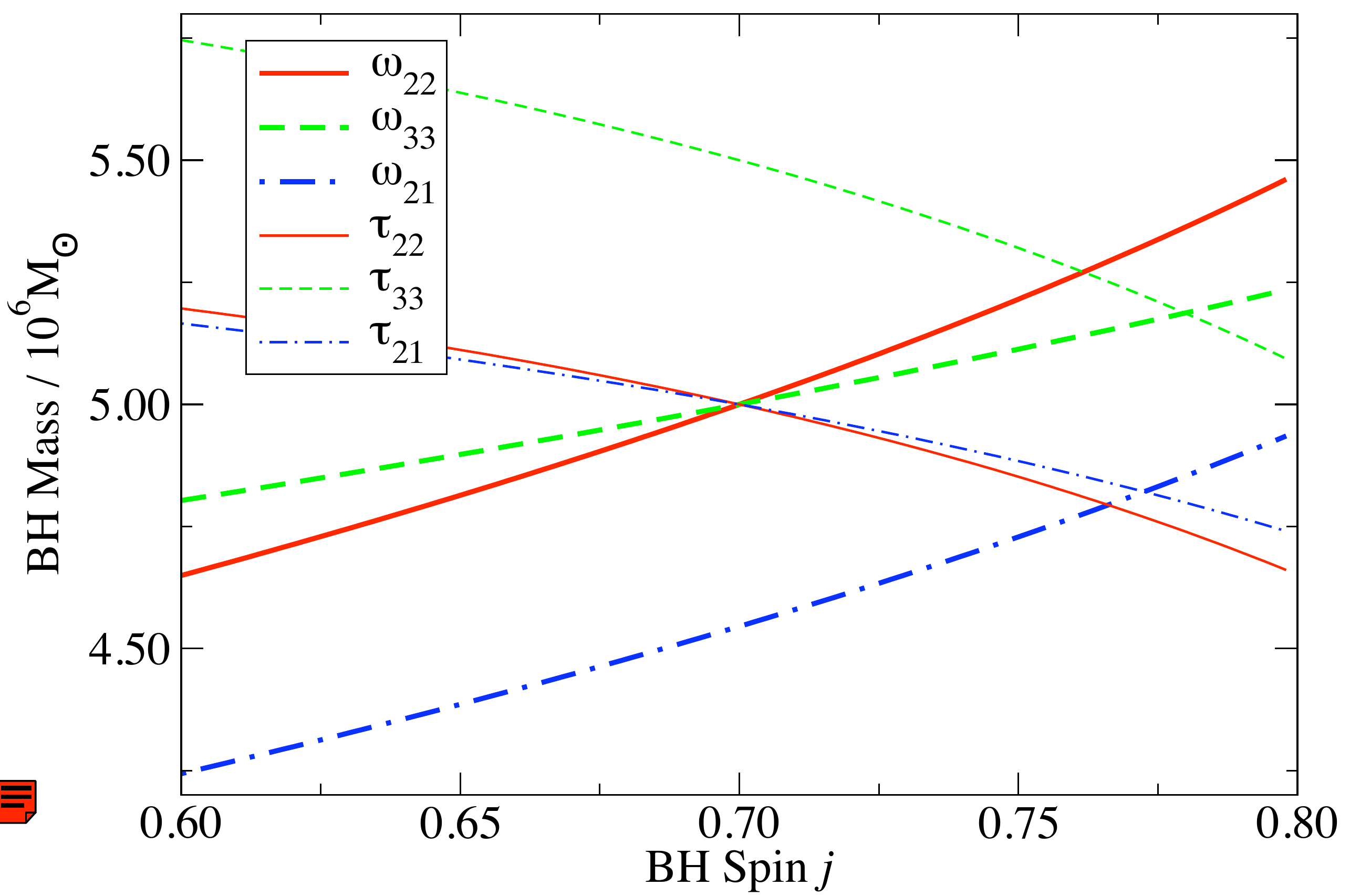}
\includegraphics[width=0.45\textwidth]{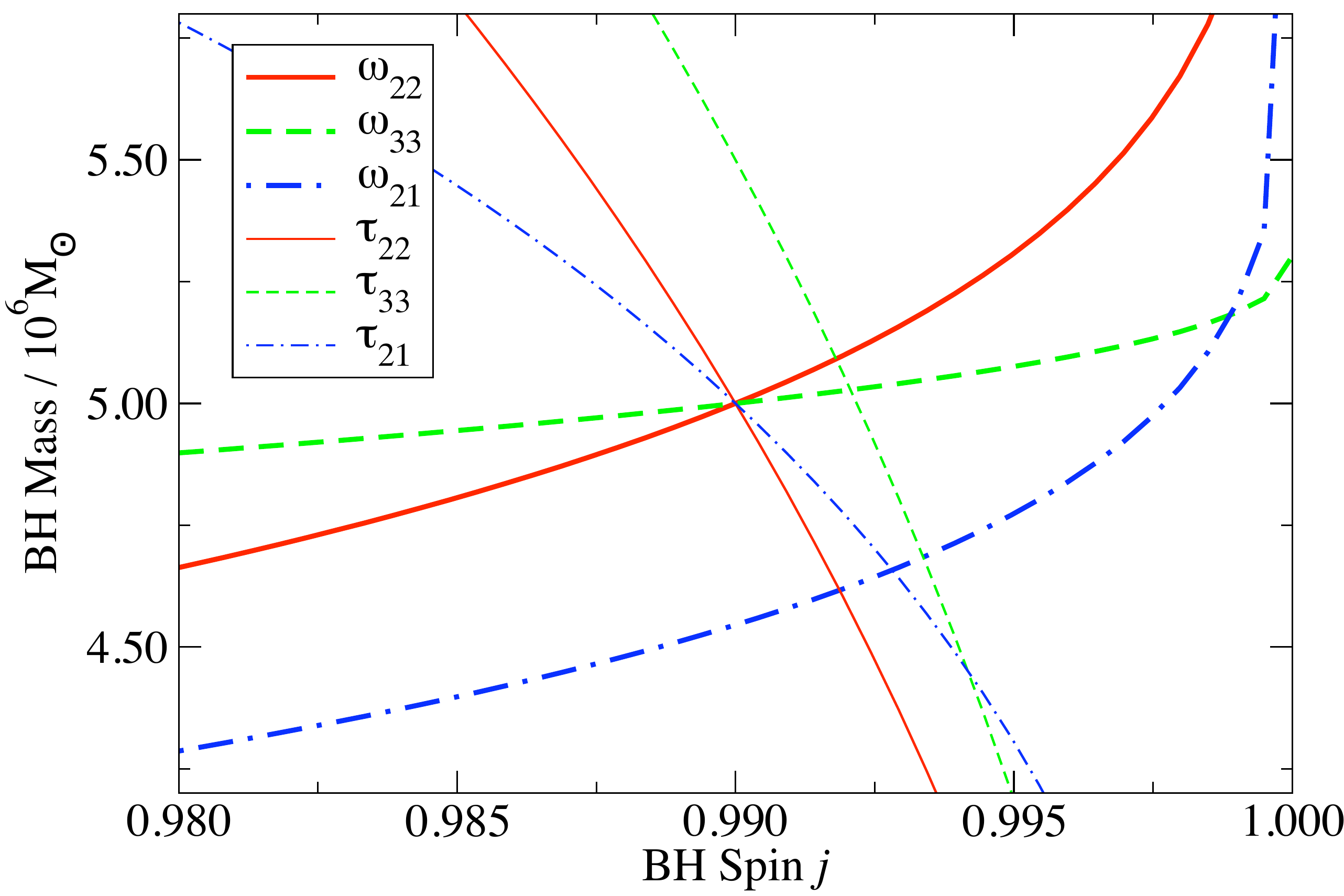}
\caption{Curves of constant mode frequencies and time-constants in the $(M,\,j)$-plane 
obtained with Eqs.\,(\ref{eq:test equations}) for the (2,2) mode and similar equations 
for other modes. Top panels correspond to the case when all measured values are exactly
as predicted by black hole perturbation theory. The bottom panels correspond to the
case where the (3,3) and (2,1) mode frequencies and time constants differ from the GR value by
10\% but the (2,2) frequency and time-constant are as in GR. An interesting thing to note is that for 
a large range of values of the black hole spin, the curves of constant $\tau_{22}$ and $\tau_{33}$ 
almost overlap, providing the opportunity for a more accurate test of the no-hair theorem. (see the text for details.)}
\label{fig:gr test max}
\end{figure*}
In this section we will consider a practical implementation of testing
general relativity using quasi normal modes.
To test general relativity with quasi-normal modes, it is 
{\em not} necessary to consider all the {\em physical} parameters but 
only those that are necessary to fully characterize the shape of the
signal.  The ringdown signal composed of a superposition of $n$ 
quasi-normal modes can be written as
\begin{equation}
h(t) = \sum_{\ell,m} A_{\ell m} e^{-t/\tau_{\ell m}} 
\cos(\omega_{\ell m} t + \gamma_{\ell m} ),
\label{eq:gr test signal}
\end{equation}
where there are $n$ each of the amplitudes $A_{\ell m},$ time-constants
$\tau_{\ell m},$ frequencies $\omega_{\ell m}$ and phases $\gamma_{\ell m}.$
The signal is, therefore, characterized by a set of $4n$ parameters in total. 

\subsection{Maximal set}
The most exhaustive test of the no-hair theorem would be to treat all $4n$
parameters to be independent. Of the $4n$ parameters, only the $2n$ mode
frequencies and time-constants would facilitate the test, the others should
be retained in order to fully capture the covariances and variances in the 
$2n$ test parameters. The consistency among every mode frequency and time constant
makes the test more stringent but the presence of a large number of parameters
(when $n$ is greater than 2) weakens the test. The reason for the latter is that
a model with too many parameters will/should be penalized for its flexibility by 
any carefully formulated test. In a covariance matrix formulation of the test,
this will be reflected by large variances in $\tau_{\ell m}$ and $\omega_{\ell m}$
(which are our test parameters) and in a Bayesian model selection a model
with a larger number of parameters will suffer from having a large 
evidence. 

How would the test work in practice?  For each measured test parameter (and 
the associated error in its measurement), one could draw a curve (or a band 
including the error) in the $(M,\,j)$ plane, by using their expressions in 
general relativity in terms of the mass and spin of the  black hole.  If the 
curves/bands fail to intersect at a single point/region in the $(M,\,j)$ plane 
then that would invalidate general relativity or, alternatively, indicate that 
the object is not a black hole. The maximal set could be weak due to the large 
variances of the various parameters but strong because many different bands
have to pass through the same region. 

Let us illustrate how the test works with some examples.  Let us suppose our 
signal model consists of a superposition of $(\ell,\,m)=(2,\,2), (2,\,1), 
(3,\,3), (4,\,4)$ modes. In this case the maximal set contains 16 parameters, 
of which 8 mode frequencies and time-constants are the test parameters. Let us 
denote by $\hat\omega_{\ell m}$ and $\hat\tau_{\ell m},$ the values of the mode 
frequencies and time-constants {\em measured} by projecting the 
data onto a superposition of quasi-normal modes as in Eq.\,(\ref{eq:gr test signal}).
For each measured parameter, we can construct an equation in $(M,\,j)$ using
the relation between the parameter and $(M,\,j)$ given in
Eqs.\,(\ref{eq:fit22})-(\ref{eq:fit44}):
\begin{eqnarray}
\hat\omega_{22} & = & \frac{1}{M}\left [ 1.5251 - 1.1568 (1 - j)^{0.1292}\right ],\nonumber\\
\hat\tau_{22} & = & \frac{2}{\omega_{22}}\left [ 0.7000 + 1.4187 (1 - j)^{-0.4990}\right ],
\nonumber\\
\label{eq:test equations}
\end{eqnarray}
and similar equations for other modes. (Note that we have used $f_{22}=M\omega_{22}$
and $2\,Q_{\ell m} =\tau_{\ell m}\omega_{\ell m}$ in rewriting these equations as
dimensionful quantities.) Measurement errors can be folded into the analysis by
using 
\begin{eqnarray}
\hat\omega_{22} \pm \sigma_{\omega_{22}} & = & \frac{1}{M} \left [ 1.5251 - 1.1568 (1 - j)^{0.1292}\right ],\nonumber\\
\hat\tau_{22} \pm \sigma_{\tau_{22}} & = & \frac{2}{\omega_{22}}\left [ 0.7000 + 1.4187 (1 - j)^{-0.4990} \right ], \nonumber\\
\label{eq:test equations with errors}
\end{eqnarray}
with similar equations for other modes.  

For a $5\times 10^6\,M_\odot$ black hole, the mode frequencies and 
time-constants for three different spin values $j=0.1,\,0.7,\,0.9,\,0.99,$
are given in Table\,\ref{tab:mode frequencies}.
\begin{table}[h]
\caption{The frequencies $F_{\ell m}=\omega_{\ell m}/(2\pi)$ (in mHz) and 
time-constants $\tau_{\ell m}$ (in s) of the first four dominant modes for
a $5\times 10^6M_\odot$ black hole of different spin magnitudes $j.$}
\label{tab:mode frequencies}
\begin{tabular}{ccccccccc}
\hline
\hline
$j$ & $F_{22}$ & $F_{21}$ & $F_{33}$ & $F_{44}$ & $\tau_{22}$ & $\tau_{21}$ & $\tau_{33}$ & $\tau_{44}$  \\
\hline
0.10 \: & 2.48  \: & 2.43  \: &3.98  \: & 5.36  \: & 282  \: & 277  \: & 269  \: & 266\:   \\
\hline
0.70 \: & 3.46  \: & 2.96  \: &5.48  \: & 7.44  \: & 303  \: & 301  \: & 295  \: & 290\:   \\
\hline
0.90 \: & 4.30  \: & 3.30  \: &6.70  \: & 9.06  \: & 383  \: & 355  \: & 379  \: & 375\:   \\
\hline
0.99 \: & 5.73  \: & 3.66  \: &8.60  \: & 11.4  \: & 823  \: & 559  \: & 828  \: & 837\:   \\
\hline
\hline
\end{tabular}
\end{table}
If the ringdown signal is consistent with the formation of a black hole 
with spin magnitude, say, $j=0.7$ ($j=0.99$) then the mode frequencies and time-constants
would be precisely as in the 2nd (respectively, 4th) row of Table \ref{tab:mode frequencies},
modulo the errors in their measurement.  Therefore, curves defined by 
$\omega_{\ell m}(M,j)=\,  \hat\omega_{\ell m}$ and $\tau_{\ell m}(M,j)=\, \hat\tau_{\ell m}$ 
will all meet at a single point in the $(M,\,j)$ plane as 
in the upper two panels of Fig.~\ref{fig:gr test max}, the point of 
intersection giving the mass and spin magnitude of the black hole.
The left panel corresponds to the formation of a black hole of spin magnitude
$j=0.7$ and the right panel to $j=0.99,$ in both cases $M=5\times 10^6\,M_\odot.$
If, however, one of the mode frequencies, say $\omega_{33}$ is different
from the general relativistic value by 10\%, then the corresponding curves
would fail to meet as shown in the lower two panels of Fig.~\ref{fig:gr test max}.

If general relativity is true then some of the curves lie almost one on top of the
other (e.g. $\omega_{22}=\rm const.$ is almost identical to $\omega_{21}=\rm const.$
in the $j=0.7$ case and $\tau_{22}=\rm const.$ is identical to $\tau_{33}=\rm const.$
both in the $j=0.7$ and $j=0.99$ cases) but even a slight departure from general
relativity will lead to big departures as demonstrated by the lower panels of 
the same figure. In any measurement the parameters are subject to statistical
and systematic errors that must be folded into the analysis which will be
taken up in a forthcoming study.

Do we need to treat all time-constants and mode frequencies to be independent 
in a test of the no-hair theorem?  We shall argue below that it is not necessary 
to treat all $4n$ parameters
to be independent; in fact, we shall see that the parametrization is not 
unique, offering a lot of flexibility in testing GR.

\subsection{Minimal set}
The minimal, or the simplest, model would consist of the smallest 
number of parameters needed to check the consistency between the
modes as predicted by the no-hair theorem, yet large enough to
capture all the variances and covariances between the parameters of
interest. Since the mode frequencies and time-constants are all 
determined in GR by the mass $M$ and spin magnitude $j$ of the black hole, 
the smallest number of parameters required to test GR would be three: these 
could be any three time-constants or two time-constants and one
mode frequency, and so on.  It would then be necessary to express the
other mode frequencies and time-constants in terms of {\em any two} of
the three parameters that were taken to be independent. 
Two of the three independent parameters could be used to solve for $(M,\,j).$
One could then see if the measured value of the third parameter is 
consistent with its predicted value based on the values of $M$ and $j.$ 
Let us note that without a prior knowledge of the amplitudes 
$A_{\ell m}$ and phases $\gamma_{\ell m}$ it will not be possible to measure 
the chosen time-constants and mode frequencies as they would induce covariances
that cannot be neglected in estimating the errors incurred in their measurement.
Thus, we (tentatively) conclude that the minimal set required for a signal model 
with $n$ modes would be $2n+3.$ However, the three test parameters can be chosen
in any way one wishes but choosing only three assures that the error in
their measurement is the smallest. In this sense the minimal set could be 
a very stringent test of GR. 

In reality, of course, the amplitudes of different modes are determined 
by the physical parameters $(M,\,j,\,q,\,D_{\rm L},\, \theta,\,\varphi,\,\psi,\,\iota)$
(cf. the discussion at the beginning of this section).
We can assume the parameters $(q,\,D_{\rm L},\, \theta,\,\varphi)$ to be known
from the inspiral phase and $(M,\,j)$ to be determined by the time-constants
and/or mode frequencies of the ringdown signal. This leaves the two angles 
$(\psi,\,\iota).$ Thus, when $n$ is greater than 2, it is not necessary to consider
all mode amplitudes to be independent but just two of them. Thus the minimal set
of parameters to be considered for testing the no-hair theorem is $n+5$ ($n$ phases,
$\gamma_{\ell m},$ the 3 test parameters and the two angles $\psi$ and $\iota$).

\section{Conclusions and future work}
\label{sec:CON}

In this paper, we have explored what information can be extracted from a 
black hole's ringdown signal, wherein the perturbation is caused by the 
tidal deformation produced during the merger of two non-spinning black holes.
To this end we used numerical simulations of the late inspiral and ringdown 
to estimate the relative amplitudes of the various modes excited.
The simulations consisted of initially non-spinning black holes in 
quasi-circular orbits for several mass ratios, ranging from 1:1 to
1:11.
  
We find that several modes have large enough luminosity --- or signal-to-noise 
ratio --- to be detectable in LISA and ET. Specifically, in order of decreasing 
power, modes (2,2), (3,3), (2,1), (3,2), (4,4), (5,5), (4,3), (6,6), (5,4) and (4,2) 
for LISA and the first four to five modes for ET, have significant luminosities.
Note that not all of these can necessarily be resolved, but it probably is the 
case for the first three (see below). In the analysis though, we decided to 
include only (2,2), (3,3) and (2,1), mainly because the available data were 
most accurate for those modes. 

We argued that the ringdown signal depends on the mass ratio of the progenitor 
binary and that this can be measured, with an error that is estimated from a 
Fisher matrix analysis. We showed how the luminosities change with the mass ratio. 
Indeed, by constructing fits to the mode amplitudes in terms of the mass ratio, 
we were able to include this effect in the analysis and estimate the errors 
involved in the measurement of various parameters. An important issue was to 
determine the epoch when the ringdown phase starts, so as to evaluate the relative 
amplitudes at that point. This was taken to be the point where each 
mode's frequency stops having an upward trend. The epoch at which the peak 
luminosity is reached is slightly different for different modes, but the mode 
amplitudes were all measured at a time $10M$ after the peak luminosity of 
the 22 mode.  

We computed the measurement errors of a number of other parameters for ringdowns 
observed in LISA and ET. These include the black hole mass, its spin, luminosity 
distance and inclination angle and how these vary with the final BH 
mass and the binary mass ratio. If we do not consider initial BH spins, 
the inclination angle can be measured solely by observing the three most 
dominant modes of the ringdown waveform, assuming that these can be resolved.  

Together, LISA and ET will be able to provide ample evidence for the 
distribution of supermassive, intermediate mass and stellar mass black 
holes, for a large part of the known Universe. For most of the parameter 
space, the reach of LISA for ringdowns is $z\sim 6,$ while for ET at least 
$z\sim 0.8$.  Also, by being able to measure the mass ratio, hints on 
the merger history and formation of black holes of a large range of masses 
could be inferred by studying ringdown signals. 

A practical implementation for testing the no-hair theorem and deciding 
the nature of the compact object that results from the merger, was presented, 
illustrating several key components. We started by providing a general 
framework, which is based on the number of parameters necessary to apply 
the test. Specifically, using $n$ modes, a generic test will 
use the $2n$ frequencies and time-constants.  This test is implemented 
by plotting the $\omega_{\ell m}$ and $\tau_{\ell m}$ curves on the 
mass-spin plane of the final black hole, where all the curves should 
intersect inside the same region if the object is a black hole. A key 
point was that some of the curves are special, as they almost overlap 
and can be thus used to check for small deviations from general relativity.

Future work should extend the study to include numerical simulations of initially 
spinning black holes, as this corresponds to a more realistic scenario. 
Additionally, there is effort to produce more accurate numerical simulations
so that less dominant modes could be studied when the mass ratio is large.

The current study has not investigated the question of decomposition of modes 
of a ringdown signal in real data.  The SNRs of different modes,
especially in the case of LISA, suggest that it should be possible to resolve
the modes and carry out the proposed tests of general relativity. 
However, a more in-depth investigation needs to be done, for instance by using
a Bayesian model selection to discriminate between different models. Given some 
prior information, the relative probability of two different multimode ringdown 
waveforms - injected in Gaussian and stationary noisy data - could be computed.  

Finally, a Fisher matrix-based parameter estimation approach (such as the one 
presented in this work) may not be robust or accurate enough, especially when 
the parameter space is large and the signal-to-noise ratios are low \cite{Vallisneri07}. 
For this reason, future studies should also involve parameter estimation in the context 
of the afore mentioned Bayesian analysis, as well as a Bayesian approach to a 
test of the no-hair theorem. In the latter case the posterior joint probabilities 
of two different models, given the `initial data', could be compared. The 
first hypothesis would be that the observed object is a black hole, while the 
second that either GR is incorrect or the merged object is not a black hole.   

\begin{acknowledgments}
We are grateful to Alessandra Buonanno, Vitor Cardoso, Thomas Dent, Ajith Parameswaran and Evan Ochsner 
for a careful reading of the manuscript and comments that helped improve parts of the paper. BSS was funded by the
Science and Technology Facilities Council (STFC) Grant No. ST/J000345/1 and MH
by STFC Grant No. ST/H008438/1.
\end{acknowledgments}
 
\bibliography{ref-list}
\end{document}